\documentclass[aps,amsmath,amssymb,nofootinbib,prd,preprintnumbers,showpacs,twocolumn,10pt]{revtex4-2}

\usepackage[dvipsnames]{xcolor}
\usepackage{bm}
\usepackage{cancel}
\usepackage{cases}
\usepackage{CJKutf8}
\usepackage{dcolumn}
\usepackage{graphicx}
\usepackage{slashed}
\usepackage{subfigure}
\usepackage{textcase}
\usepackage{ytableau}
\usepackage[mathlines]{lineno}
\usepackage[colorlinks=true,pdfstartview=FitV,linkcolor=blue,citecolor=blue,urlcolor=blue]{hyperref}
\usepackage[compat=1.1.0]{tikz-feynhand}

\DeclareMathOperator{\tr}{tr}

\let\Im\relax
\DeclareMathOperator{\Im}{Im}

\begin{document}

\newcommand{\beq}{\begin{equation}}
\newcommand{\eeq}{\end{equation}}
\newcommand{\beqs}{\begin{eqnarray}}
\newcommand{\eeqs}{\end{eqnarray}}
\newcommand{\kh}[1]{{\textcolor{blue}{#1}}}
\newcommand{\rj}[1]{{\textcolor{Green}{#1}}}
\newcommand{\rt}[1]{{\textcolor{Magenta}{#1}}}
\newcommand{\del}[1]{{\color{red}\sout {#1}}}
\newcommand{\cjk}[1]{\begin{CJK}{UTF8}{ipxm}#1\end{CJK}}

\title{
Leptogenesis in the presence of density perturbations
}

\author{Kenta Hotokezaka$^{1}$}
\thanks{{\color{blue}kentah@resceu.s.u-tokyo.ac.jp}}

\author{Ryusuke Jinno$^{2}$}
\thanks{{\color{blue}jinno@phys.sci.kobe-u.ac.jp}}

\author{Rin Takada$^{1}$}
\thanks{{\color{blue}takada-rin@resceu.s.u-tokyo.ac.jp}}

\affiliation{$^{1}$
Research Center for the Early Universe (RESCEU), Graduate School of Science, The University of Tokyo, 7-3-1 Hongo, Bunkyo, Tokyo 113-0033, Japan
}

\affiliation{$^{2}$
Department of Physics, Graduate School of Science, Kobe University, 1-1 Rokkodai, Kobe, Hyogo 657-8501, Japan
}

\preprint{RESCEU-1/25}
\preprint{KOBE-COSMO-25-01}

\date{\today}

\begin{abstract}
We point out a new effect on the freeze-out process of heavy particles induced by density perturbations in the early universe, which we call ``acoustically driven freeze-out.''
This beyond-linear effect is caused by the exponential decoupling of heavy particles from the thermal bath in the presence of density perturbations, and already at moderately large values $\delta T / \bar{T} = \mathcal{O} (10^{-2})$ it cannot be captured by linear perturbation theory.
We illustrate this effect with leptogenesis taking the decay and inverse decay of heavy neutrinos into account, and discuss its phenomenological implications.
We found that perturbations always enhance the (spatially averaged) values of the final lepton asymmetry, and as a result, constraints on the mass of heavy neutrinos are found to be relaxed in the presence of perturbations.
\end{abstract}

\maketitle
\section{Introduction}

Inflation provides a successful explanation for the homogeneity, isotropy, and flatness of the universe~\cite{Starobinsky:1979ty,Sato:1981qmu,Kazanas:1980tx,Guth:1980zm,Linde:1981mu} as well as the origin of the large-scale structure~\cite{HAWKING1982295,Guth:1982ec,Starobinsky:1982ee,PhysRevD.28.679}.
One of its characteristic predictions is that it generates density perturbations over different scales.
While density perturbations are well constrained at large-scales from the Cosmic Microwave Background (CMB)~\cite{Planck:2018vyg} and large-scale structure observations~\cite{eBOSS:2020yzd}, those at small scales are yet to be explored.
These scalar fluctuations re-enter the Hubble horizon after inflation, driving acoustic oscillation of density and temperature.

Out-of-equilibrium phenomena play central roles in the history of the universe.
One of the prominent examples is the production of baryon asymmetry, which requires out-of-equilibrium interactions among the Sakharov's three conditions~\cite{Sakharov:1967dj}.
Other examples include preheating after inflation~\cite{Kofman:1994rk, Kofman:1997yn, Greene:1997fu} and freeze-out/freeze-in of dark matter~\cite{McDonald:2001vt,Hall:2009bx},
as well as Standard Model (SM) processes such as big bang nucleosynthesis (BBN) and recombination.
Among these out-of-equilibrium processes, the effect we point out occurs when heavy particles decouple from the thermal bath.
In the presence of temperature fluctuation $\delta T$, the Boltzmann factor for the equilibrium abundance of such particles with mass $M$ is given by
\begin{align}
\exp (- E / T) 
&= \exp [- E / (\bar{T} + \delta T)]\notag\\
&= \exp (- E / \bar{T}) \exp (E \delta T / \bar{T}^2).
\label{eq:I-0}
\end{align}
When $E \simeq M$ is much larger than $\bar{T}$, the exponent of the second factor in the right-hand side of equation (\ref{eq:I-0}) is not necessarily smaller than unity, invalidating truncation of the system at the first order in perturbation.
The actual abundance of this particle species also closely follows this equilibrium value around the decoupling time, and hence the final abundance of this species itself or its decay product cannot also be predicted with truncation at the first order.

Baryogenesis is one of the phenomena in which this beyond-linear effect can potentially play an important role, and in this paper we illustrate the effect with one the simplest baryogenesis scenarios.
{\it Leptogenesis} \cite{Fukugita:1986hr} is among the most successful scenarios, in which the Sakharov's three conditions are satisfied by the CP-violating decay of right-handed neutrinos together with the sphaleron process in the SM.
These right-handed neutrinos naturally exist in the UV completion of the SM such as grand unification theories, and can also naturally explain the masses of light neutrinos through the seesaw mechanism~\cite{Minkowski:1977sc,Yanagida:1979as,Gell-Mann:1979vob,Mohapatra:1979ia}.
The masses, mixing angles, and CP phase(s) of these light neutrinos are being explored with solar, reactor, atmospheric and accelerator neutrino oscillation experiments (see e.g., Ref.~\cite{Esteban:2020cvm}), while their Dirac or Majorana nature may be determined in the current or future $0\nu\beta\beta$ decay experiments (see e.g., Ref.~\cite{Dolinski:2019nrj}).

Density perturbations at small scales, on the other hand, are gaining interest in view of gravitational-wave (GW) observations~\cite{LIGOScientific:2014pky,VIRGO:2014yos,KAGRA:2020tym,LISA:2017pwj,Punturo:2010zz,LIGOScientific:2016wof}.
They give rise to multipole structures in the matter distribution, and for sufficiently large amplitude they produce secondary GWs~\cite{Ananda:2006af,Baumann:2007zm} and lead to the formation of primordial black holes~\cite{Carr:1974nx,Hawking:1971ei,Zeldovich:1967lct}.
Part of the motivations of the current paper is re-assessing the effect of inhomogeneous universe on particle physics processes, demonstrating the interplay of micro- and macro-physics.
In this perspective, we investigate the influence of density perturbations on the lepton and baryon asymmetries within the simplest leptogensis scenarios.
As we see in the following, sound waves induce a sudden freeze-out of heavy particles, resulting in a net enhancement in the particle abundance after freeze-out even after spatial average.\footnote{
The effects of density perturbations on baryon asymmetry is partially studied in Ref.~\cite{Kartavtsev:2008fp}, where the authors consider perturbations of the same order of magnitude as CMB and find the effect to be negligibly small.
The present analysis is different from theirs in that we calculate the spatial average of the resulting baryon asymmetry, which should be the relevant quantity for the observed baryon asymmetry.
Another difference is that we consider relatively large density perturbations, which we will find to have a net effect on the final averaged baryon abundance.
}
We call this process ``acoustically driven freeze-out.''

This paper is organized as follows.
In Section \ref{sec:II}, we derive the Boltzmann equations for the heavy neutrino $N_1$ and for the baryon-minus-lepton number $B-L$, in the presence of density perturbations.
We also review various parameters related to the neutrino masses.
In Section \ref{sec:III}, we discuss the intuitive picture for the acoustically driven freeze-out, and then present the time evolution of the right-handed neutrino and $B-L$ as well as the parameter space for successful baryogenesis.
Finally in Section \ref{sec:V} we discuss the implications of our results and present possible future directions.

\section{Leptogenesis in the presence of cosmological perturbations}
\label{sec:II}

\subsection{Prerequisites for leptogenesis}

{\it Leptogenesis}~\cite{Fukugita:1986hr} is a scenario in which the CP-violating decays of heavy right-handed neutrinos, together with the sphaleron process in the SM, generate the baryon asymmetry of the universe.
In leptogenesis, the SM is extended with right-handed neutrinos with heavy Majorana masses that violate the lepton number.
These right-handed neutrinos have Yukawa interactions with the Higgs and light leptons, whose complex phases give rise to CP violation.
Their CP-violating decays produce lepton asymmetry, part of which is then converted into baryon asymmetry through the SM sphaleron process.
The Majorana masses of the right-handed neutrinos at the same time explain the observed light masses for left-handed neutrinos via the seesaw mechanism.

Leptogenesis scenarios can be classified into two classes: thermal and non-thermal (see, e.g., \cite{Buchmuller:2004nz,Giudice:1999fb}).
The former assumes thermal abundance for the right-handed neutrinos as one of the initial conditions, while in the latter the right-handed neutrinos are produced via various non-thermal processes.
One of the attractive features of the former class is its robust predictivity: the number of free parameters is significantly smaller, thereby allowing for more robust experimental predictions.
For this reason we focus on thermal leptogenesis in this paper.

We consider a simple model so-called vanilla leptogenesis, in which heavy right-handed neutrinos $N_1$, $N_2$, and $N_3$ with masses in the order $M_1 < M_2 < M_3$ are added to the SM.
In this model, the Majorana mass terms and Yukawa couplings of the right-handed neutrinos
\beq
\label{eq:I-4}
\mathcal{L}=-\dfrac{M_{\alpha}}{2}\bar{N}_{\alpha}N_{\alpha}
-y_{\alpha\beta}\phi^{\ast}\bar{l}_{\alpha}N_{\beta}
+{\rm h.c.},
\eeq
are added to the SM Lagrangian together with their kinetic terms.
Here $M_{\alpha}$ and $y_{\alpha\beta}$ denote the Majorana masses and Yukawa couplings, respectively, while $\phi$ and $l_{\alpha}$ are the Higgs and lepton doublets of the SM, respectively.
The summation over lepton flavors $\alpha,\beta=1,2,3$ is implicit.
The light neutrino masses are generated from the type-I seesaw mechanism~\cite{Minkowski:1977sc,Yanagida:1979as,Gell-Mann:1979vob,Mohapatra:1979ia}.

The lightest right-handed neutrino $N_1$ decays into a lepton doublet and Higgs doublet pair or its CP-conjugate:
\beq
\label{eq:I-5}
N_1\to l_{\alpha}\phi,\qquad
N_1\to\bar{l}_{\alpha}\phi^{\ast}.
\eeq
In this paper, only the decay and inverse decay of $N_1$ are included, and scattering processes are ignored.
The decays of $N_2$ and $N_3$ are assumed to be negligible in the final asymmetry, and they appear only as internal lines in the diagrams of $N_1$ decay.
This simplified model is sufficient to capture the beyond-linear effect we point out in the present study.
The $B-L$ asymmetry is generated from the CP-violating decays (\ref{eq:I-5}) as
\beq
\label{eq:I-6}
\begin{array}{l}
\dfrac{\Gamma(N_1\to l_{\alpha}\phi)}
{\Gamma(N_1\to l_{\alpha}\phi)+\Gamma(N_1\to\bar{l}_{\alpha}\phi^{\ast})}
=\dfrac{1+\varepsilon_1}{2},\\[1.25em]
\dfrac{\Gamma(N_1\to\bar{l}_{\alpha}\phi^{\ast})}
{\Gamma(N_1\to l_{\alpha}\phi)+\Gamma(N_1\to\bar{l}_{\alpha}\phi^{\ast})}
=\dfrac{1-\varepsilon_1}{2}.
\end{array}
\eeq
Here the parameter $\varepsilon_1 (<0)$ describing CP asymmetry is given by (see Appendix~\ref{Appendix:C})
\beq
\label{eq:II-CPA}
\begin{array}{l}
\varepsilon_1
=\displaystyle
\sum_{\gamma=2,3}\dfrac{\Im(y_{\alpha 1}y_{\beta 1}y_{\alpha\gamma}^{\ast}y_{\beta\gamma}^{\ast})}
{8\pi\sum_{\alpha}|y_{\alpha 1}|^2}
f(x),
\qquad
x\equiv\dfrac{M_{\gamma}^2}{M_1^2}
\\[1.75em]
f(x)\equiv\sqrt{x}\,\biggl[1+\dfrac{1}{1-x}
-(1+x)\ln\biggl(1+\dfrac{1}{x}\biggr)\biggr].
\end{array}
\eeq

In the following, we consider the time evolution of the ratios $N_{N_1}\equiv n_{N_1}/n_{\gamma}$ and $N_{B-L}\equiv (n_B-n_L)/n_{\gamma}$, where $n_{N_1}$, $n_{\gamma}$, $n_{B}$, and $n_L$ denote the number densities of right-handed neutrinos, photons, baryons, and leptons, respectively.
Their initial conditions are given at sufficiently high temperatures $T\gg M_1$ by
\beq
\label{eq:I-7}
N_{N_1}^{\rm eq}(T\gg M_1)
=\dfrac{n_{N_1}^{\rm eq}}{n_{\gamma}^{\rm eq}}
=\dfrac{3}{4},
\qquad
N_{B-L}^{\rm eq}(T\gg M_1)
=0.
\eeq
After the decay of right-handed neutrinos, the negative lepton number $L$ is converted into positive $B$ and negative $L$ through the sphaleron process~\cite{Manton:1983nd,PhysRevD.30.2212}.
Since $B-L$ is conserved in this process, the final baryon number $B$ is related to the value of $B-L$ generated by $N_1$ decay as~\cite{Khlebnikov:1988sr,Harvey:1990qw}
\beq
\label{eq:I-sphaleron}
B=\dfrac{8\nu_f+4\nu_s}{22\nu_f+13\nu_s}(B-L)=\dfrac{28}{79}(B-L),
\eeq
where $\nu_f = 3$ is the number of fermion generations and $\nu_s = 1$ is the number of Higgs doublets.

\subsection{Prerequisites for cosmological perturbations}

In an inhomogeneous universe, the time evolution of the right-handed neutrinos and baryon and/or lepton numbers occurs in the presence of cosmological perturbations.
In this paper, we follow Refs.~\cite{Kodama:1984ziu,Baumann:2022mni} for the formulation of cosmological perturbations.

The background metric is given by the Friedmann-Lema\^{i}tre-Robertson-Walker (FLRW) metric
\beq
\label{eq:II-1}
ds^2=g_{\mu\nu}dx^{\mu}dx^{\nu}
=-dt^2+a^2\delta_{ij}dx^idx^j,
\eeq
where $g_{\mu\nu}$ is the metric tensor and $a = a(t)$ is the scale factor of the universe.
We assume that the universe is radiation-dominated in the relevant epoch, and in this case the Friedmann equation is
\begin{align}
3 M_{\rm P}^2 H^2 = \rho_r,
\end{align}
with
\begin{align}
H = \frac{1}{2 t}, \qquad \rho_r = \frac{\pi^2}{30} g_* T^4.
\end{align}
Here $M_{\rm P} = 1/\sqrt{8 \pi G}$ is the reduced Planck mass, $H\equiv\dot{a}/a=a'/a^2$ is the Hubble parameter, and $\rho_r$ is the radiation energy density.
During the radiation-dominated epoch, the conformal time $\eta$ defined by $d\eta=dt/a$ is related to the cosmological time as $\eta\propto t^{1/2}$.
In what follows the dot denotes differentiation with respect to the physical time $t$, and the prime denotes differentiation with respect to the conformal time $\eta$.

The scalar perturbations of the metric tensor $g_{\mu\nu}$ are expressed in terms of four independent functions of time and space, $A$, $B$, $C$, and $E$, as~\cite{Baumann:2022mni}
\begin{align}
g_{00}
&=-a^2(1+2A),\label{eq:II-8}\\
g_{0i}
&=-a^2B_{,i},\label{eq:II-9}\\
g_{ij}
&=a^2\biggl[(1+2C)\delta_{ij}
+2\biggl(
E_{,ij}-\dfrac{1}{3}\delta_{ij}\nabla^2E
\biggr)
\biggr].\label{eq:II-10}
\end{align}
On the other hand, perturbation in the matter sector appears in the temperature fluctuation $\delta_T$ defined by
\beq
\label{eq:II-11}
T=\bar{T}+\delta T,\qquad
\delta_T\equiv\dfrac{\delta T}{\bar{T}},
\eeq
where $\bar{T}$ is the spatially averaged temperature of the universe.
With these definitions, it is useful to introduce new variables:
\begin{align}
\bar{z}&\equiv\dfrac{M_1}{\bar{T}},\label{eq:II-12}\\
z_T&\equiv\dfrac{M_1}{\bar{T}+\delta T}
=\dfrac{\bar{z}}{1+\delta_T}.\label{eq:II-13}
\end{align}
In the following sections, we will use $\bar{z}$ and $z_T$ instead of physical time $t$ and physical temperature $T$, respectively.

\subsection{Boltzmann equation}

The evolution of the number density of each particle species in the presence of collision processes is governed by the Boltzmann equation.
After integrating over the momentum space, it takes the form (see Appendix~\ref{Appendix:A})
\beq
\label{eq:II-103}
{\mathcal{N}^{\mu}}_{;\mu}
=\dfrac{g_{\rm deg}}{(2\pi)^3}\int\dfrac{d^3p}{E}\;C[f],
\eeq
where $\mathcal{N}^{\mu}$ is the number density current defined by $\mathcal{N}^{\mu}\equiv nU^{\mu}$ with $n$ and $U^{\mu}$ being the number density and the 4-velocity, respectively, $g_{\rm deg}$ is the degrees of freedom of the species, $f$ is the distribution function, and $C[f]$ is the collision term.
The left-hand side of equation (\ref{eq:II-103}) can be written as
\beq
\label{eq:II-105}
(nU^{\mu})_{;\mu}
=\bar{z}H(1-A)
\biggl(\dfrac{dn}{d\bar{z}}+\dfrac{3}{\bar{z}}n\biggr)
+\dfrac{n}{a}({v^i}_{,i}+3C'),
\eeq
where $v^i$ is the 3-velocity, the latter of which is first order in perturbation.
Here we used
\beq
\label{eq:II-104}
U^0=a^{-1}(1-A),
\qquad
U^i=a^{-1}v^i,
\eeq
and equations (\ref{eq:B-1})--(\ref{eq:B-9}).

For the right-handed neutrinos $N_1$, we take only the decay and inverse decay into account.
In this case, assuming that the right-handed neutrinos are in kinetic equilibrium, the thermal average of the right-hand side of equation (\ref{eq:II-103}) yields
\begin{align}
&\biggl\langle\dfrac{g_{N_1}}{(2\pi)^3}\int\dfrac{d^3p_{N_1}}{E_{N_1}}\;
C\biggr\rangle\notag\\
&=-\dfrac{4\pi^3g_lg_{\phi}|\mathcal{M}|^2}{M_1}\cdot
\dfrac{K_1(z_T)}{K_2(z_T)}\cdot
\bigl[n_{N_1}(\bar{z})-n_{N_1}^{\rm eq}(z_T)\bigr]\notag\\
&\equiv -\Gamma_D(\bar{z}=\infty)
\biggl\langle\dfrac{1}{\gamma}\biggr\rangle(z_T)
\bigl[n_{N_1}(\bar{z})-n_{N_1}^{\rm eq}(z_T)\bigr].\label{eq:II-106}
\end{align}
The collision term (\ref{eq:II-106}) is invariant under coordinate transformations: indeed, when the collision term is in the Local Inertial Frame Instantaneously at Rest with respect to the Comoving Observer (LIFIRCO), the cross sections have the same expressions as in the Minkowski space, and the collision term contains {\it no metric fluctuations}~\cite{Senatore:2008vi}.
Any metric fluctuation is contained in, and arises from, the Liouville term, i.e., the left-hand side of equation (\ref{eq:II-103}).
As a result, we obtain
\begin{align}
&\biggl(\dfrac{dn_{N_1}}{d\bar{z}}+\dfrac{3}{\bar{z}}n_{N_1}\biggr)
+\dfrac{n_{N_1}}{a\bar{z}H}({v^i}_{,i}+3C')\notag\\
&\simeq -(1+A)\dfrac{\Gamma_D(\bar{z}=\infty)}{\bar{z}H}
\biggl\langle\dfrac{1}{\gamma}\biggr\rangle(z_T)
\bigl[n_{N_1}(\bar{z})-n_{N_1}^{\rm eq}(z_T)\bigr]\notag\\
&=-(1+A)\bar{z}K\biggl\langle\dfrac{1}{\gamma}\biggr\rangle(z_T)
\bigl[n_{N_1}(\bar{z})-n_{N_1}^{\rm eq}(z_T)\bigr],\label{eq:II-107}
\end{align}
where $K\equiv\Gamma_D(\bar{z}=\infty)/\bar{z}^2H$ is the so-called decay parameter.
The evolution of the number density ratio $N_{N_1}=n_{N_1}(\bar{z})/n_{\gamma}^{\rm eq}(z_T)$ becomes
\beq
\label{eq:II-108}
\dfrac{dN_{N_1}(\bar{z})}{d\bar{z}}
=-(1+A)D(z_T)
\bigl[N_{N_1}(\bar{z})-N_{N_1}^{\rm eq}(z_T)\bigr],
\eeq
where
\beq
\label{eq:II-D}
D(z_T)\equiv K\bar{z}\biggl\langle\dfrac{1}{\gamma}\biggr\rangle(z_T)
=K\bar{z}\dfrac{K_1(z_T)}{K_2(z_T)},
\eeq
and
\beq
\label{eq:II-equilibrium}
N_{N_1}^{\rm eq}(z_T)=\dfrac{3}{8}z_T^2K_2(z_T).
\eeq

For $N_{B-L}$, the Liouville term is evaluated in the same way as before.
The collision term contains the sourcing of lepton asymmetry from the CP asymmetry $\varepsilon_1$ through the difference between the non-equilibrium and equilibrium abundances of right-handed neutrinos.
As a result, the evolution of $N_{B-L}$ is governed by
\begin{align}
\dfrac{dN_{B-L}(\bar{z})}{d\bar{z}}
&=-\varepsilon_1(1+A)D(z_T)
\bigl[N_{N_1}(\bar{z})-N_{N_1}^{\rm eq}(z_T)\bigr]\notag\\
&\quad
-(1+A)W_{\rm ID}N_{B-L},\label{eq:II-1010}
\end{align}
where
\beq
\label{eq:II-add-1}
W_{\rm ID}(z_T)
=\dfrac{1}{4}K\bar{z}z_T^2K_1(z_T),
\eeq
is the washout factor without scattering~\cite{Buchmuller:2004nz}.

To solve equations (\ref{eq:II-108}) and (\ref{eq:II-1010}), we choose the conformal Newtonian gauge $A=\Psi$, $C=-\Phi$ and $B = E = 0$.
The anisotropic stress is assumed to be negligible, and in this case $\Phi=\Psi$ holds.
Equations (\ref{eq:II-108}) and (\ref{eq:II-1010}) then respectively become
\beq
\label{eq:II-109}
\dfrac{dN_{N_1}(\bar{z})}{d\bar{z}}
=-(1+\Psi)D(z_T)\bigl[N_{N_1}(\bar{z})-N_{N_1}^{\rm eq}(z_T)\bigr],
\eeq
and 
\begin{align}
\dfrac{dN_{B-L}(\bar{z})}{d\bar{z}}
&=-\varepsilon_1(1+\Psi)D(z_T)
\bigl[N_{N_1}(\bar{z})-N_{N_1}^{\rm eq}(z_T)\bigr]\notag\\
&\quad
-(1+\Psi)W_{\rm ID}N_{B-L}.\label{eq:II-1010-add}
\end{align}
The time evolution of $\Psi=\Phi$ in a radiation dominated universe is given by~\cite{Baumann:2022mni}
\beq
\Psi(\eta,\bm{k})=2\mathcal{R}_i \dfrac{\sin\varphi-\varphi\cos\varphi}{\varphi^3},
\eeq
where $\mathcal{R}_i$ is the primordial curvature perturbation in wavenumber space, and $\bm{k}$ is the wavenumber with $k$ being its absolute value.
In a radiation-dominated epoch, $\varphi$ is expressed as
\beq
\label{eq:II-1015}
\varphi=\dfrac{k\eta}{\sqrt{3}}
\equiv\dfrac{\bar{z}}{\bar{z}_H}.
\eeq
The parameter $\bar{z}_H$ defined here can be regarded as specifying the time of horizon entry for each wavenumber.
Since $\Psi(\eta,\bm{x})$ is real, the Fourier transform can be written as
\begin{align}
\Psi(\eta,\bm{x})
&=\int\frac{d^3k}{(2\pi)^3}\mathrm{e}^{\mathrm{i}\bm{k}\cdot\bm{x}}\Psi(\eta,\bm{k})\notag\\
&=\int\dfrac{d^3k}{(2\pi)^3}\dfrac{1}{2}\left[\mathrm{e}^{\mathrm{i}\bm{k}\cdot\bm{x}}\Psi(\eta,\bm{k})+\mathrm{e}^{-\mathrm{i}\bm{k}\cdot\bm{x}}\Psi^*(\eta,\bm{k})\right]\notag\\
&=\int\dfrac{d^3k}{(2\pi)^3}\left[2|\mathcal{R}_i|\cos\delta\,\dfrac{\sin\varphi-\varphi\cos\varphi}{\varphi^3}\right].
\end{align}
Here $\delta\equiv\delta'+\bm{k}\cdot\bm{x}$ is the phase of the perturbation with $\delta'$ being the complex phase of $\mathcal{R}_i=|\mathcal{R}_i|\,\mathrm{e}^{\mathrm{i}\delta'}$.
Written this way, the time evolution of $\Psi=\Phi$ can be understood in terms of real functions only
\beq
\label{eq:II-1011}
\Psi(\eta,\bm{k})=2|\mathcal{R}_i|\cos\delta\,\dfrac{\sin\varphi-\varphi\cos\varphi}{\varphi^3}.
\eeq
For each wavenumber $\bm{k}$, different values of the phase parameter $\delta$ can be regarded as specifying different spatial points.
Since the final baryon asymmetry we observe is the spatial average, in our numerical analysis we take average of the final asymmetry with respect to the phase $\delta$.

The temperature fluctuation $\delta_T$ is related to 
the fluctuation in the radiation energy density $\delta_r$ as
\beq
\delta_r
=\dfrac{\delta\rho_r}{\bar{\rho}}
=4\dfrac{\delta T}{\bar{T}}
=4\delta_T.
\eeq
The evolution of $\delta_r$ is given by \cite{Baumann:2022mni}
\beq
\label{eq:II-1012}
\delta_r=-\dfrac{2}{3}(k\eta)^2\Psi-2\eta\Psi'-2\Psi,
\eeq
where we used $\Psi=\Phi$.
Substituting equation (\ref{eq:II-1011}) into equation (\ref{eq:II-1012}), we obtain
\begin{align}
\delta_r
&=8|\mathcal{R}_i|\cos\delta\,
\dfrac{\sin\varphi-\varphi\cos\varphi-\varphi^2\sin\varphi+\frac{1}{2}\varphi^3\cos\varphi}{\varphi^3},
\label{eq:II-1013}
\end{align}
and correspondingly
\begin{align}
\delta_T
&=2|\mathcal{R}_i|\cos\delta\,\dfrac{\sin\varphi-\varphi\cos\varphi-\varphi^2\sin\varphi+\frac{1}{2}\varphi^3\cos\varphi}{\varphi^3}.
\label{eq:II-1014}
\end{align}

\subsection{Neutrino mass parameters}

In our setup, the amount of baryon asymmetry depends on the heavy neutrino mass $M_1$, the decay parameter $K$, and the the degree of CP asymmetry $\varepsilon_1$. $M_1$ determines the temperature at which the abundance of $N_1$ starts to decline. $K$ determines the time (or, the value of $\bar{z}$) when $N_1$ decouples from the thermal bath. $\varepsilon_1$ controls the efficiency of the generation of $B-L$ via $N_1$ decay.
The latter two parameters $K$ and $\varepsilon_1$ depend on the neutrino masses, Yukawa couplings, and other parameters in the Lagrangian.
In this subsection, we introduce various neutrino mass parameters that are useful to interpret the results (see, e.g., Ref.~\cite{Buchmuller:2004nz} for details). 

The masses of light neutrinos are denoted by $m_1$, $m_2$, and $m_3$ ($m_1<m_2<m_3$).
The {\it effective neutrino mass} is defined by
\beq
\label{eq:III-0}
\tilde{m}_1\equiv\dfrac{(m_{\rm D}^{\dagger}m_{\rm D})_{11}}{M_1},
\eeq
where $(m_{\rm D})_{\alpha\beta}=y_{\alpha\beta}v$ with $v\simeq 174\,{\rm GeV}$ being the Higgs vacuum expectation value (VEV).
The decay parameter $K$ is proportional to $\tilde{m}_1$:
\beq
\label{eq:III-15}
K\equiv\dfrac{\tilde{m}_1}{m_{\ast}},\qquad
m_{\ast}=\sqrt{\dfrac{64\pi^3g_{\ast}}{45}}\dfrac{v^2}{M_{\rm P}}\simeq 1.08\times 10^{-3}\,{\rm eV},
\eeq
where $g_*=106.75$ is the number of degrees of freedom in the SM.
For instance, $K=0.01,1,100$ corresponds to $\tilde{m}_1\simeq 10^{-5},10^{-3},10^{-1}\,{\rm eV}$, respectively.
The two regimes $\tilde{m}_1>m_{\ast}$ and $\tilde{m}_1<m_{\ast}$ are referred to as the {\it strong washout} and {\it weak washout} regimes, respectively~\cite{Buchmuller:2004nz}.
Note that $\tilde{m}_1$ must lie in the range of $m_1\leqslant\tilde{m}_1\lesssim m_3$~\cite{Buchmuller:2003gz}.

The Yukawa coupling has an upper bound from perturbativity, which in turn imposes a corresponding upper limit on the absolute value of the CP asymmetry.
The maximal absolute CP asymmetry is given by (see Appendix~\ref{Appendix:D})
\beq
\label{eq:III-1}
|\varepsilon_1|^{\rm max}
=\dfrac{3}{16\pi}\dfrac{M_1m_3}{v^2}\Biggl[1-\dfrac{m_1}{m_3}\biggl(1+\dfrac{m_3^2-m_1^2}{\tilde{m}_1^2}\biggr)^{1/2}\Biggr].
\eeq
 In this paper, we use the maximum value of the CP asymmetry
(\ref{eq:III-1}) for $|\varepsilon_1|$ to calculate the maximum baryon-to-photon ratio $\eta_b^{\rm max}$.

In the subsequent section, we compute the evolution of $N_{N_1}$ and $N_{B-L}$ for given $M_1$, $K$, and $\varepsilon_1$.
Then we map the result to the $(\tilde{m}_1,M_1)$ plane and discuss the allowed region, where the generated $B-L$ asymmetry is large enough to explain the observed baryon asymmetry.
Since entropy injection may occur after the production of asymmetry, we only require that the baryon asymmetry exceed the observed value.
For the latter we use the result from the Planck collaboration~\cite{Planck:2018vyg}
\beq
\label{eq:III-2}
\eta_b^{\rm CMB}=(6.144\pm 0.038)\times 10^{-10},
\eeq
and require $\eta_b^{\rm max}\geqslant (\eta_b^{\rm CMB})_{\rm low}$, where 
\beq
\bigl(\eta_b^{\rm CMB}\bigr)_{\rm low}=6.0\times 10^{-10},
\eeq
is the $3\sigma$ lower limit.

Another useful mass is the {\it absolute mass scale} defined by 
\beq
\label{eq:mbar}
\overline{m}=\sqrt{m_1^2+m_2^2+m_3^2}.
\eeq
As discussed in Ref.~\cite{Buchmuller:2003gz}, the allowed region on the $(\tilde{m}_1,M_1)$ plane shrinks with increasing $\overline{m}$ because the lower limit of $\tilde{m}_1$ increases, as we sketch below.

The value of $\overline{m}$ can be inferred from the mass measurements of neutrino oscillation experiments and cosmological observations.
Given the neutrino mass patterns, i.e., either $m_3^2-m_2^2>m_2^2-m_1^2$ or $m_3^2-m_2^2<m_2^2-m_1^2$, the dependence of $m_3$ on $m_1$ is fixed.
In this paper we analyze the normal hierarchy $m_1 < m_2 < m_3$ only.
In this case, the relations $m_3^2-m_2^2=\Delta m_{\rm atm}^2$ and $m_2^2-m_1^2=\Delta m_{\rm sol}^2$ lead to
\beq
\label{eq:III-3}
\begin{array}{l}
m_3^2=m_1^2+\Delta m_{\rm atm}^2+\Delta m_{\rm sol}^2,\\[0.5em]
m_2^2=m_1^2+\Delta m_{\rm sol}^2,\\[0.5em]
\overline{m}^2=3m_1^2+\Delta m_{\rm atm}^2+2\Delta m_{\rm sol}^2.
\end{array}
\eeq
These relations can be rewritten as 
\beq
\label{eq:III-6}
\begin{array}{l}
m_1^2=\dfrac{1}{3}
(\overline{m}^2-\Delta m_{\rm atm}^2-2\Delta m_{\rm sol}^2),\\[1em]
m_2^2=\dfrac{1}{3}
(\overline{m}^2-\Delta m_{\rm atm}^2+\Delta m_{\rm sol}^2),\\[1em]
m_3^2=\dfrac{1}{3}
(\overline{m}^2+2\Delta m_{\rm atm}^2+\Delta m_{\rm sol}^2).
\end{array}
\eeq
According to the neutrino oscillation experiments (see Ref.~\cite{Workman:2022ynf} and references therein), the mass differences are known to be
\beq
\label{eq:III-9}
\begin{array}{l}
\Delta m_{\rm atm}^2=(2.453\pm 0.033)\times 10^{-3}\,{\rm eV^2},
\\[0.5em]
\Delta m_{\rm sol}^2=(7.53\pm 0.18)\times 10^{-5}\,{\rm eV^2}.
\end{array}
\eeq
Thus $\overline{m}$ takes its minimum value
\beq
\label{eq:III-11}
\overline{m}=\sqrt{\Delta m_{\rm atm}^2+2\Delta m_{\rm sol}^2}
\simeq 0.051\,{\rm eV},
\eeq
for $m_1=0$.
Furthermore, CMB observations together with baryon acoustic oscillation (BAO) measurements constrain the sum of the mass eigenvalues of light neutrinos from above as~\cite{eBOSS:2020yzd}
\beq
\label{eq:III-12}
m_1+m_2+m_3<0.13\,{\rm eV}
\eeq
at the $2\sigma$ level.
Substituting equation (\ref{eq:III-6}) and using equation (\ref{eq:III-9}), we find
\beq
\label{eq:III-13}
\overline{m}<0.078\,{\rm eV}.
\eeq
This upper limit is more relaxed when CMB alone is used without BAO~\cite{Planck:2018vyg}
\beq
\label{eq:III-14-add}
m_1+m_2+m_3<0.26\,{\rm eV}
\eeq
at the $2\sigma$ level.
In this case, we obtain
\beq
\label{eq:III-14}
\overline{m}<0.15\,{\rm eV}.
\eeq
In the following we present the allowed regions in the $(\tilde{m}_1,M_1)$ plane
for the lower limit on $\overline{m}$ (\ref{eq:III-11}),
the two types of upper limits (\ref{eq:III-13}) and (\ref{eq:III-14}), and also $\overline{m}=\overline{m}_{\rm max}\equiv 0.19\,{\rm eV}$, the last of which corresponds to the value just before the allowed region disappears when scattering is taken into account.

\section{Implications to leptogenesis}
\label{sec:III}

\subsection{Intuitive picture}
\label{sec:III-intuitive}

Before proceeding to the numerical results, we describe the qualitative behavior of the evolution of right-handed neutrinos in the presence of density perturbations.
When the right-handed neutrinos decouple from the thermal bath at temperature $T$, the equilibrium abundance of right-handed neutrinos $N_1$ can be approximated by the Boltzmann distribution
\beq
\label{eq:IV-1}
N_{N_1}^{\rm eq}\sim\exp\biggl(-\dfrac{M_1}{T}\biggr).
\eeq
Substituting equation (\ref{eq:II-11}) into (\ref{eq:IV-1}), $N_{N_1}^{\rm eq}$ behaves as
\beq
\label{eq:IV-2}
N_{N_1}^{\rm eq}\sim\exp\biggl(-\dfrac{M_1}{\bar{T}}\biggr)
\exp\biggl(\dfrac{M_1}{\bar{T}}\delta_T\biggr).
\eeq
Therefore, if the temperature fluctuation is sufficiently large
\beq
\label{eq:IV-3}
\delta_T\gtrsim\dfrac{\bar{T}}{M_1},
\eeq
the linear expansion of $\exp(M_1\delta_T/\bar{T})$ causes a significant error.
For the strong washout regime, the actual abundance $N_{N_1}$ closely follows this equilibrium abundance, and hence the linear approximation also breaks down for $N_{N_1}$.
The condition (\ref{eq:IV-3}) translates to $\delta_T\gtrsim 0.05$ given that the freeze-out of $|N_{B-L}|$ occurs at $\bar{z}=M_1/\bar{T}\sim 10-20$.

If we linearize with respect to $\delta_T$, the baryon number would never increase because the average over the phase $\delta$ will cancel out the contributions from positive $\delta_T$ and those from negative $\delta_T$.
Therefore, the increase in the baryon number (Figure~\ref{fig:2}) represents a {\it beyond-linear} effect related to $\delta_T$, which we call ``acoustically driven freeze-out.''

\subsection{Time evolution of \texorpdfstring{$N_1$}{N1} and \texorpdfstring{$B-L$}{B-L}}

\begin{figure}[tbhp]
\centering
\includegraphics[width=\linewidth]{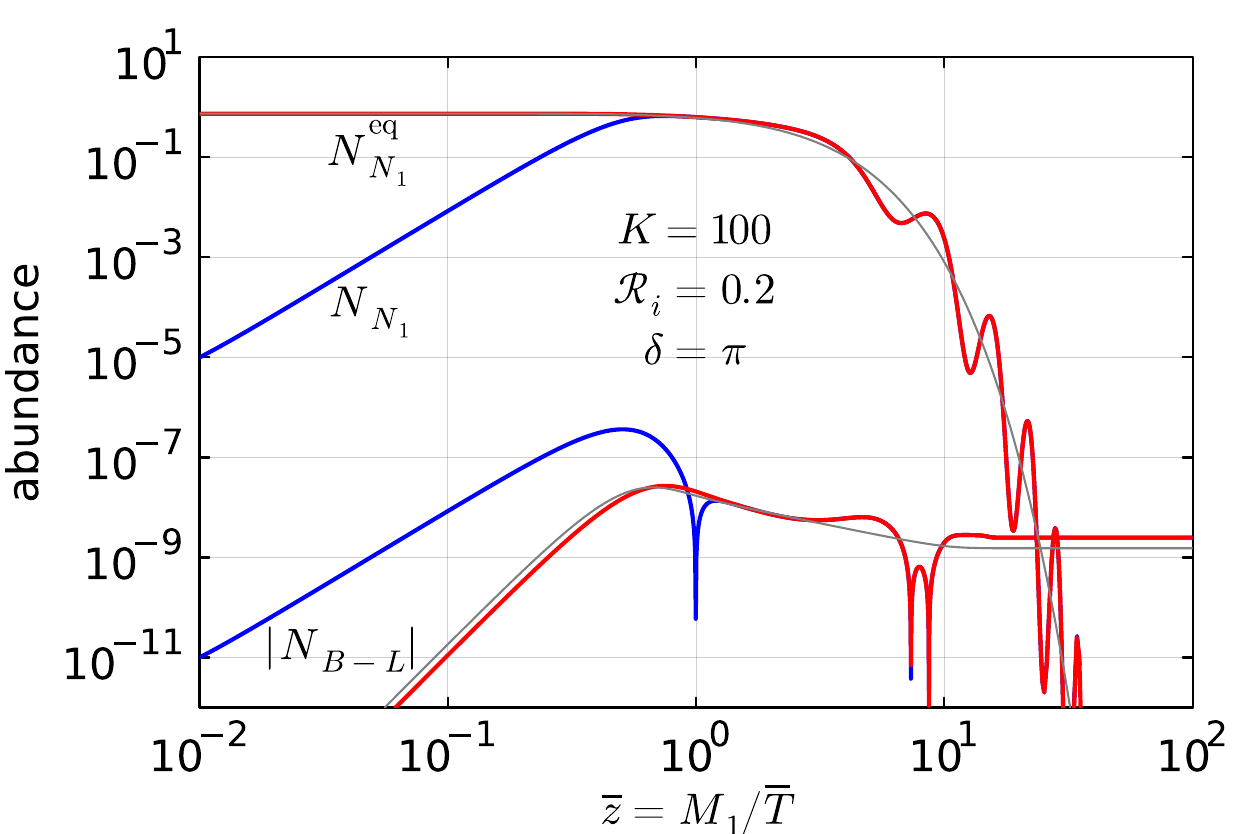}
\includegraphics[width=\linewidth]{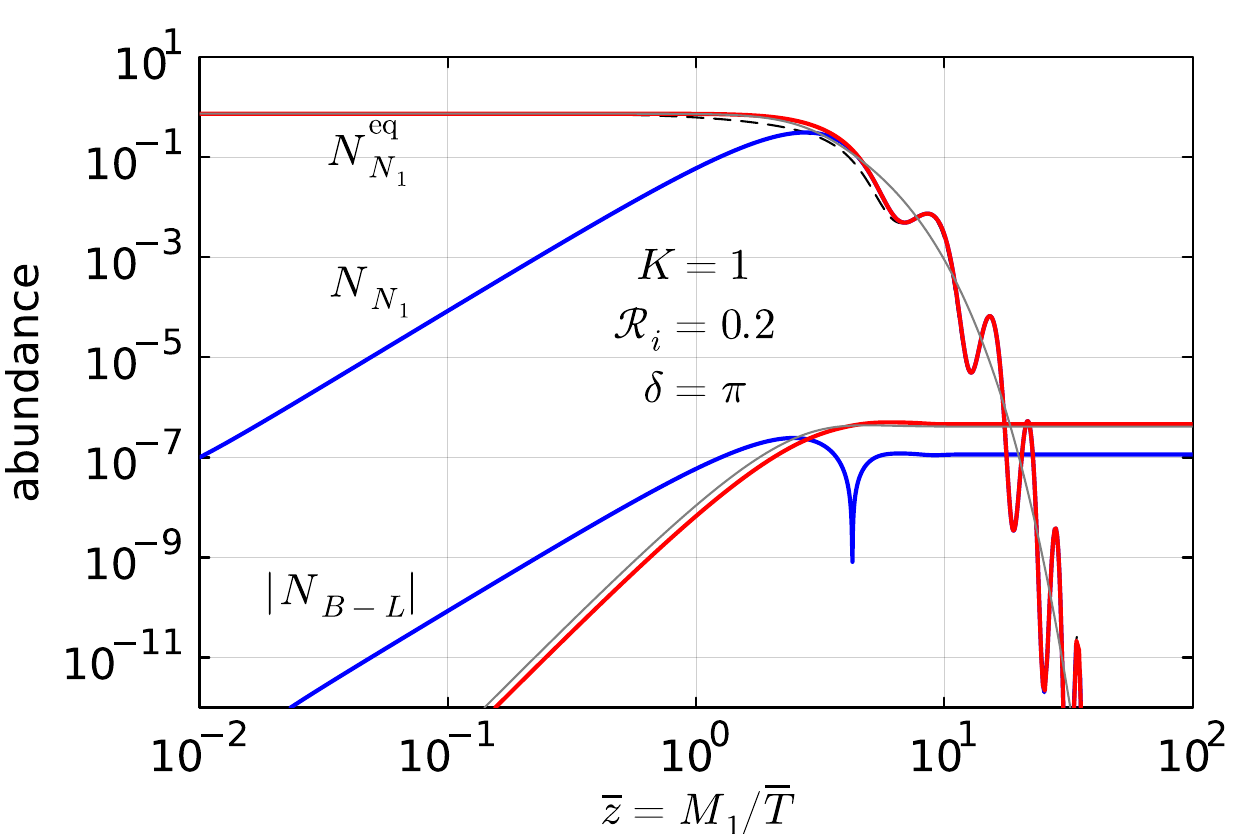}
\includegraphics[width=\linewidth]{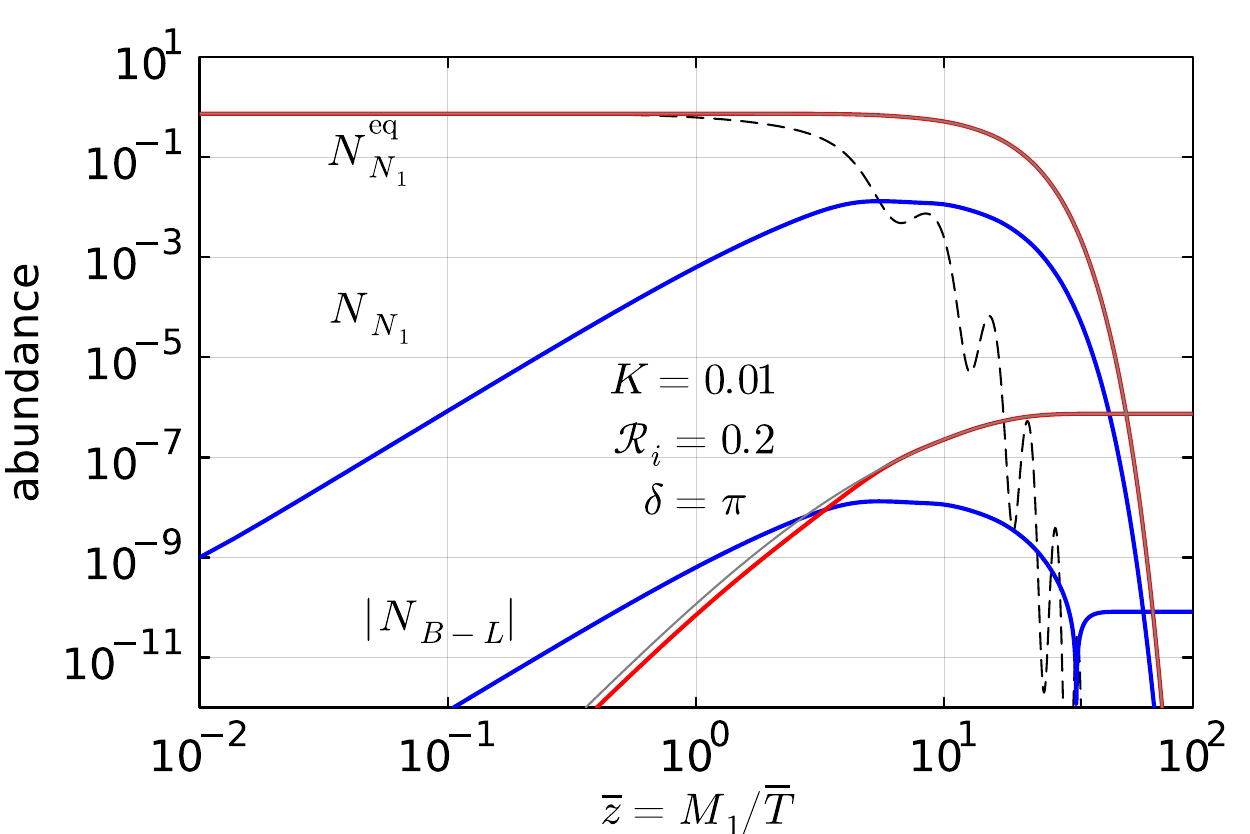}
\caption{\small
{\it Top}:
evolution of the abundance of right-handed neutrinos $N_{N_1}$ and the absolute value of the baryon-minus-lepton number $|N_{B-L}|$ for the decay parameter $K=100$, the amplitude of primordial curvature perturbation $|\mathcal{R}_i|=0.2$, and the phase $\delta=\pi$.
The red and blue curves show the thermal and zero initial abundances for $N_{N_1}$, respectively.
These curves merge into a single curve at later times.
The thin gray curves depict the results for $|\mathcal{R}_i|=0$ with the thermal initial abundance.
The dashed line represents the equilibrium number density of right-handed neutrinos $N_{N_1}^{\rm eq}$.
{\it Middle}: same as the top panel but for the decay parameter $K=1$.
{\it Bottom}: same as the top panel but for the decay parameter $K=0.01$.
}
\label{fig:1}
\end{figure}

\begin{figure}
\centering
\includegraphics[width=\linewidth]{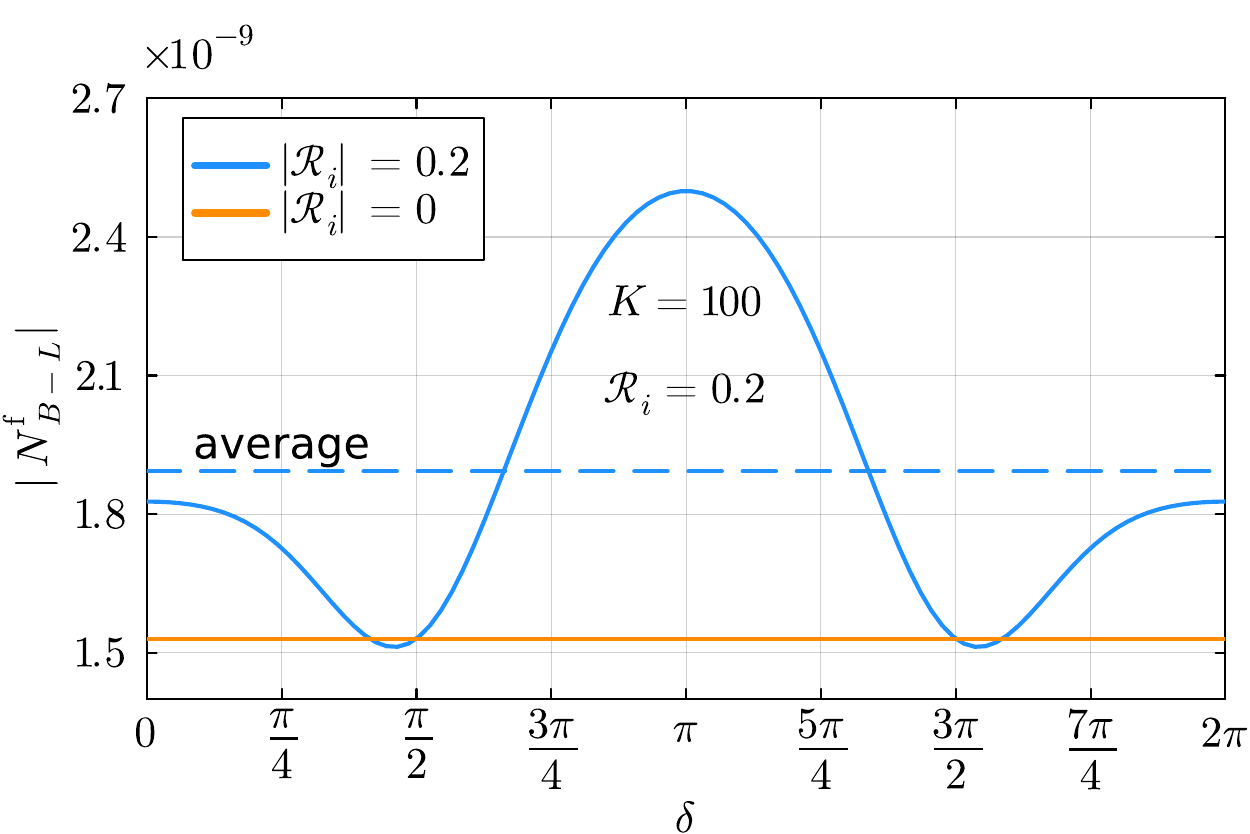}
\caption{\small
Baryon-minus-lepton number $|N_{B-L}|$ after freeze-out, denoted by $|N_{B-L}^{\rm f}|$, for different values of $\delta$.
The blue solid curve shows the result with temperature fluctuations, while the orange solid line shows that without fluctuations.
In the presence of perturbations, temperature oscillations at different spatial locations have different values for the phase $\delta$, resulting in the difference in $|N_{B-L}^{\rm f}|$ at different spatial points.
Averaging this over $\delta$ yields the spatial average shown by the blue dashed line, which is found to be greater than the value without perturbations.
Here the zero initial abundance is assumed for $N_{N_1}$, but the result with the thermal initial abundance is almost the same.
}
\label{fig:2}
\end{figure}

\begin{figure}
\centering
\includegraphics[width=\linewidth]{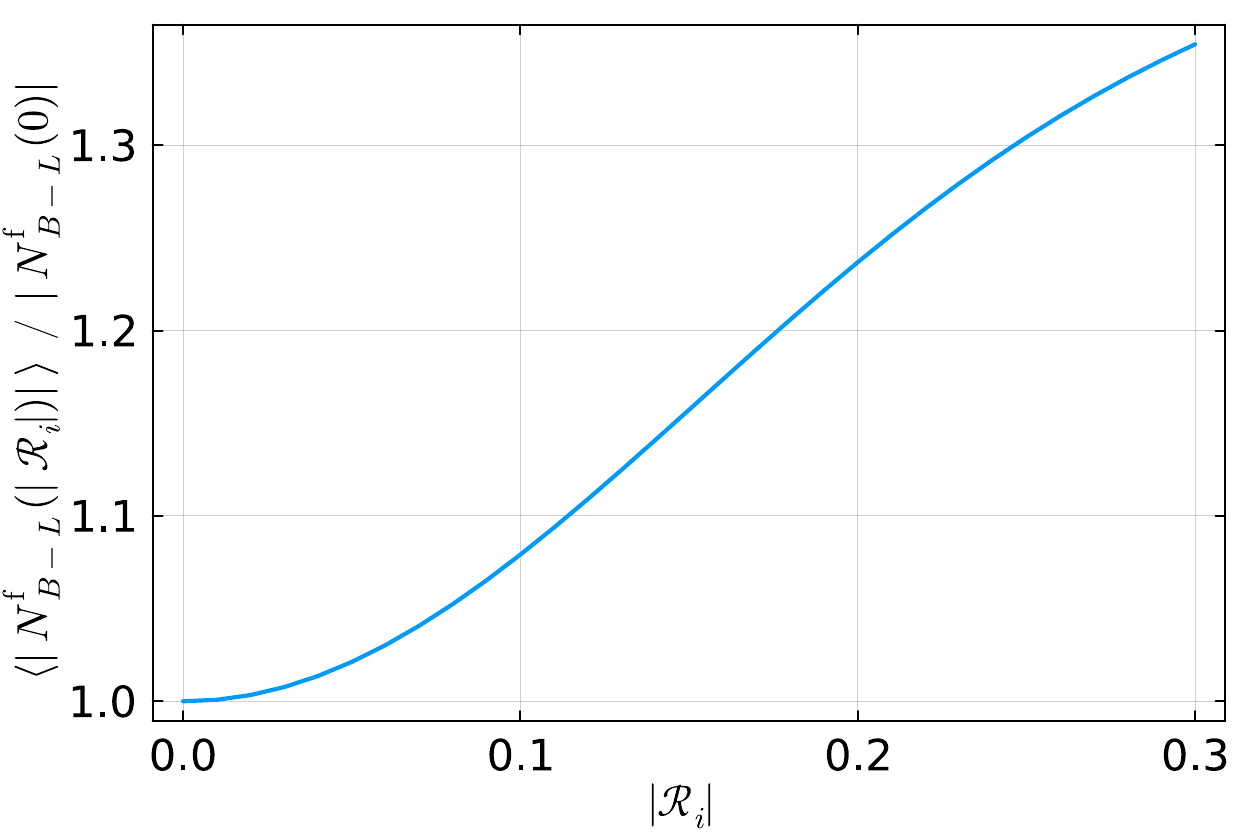}
\caption{\small
Ratio between the spatially averaged freeze-out value $\langle |N_{B-L}^{\rm f}|\rangle_{\rm space}$ with and without perturbations, as a function of the amplitude of the primordial curvature perturbation $|\mathcal{R}_i|$.
}
\label{fig:Ri-zH1}
\end{figure}

Figure~\ref{fig:1} compares the time evolutions of the abundance of the right-handed neutrino $N_{N_1}$ and the baryon-minus-lepton number $N_{B-L}$ in the presence (thick lines) and absence (thin lines) of perturbations.
The decay parameter $K=\tilde{\Gamma}_{\rm D}/\bar{z}^2H$ is taken to be $100$, $1$, and $0.01$ from top to bottom.
The curvature perturbation is set to $|\mathcal{R}_i|=0.2$ and the phase is chosen to $\delta=\pi$ that corresponds to a specific spatial point.
Also, a monochromatic wave $\bar{z}_H=1$ is assumed, corresponding to the horizon re-entry somewhat before the freeze-out of $|N_{B-L}|$.
For the time evolution with other values of $\bar{z}_H=1$, see Appendix~\ref{Appendix:plot}.
Also, for more realistic density fluctuations described by a superposition of monochromatic waves, see the discussion in Section~\ref{sec:V}.

In the top panel of Figure~\ref{fig:1} ($K=100$), the actual and equilibrium abundances of right-handed neutrinos $N_{N_1}$ and $N_{N_1}^{\rm eq}$ oscillate significantly for $\bar{z}\gtrsim \mathcal{O}(1)$.
As a result, $|N_{B-L}|$ also oscillates around $\bar{z}\sim 10$, leading to the increase in the value of $|N_{B-L}|$ after the freeze-out denoted by $|N^{\rm f}_{B-L}|$.
This increase occurs for both initial conditions $N_{N_1}=3/4$ or $N_{N_1}=0$.
On a closer look, we observe that the freeze-out value $|N^{\rm f}_{B-L}|$ is determined by the downstroke of the temperature and of the corresponding equilibrium value $N_{N_1}^{\rm eq}$ around $\bar{z}\simeq 15$.
For this parameter point, the freeze-out is dominantly driven by the sudden decrease in the temperature caused by sound waves, not by the slow decrease in the average temperature.
Of course, different spatial points have different oscillation phases, and thus we have to take average over the phase $\delta$ as performed below.

In the middle panel of Figure~\ref{fig:1} ($K=1$), we still observe oscillations in the abundance of right-handed neutrinos $N_{N_1}$ and its equilibrium value $N_{N_1}^{\rm eq}$.
The freeze-out value $|N^{\rm f}_{B-L}|$ is thus enhanced in the same way as the top panel, though the enhancement is smaller.
Such enhancement is observed for the strong washout regime $1\leq K\leq 100$.
For the weak washout regime $K< 1$, in contrast, $N_{N_1}$ does not oscillate any more and the freeze-out value $|N^{\rm f}_{B-L}|$ is almost identical to the case without perturbations, as seen from the bottom panel of Figure~\ref{fig:1} ($K=0.01$).
This tendency is simply because $N_{N_1}$ is not tightly coupled with $N_{N_1}^{\rm eq}$ any more and thus the oscillations do not play any role in this parameter range.

To evaluate the effect of density fluctuations on the final baryon asymmetry, the freeze-out value $|N_{B-L}^{\rm f}|$ must be averaged over different values of $\delta$.
Figure~\ref{fig:2} depicts the freeze-out value $|N_{B-L}^{\rm f}|$ as a function of $\delta$.
Interestingly, the freeze-out value $|N_{B-L}^{\rm f}|$ is found to increase for almost all values of $\delta$.
This is because, oscillations that happen with the ``right phase'' boost the freeze-out of $N_{N_1}$ and $N_{B-L}$ as seen in the top panel of Figure~\ref{fig:1}, while oscillations with the ``wrong phase'' (i.e., oscillations that increase the temperature around $\bar{z} \simeq 15$) simply push the system back to equilibrium again, resulting in the system waiting for the freeze-out that occurs on the next occasion of temperature decrease.
Thus the ``wrong phase'' does not lead to the decrease in the freeze-out value.

Based on these observations, we compare the spatially averaged value $\langle |N_{B-L}^{\rm f}|\rangle_{\rm space}$ with the freeze-out value in the absence of fluctuations.
We find 
\beq
\label{eq:II-1017}
\dfrac{\langle |N_{B-L}^{\rm f}(|\mathcal{R}_i|=0.2)|\rangle_{\rm space}}{|N_{B-L}^{\rm f}(|\mathcal{R}_i|=0)|}
\sim 1.25.
\eeq
for the parameter point $|\mathcal{R}_i|=0.2$ and $\bar{z}_H=1$, as seen from Figure~\ref{fig:2}.
We emphasize that the ratio is greater than unity, because the increase in $|N_{B-L}^{\rm f}|$ for the phase $\delta\sim\pi$ is not canceled out by the contribution from $\delta\sim 0$ due to the reason explained just above.
This enhancement would vanish if one linearizes the system in the temperature fluctuation $\delta_T$, and therefore this effect is a beyond-linear effect, which might be called ``acoustically enhanced freeze-out.''

In Figure \ref{fig:Ri-zH1} we plot the ratio between the spatially averaged freeze-out value $\langle |N_{B-L}^{\rm f}(|\mathcal{R}_i|)|\rangle_{\rm space}$ and the freeze-out value without fluctuations as a function of the primordial curvature perturbation $|\mathcal{R}_i|$.
We find that the ratio exceeds unity for any $|\mathcal{R}_i|>0$, and that it increases monotonically for $0\leqslant|\mathcal{R}_i|\lesssim 0.3$.
Figures with other values of $\bar{z}_H$ are shown in Appendix~\ref{Appendix:plot}.

\subsection{Parameter regions for successful leptogenesis}

\begin{figure*}[tbhp]
\centering
\includegraphics[width=\linewidth]{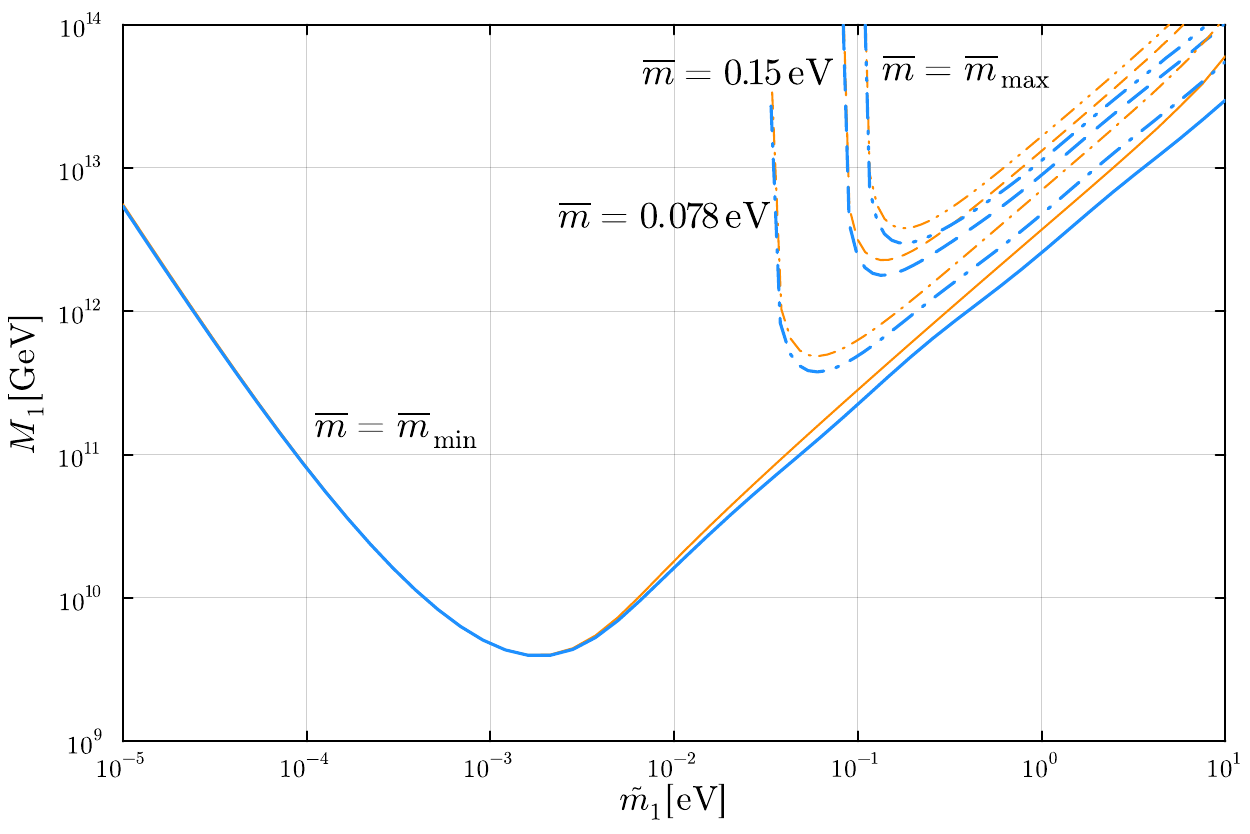}
\caption{\small
Allowed regions for successful leptogenesis.
The normal hierarchy and zero initial abundance for $N_{N_1}$ are assumed. 
The curves show $(\tilde{m}_1,M_1)$ where $\eta_b^{\rm max}=(\eta^{\rm CMB}_b)_{\rm low}=6.0\times 10^{-10}$ is satisfied for $\overline{m}=(0.051,\,0.078,\,0.15,\,0.19)\,{\rm eV}$.
For a given $\overline{m}$, the region above the curves is the one for successful leptogenesis.
The blue and orange curves represent the case with and without temperature fluctuations $\delta_T$, respectively.
The primordial curvature perturbation is taken to be $|\mathcal{R}_i|=0.2$.
We can see that the allowed region in $(\tilde{m}_1,M_1)$ is expanded by the effect of $\delta_T$.
The solid lines correspond to $m_1=0$ such that $\overline{m}=\overline{m}_{\rm min}=0.051\,{\rm eV}$.
The dash-dot lines are for $\overline{m}=0.078\,{\rm eV}$, corresponding to the upper limit from CMB and BAO observations~\cite{eBOSS:2020yzd}.
The dashed lines are for $\overline{m}=0.15\,{\rm eV}$, corresponding to the upper limit from CMB alone~\cite{Planck:2018vyg}.
The dot-dot-dashed lines correspond to $\overline{m}=0.19\,{\rm eV}$.
If scatterings are included, the upper boundary appears, and the allowed region would disappear for $\overline{m}>0.19\,{\rm eV}$~\cite{Buchmuller:2003gz}.
}
\label{fig:3}
\end{figure*}

Figure~\ref{fig:3} shows the parameter regions for successful leptogenesis on the $(\tilde{m}_1,M_1)$ plane.
For a given value of $\bar{m}$, there are two curves with (blue) and without (orange) density fluctuations.
The primordial curvature perturbation is taken to be $|\mathcal{R}_i|=0.2$.
In the region above each curve, the produced amount of baryons exceeds the lower limit given by the CMB data (\ref{eq:III-2}).
In the previous subsection, we considered the case of $K=100$, $1$, and $0.01$ corresponding to $\tilde{m}_1\simeq 0.1$, $10^{-3}$, and $10^{-5}\,{\rm eV}$, respectively.
While the contours are not closed in this plot, it is known that upper limits on $M_1$ arise once $\Delta L=2$ scatterings are taken into account (see e.g., Ref.~\cite{Buchmuller:2004nz}).
The reason we do not take account of these scatterings is simply to illustrate the effect of acoustically driven freeze-out in the simplest setup (see also discussion in Section \ref{sec:V}).

In the case of $m_1=0$, the absolute mass (\ref{eq:mbar}) takes its minimum value $\overline{m}=0.051\,{\rm eV}$.
The resulting constraint is shown in the solid lines.
In the weak washout regime $\tilde{m}_1\lesssim 10^{-3}\,{\rm eV}$, the constraint with temperature fluctuations is almost identical to the one without temperature fluctuations.
In the strong washout regime $\tilde{m}_1 \gtrsim 10^{-3}\,{\rm eV}$, the allowed region with fluctuations gets slightly extended from that without fluctuations due to the enhancement by the acoustically driven freeze-out.

The allowed parameter space shrinks as $\overline{m}$ increases, as discussed in Ref.~\cite{Buchmuller:2003gz}.
For instance, in the case of $\overline{m}=0.078\,{\rm eV}$, the region for $\tilde{m}_1\gtrsim 0.03\,{\rm eV}$ is ruled out as shown in the dot-dashed lines.
As in the case of $\bar{m}=0.051\,{\rm eV}$, including fluctuations enlarges the allowed region.
Similar behavior is observed for $\overline{m}=0.15\,{\rm eV}$ (dashed lines) and $\overline{m}_{\rm max}=0.19\,{\rm eV}$ (dot-dot-dashed lines).

We thus conclude that the beyond-linear effect can enlarge the parameter space for successful leptogenesis if the amplitudes of primordial density fluctuations are sufficiently large.
The effect of perturbations is important in the strong washout regime
\beq
\label{eq:IV-16}
8.7\times 10^{-3}\,{\rm eV}\lesssim\tilde{m}_1
\lesssim 5.0\times 10^{-2}\,{\rm eV}.
\eeq
The dependence on the primordial curvature perturbation $|\mathcal{R}_i|$ can be inferred from Figure~\ref{fig:Ri-zH1}.
Note that the neutrino oscillation experiments suggest that the masses of light neutrinos likely lie in this range, see equation~(\ref{eq:III-9}).
In the future, as neutrinoless double beta decay experiments such as CANDLES~\cite{CANDLES:2020iya}, CUORE~\cite{CUORE:2017tlq}, EXO~\cite{EXO-200:2019rkq}, KamLAND-Zen~\cite{KamLAND-Zen:2022tow}, LEGEND~\cite{LEGEND:2017cdu}, NEMO-3 and SuperNEMO~\cite{NEMO-3:2015jgm,NEMO-3:2016qxo} (see also Ref.~\cite{Dolinski:2019nrj} and references therein) achieve better sensitivities, not only will the upper limits on the mass eigenvalues of the light neutrinos improve, but the absolute mass scale $\overline{m}$ may finally be determined.
If the determined mass scale falls within the ballpark of $\overline{m} \gtrsim 0.1\,{\rm eV}$, the effect discussed in the present paper turns out to be relevant, as seen from Figure~\ref{fig:3}.

\section{Discussion and conclusions}
\label{sec:V}

In this paper, we studied the effect of primordial density fluctuations on the freeze-out of heavy particles.
Primordial density fluctuations within the Hubble horizon propagate as sound waves during the freeze-out of these particles.
Sound waves cause temperature oscillation at each spatial point, driving oscillation in the abundance of massive particles with respect to radiation.
Using leptogenesis as an illustrative framework, we have derived the governing equations that describe the evolution of these heavy particles in the presence of fluctuations, and found that the exponential dependence of the abundance of heavy particles on the local temperature causes a beyond-linear effect in the system,
see equations~(\ref{eq:IV-1}) and (\ref{eq:IV-2}).

As a proof of concept, we applied the formulation to the scenario of vanilla leptogenesis and demonstrated how temperature fluctuations affect the abundance of the right-handed neutrino and the resulting $B-L$ asymmetry.
We indeed observed oscillations in the equilibrium abundance of the right-handed neutrino $N_1$ and in the baryon-minus-lepton number $N_{B-L}$ (Figure~\ref{fig:2}).
Interestingly, the density fluctuations always increase the final baryon-to-photon ratio compared to the case without fluctuations, even after taking spatial average (Figure~\ref{fig:Ri-zH1}).
This suggests that the effect of oscillations cannot be captured within linear perturbation.
Qualitatively, the equilibrium value of the right-handed neutrino is governed by the Boltzmann factor $\exp (- E / T) \simeq \exp (- E / \bar{T}) \exp (E \delta T / \bar{T}^2)$, and the fast temperature oscillation makes it harder for the non-equilibrium abundance of $N_1$ to follow this equilibrium value.
If the temperature oscillation occurs with the right phase, the freeze-out of $N_1$ is dominantly driven by the oscillation, not by the slow decrease in the average temperature, and the resulting $|N_{B-L}|$ is enhanced (top panel of Figure~\ref{fig:Ri-zH1}, around $\bar{z} \simeq 15$).
We call this effect ``acoustically driven freeze-out''.
Of course, each spatial point has its own oscillation phase, and thus the enhancement in $|N_{B-L}|$ does not occur uniformly.
However, the exponential enhancement of $|N_{B-L}|$ in some regions cannot be fully canceled out by the contributions from other regions.
As a result, we get a net enhancement of the baryon asymmetry even after spatial average.

From the consideration above, the allowed parameter space for leptogenesis gets enlarged in the presence of fluctuations.
We presented the allowed regions $\eta_b \geq (\eta^{\rm CMB}_b)_{\rm low}$ in the $(\tilde{m}_1,M_1)$ plane for a given absolute neutrino mass scale $\overline{m}=\sqrt{m_1^2+m_2^2+m_3^2}$ (Figure~\ref{fig:3}).
The effective neutrino mass $\tilde{m}_1$ is linearly proportional to the decay parameter $K$ and thus characterizes the weak and strong washout regimes (Figure~\ref{fig:3}, $\tilde{m} \lesssim 10^{- 3} \, 
{\rm eV}$ and $\tilde{m} \gtrsim 10^{- 3} \, 
{\rm eV}$, respectively).
When temperature fluctuations are included, we indeed found an enlarged allowed regions.
The difference arises mainly in the strong washout regime, because in the weak washout regime the actual $N_1$ abundance is almost unaffected by its oscillating equilibrium value (bottom panel of Figure~\ref{fig:Ri-zH1}).
While the quantitative difference is not dramatically large, given that the neutrino oscillation experiments point toward the strong washout regime~\cite{Workman:2022ynf}, our findings will be relevant when we find relatively large $\bar{m}$ from cosmological observations~\cite{eBOSS:2020yzd,Planck:2018vyg}.

We conclude by mentioning several possible future directions.
For the microphysics side, we took only decay and inverse decay into account in this study.
However, it is well known that $\Delta L=2$ scattering imposes an ($\overline{m}$-dependent) upper bound on $M_1$ and thus closes the allowed region in the $(M_1,\tilde{m}_1)$ plane~\cite{Buchmuller:2003gz}.
Such scattering processes will not bring qualitative changes in the behavior of acoustically driven freeze-out since they do not affect the oscillating equilibrium abundance itself, which is the main reason for the net enhancement in the asymmetry.
Nevertheless, identifying the allowed parameter space including all the known microphysical effect will be important in preparing for future precision measurements.
For the macrophysics side, in this paper we injected monochromatic sound waves for simplicity.
In reality, however, fluctuations are a superposition of waves with various wavenumbers characterized by the power spectrum of the primordial curvature perturbation.
Thus the effect of fluctuations on freeze-out will ultimately be described by the power spectrum, and it would be interesting to construct analytical formulations for it.
In addition, it has been pointed that the sound waves can evolve to shock waves in the long run~\cite{Pen:2015qta}.
Shock waves by definition have discontinuities in thermodynamic parameters and hence are expected to enhance the sudden freeze-out of heavy particles.
Incorporating all these aspects will be important and interesting future steps, to which we will come back elsewhere.

\begin{acknowledgments}
The work of K.H. is supported by the JST FOREST Program (JPMJFR2136) and the JSPS Grant-in-Aid for Scientific Research (20H05639, 20H00158, 23H01169, 23H04900).
The work of R.J. is supported by JSPS KAKENHI Grant Numbers 23K17687, 23K19048, and 24K07013.
\end{acknowledgments}

\appendix

\section{Time evolution for different values of \texorpdfstring{$\bar{z}_H$}{zbarH}}
\label{Appendix:plot}

In this Appendix, we plot the time evolution of $N_1$ and $B-L$, the freeze-out values of $B-L$, and their dependence on $|\mathcal{R}_i|$ for $\bar{z}_H=0.5$ and $2$.

\begin{figure}[tbhp]
\centering
\includegraphics[width=\linewidth]{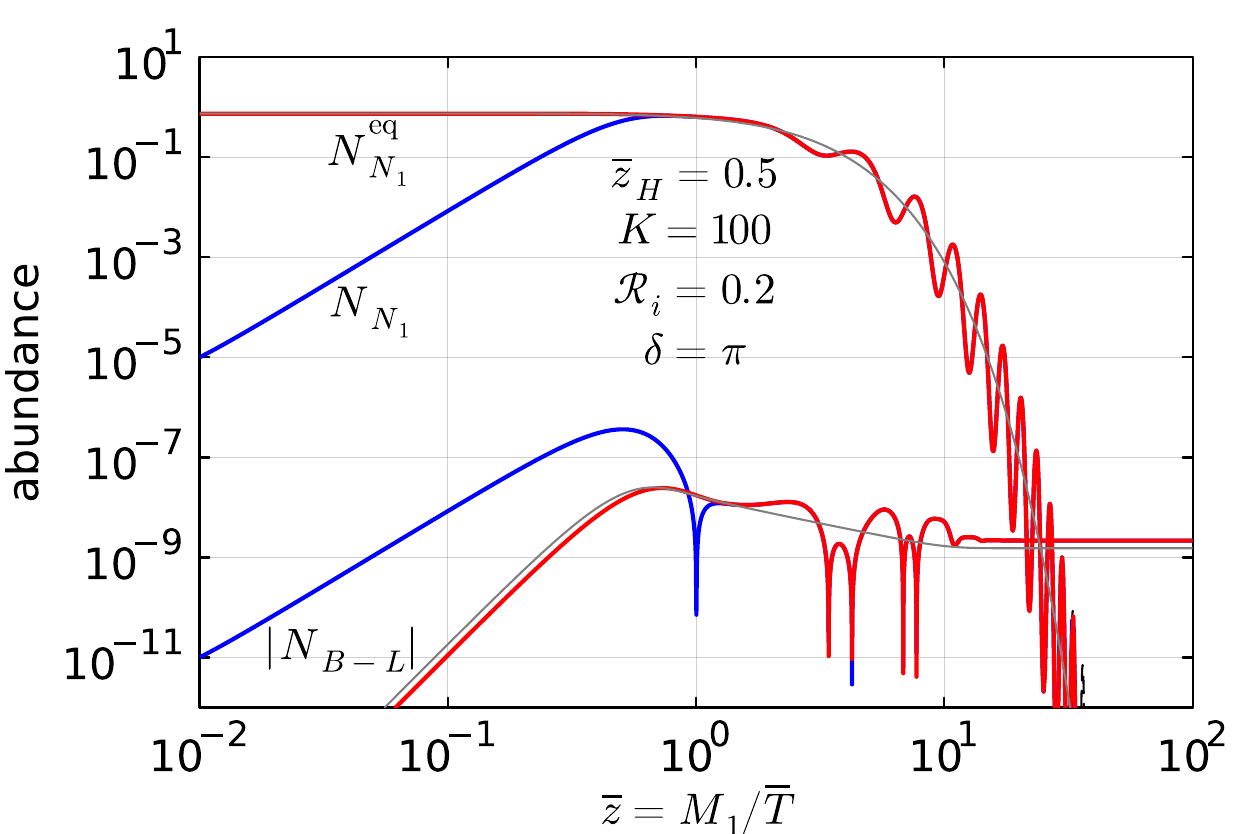}
\includegraphics[width=\linewidth]{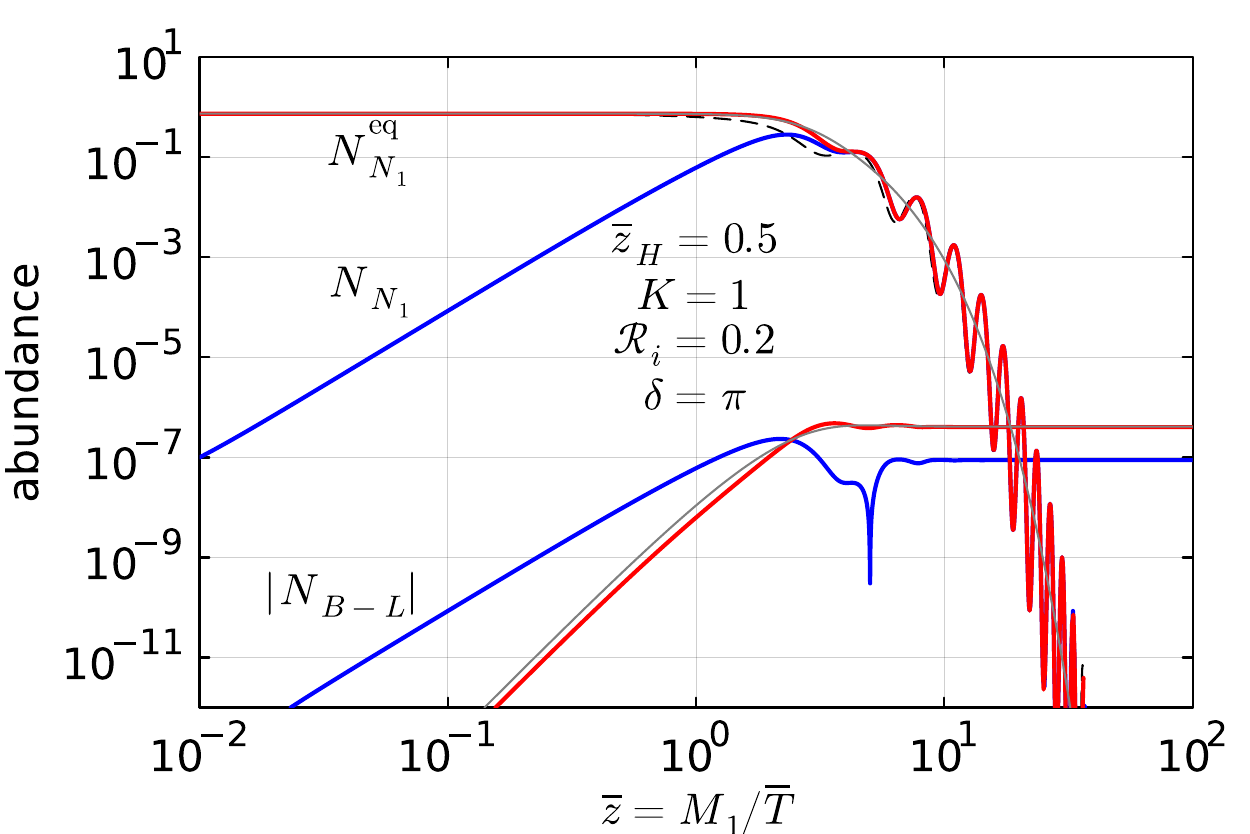}
\includegraphics[width=\linewidth]{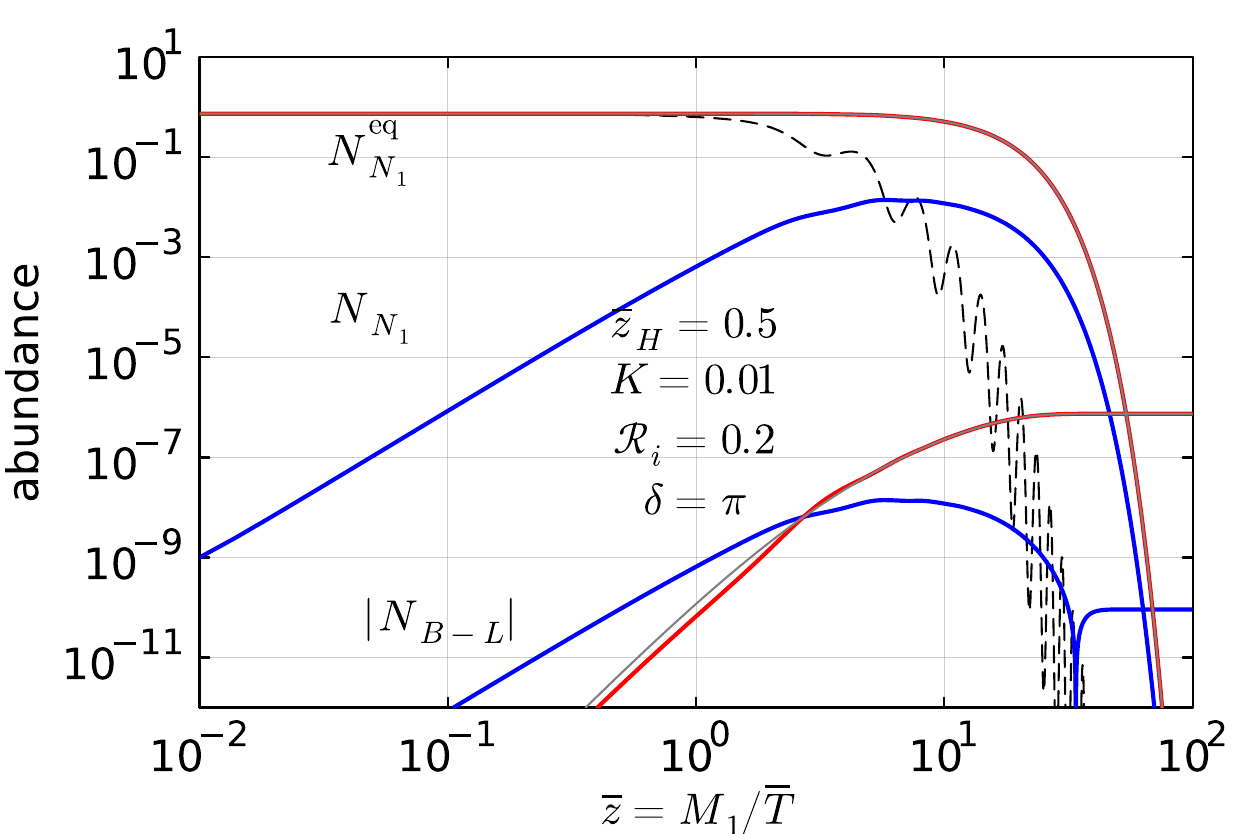}
\caption{\small
Time evolution of $N_{N_1}$ and $|N_{B-L}|$.
Parameter values are the same as Figure~\ref{fig:1} except for $\bar{z}_H=0.5$.
}
\label{fig:zH05}
\end{figure}

\begin{figure}[tbhp]
\centering
\includegraphics[width=\linewidth]{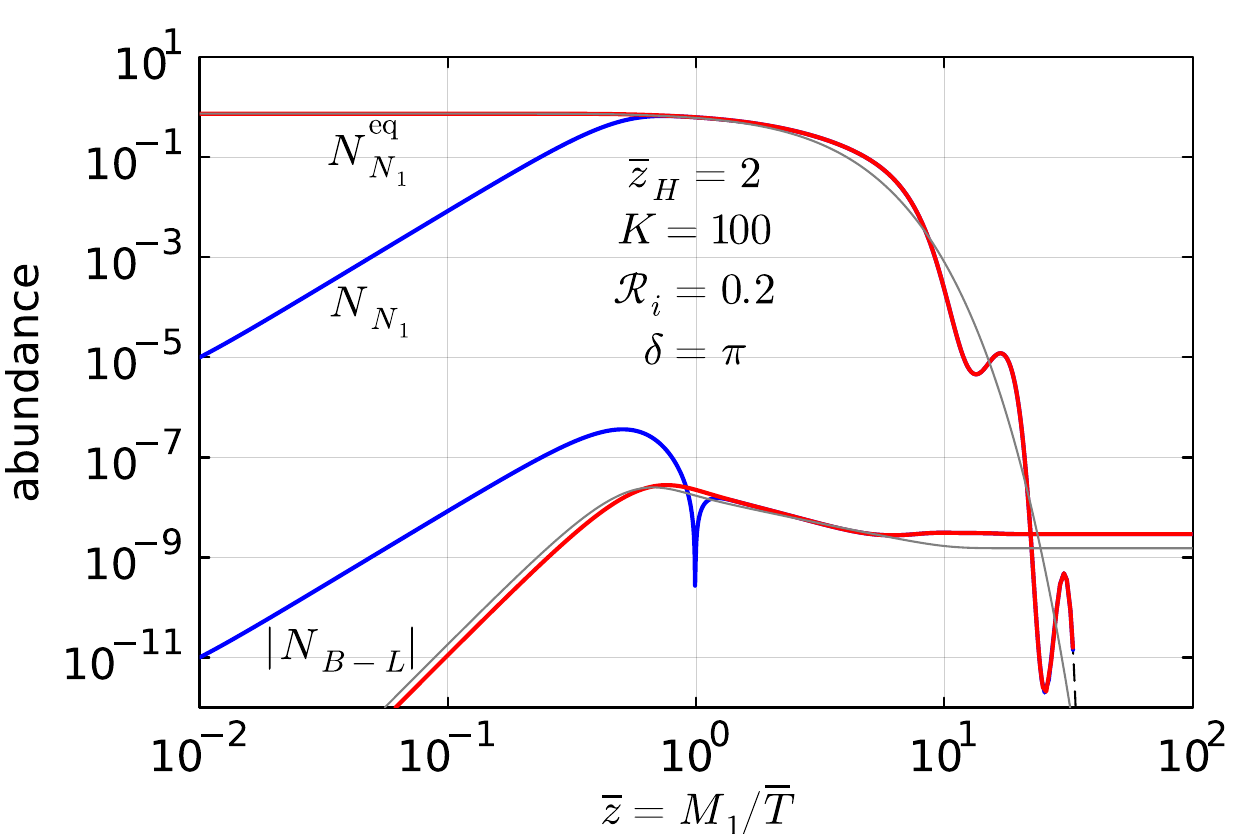}
\includegraphics[width=\linewidth]{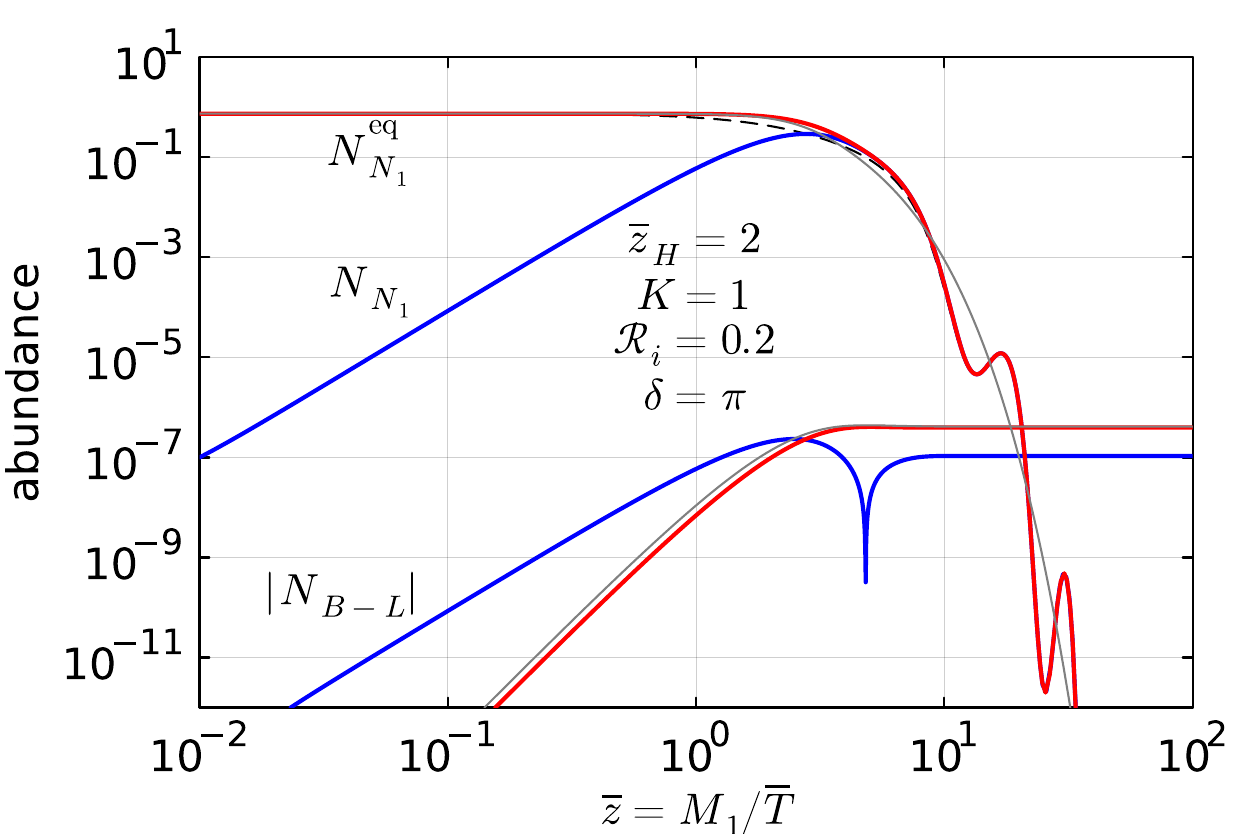}
\includegraphics[width=\linewidth]{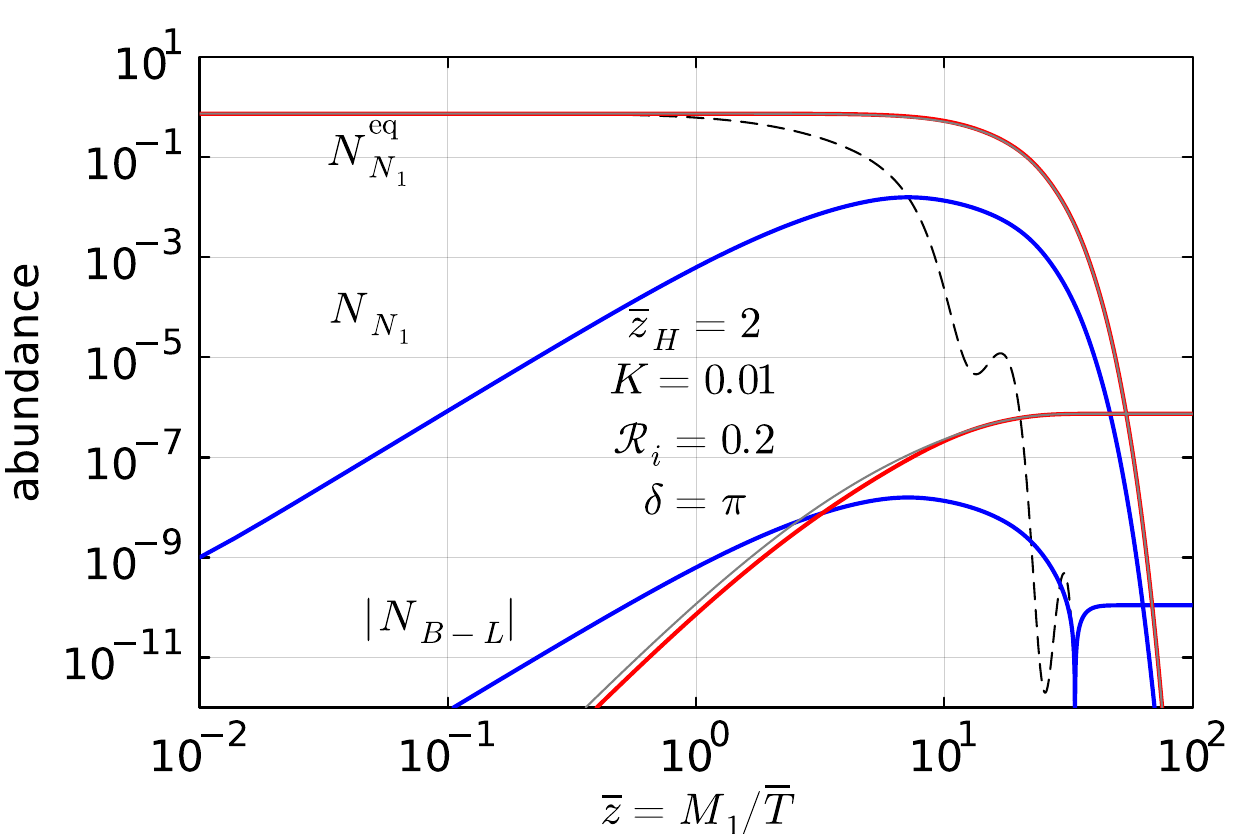}
\caption{\small
Time evolution of $N_{N_1}$ and $|N_{B-L}|$.
Parameter values are the same as Figure~\ref{fig:1} except for $\bar{z}_H=2$.
}
\label{fig:zH2}
\end{figure}

\begin{figure}
\centering
\includegraphics[width=\linewidth]
{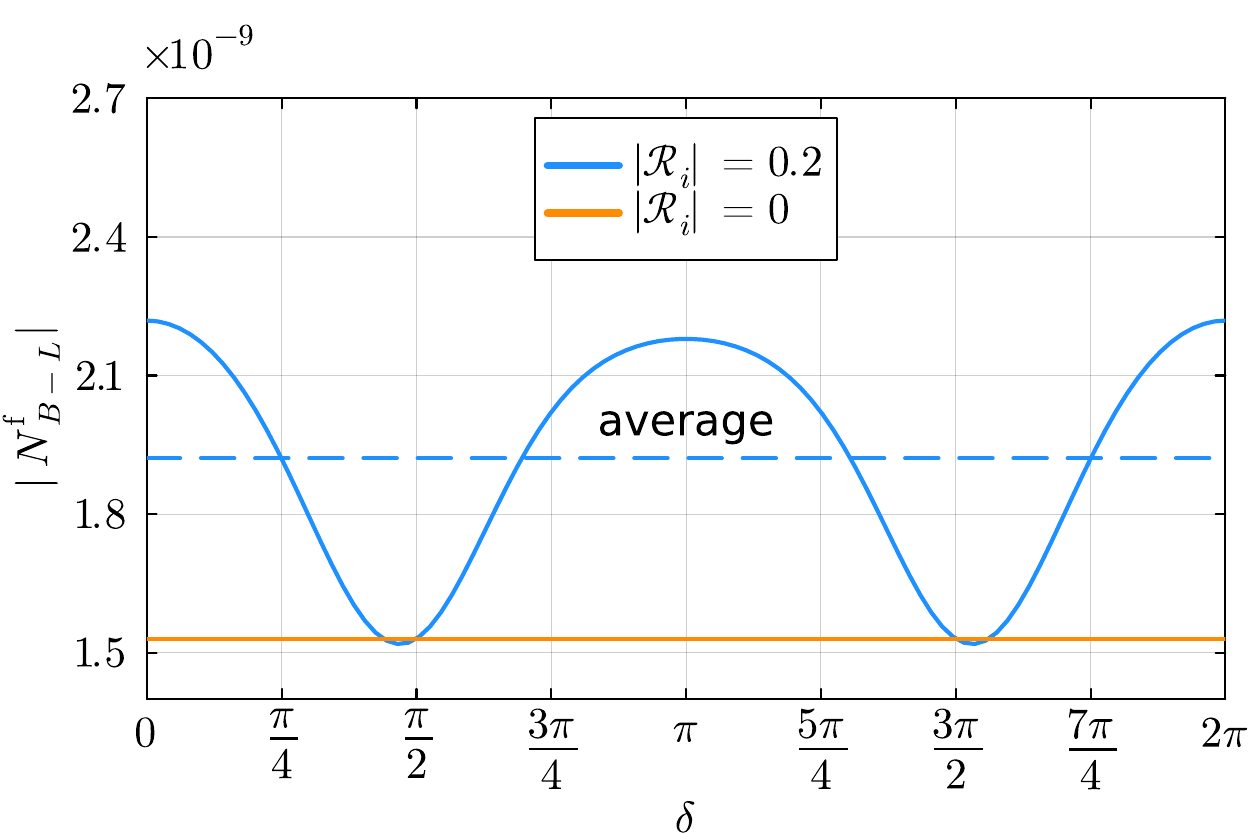}
\caption{\small
Freeze-out value of $|N_{B-L}|$ for different values of $\delta$.
Parameter values are the same as Figure~\ref{fig:2} except for $\bar{z}_H=0.5$.
}
\label{fig:2-zH05}
\end{figure}

\begin{figure}
\centering
\includegraphics[width=\linewidth]
{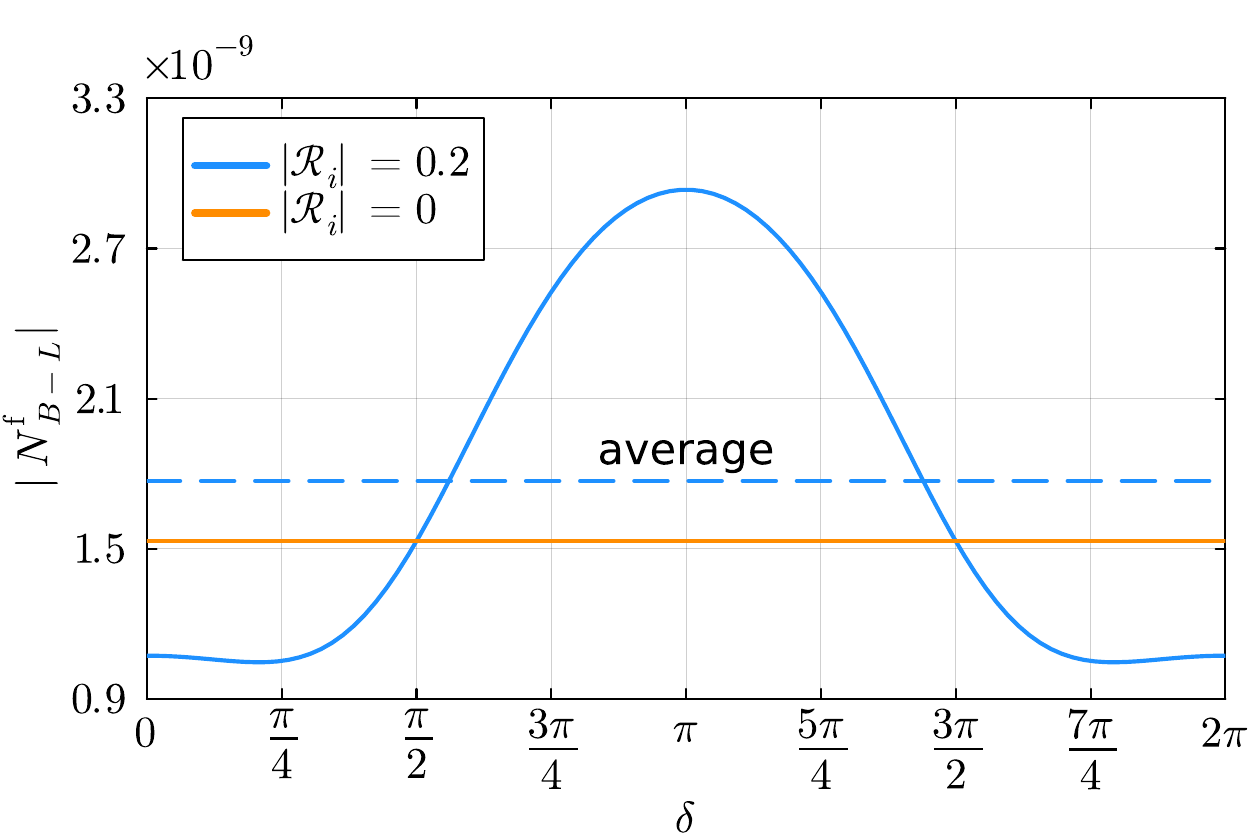}
\caption{\small
Freeze-out value of $|N_{B-L}|$ for different values of $\delta$.
Parameter values are the same as Figure~\ref{fig:2} except for $\bar{z}_H=2$.
}
\label{fig:2-zH2}
\end{figure}

\begin{figure}
\centering
\includegraphics[width=\linewidth]{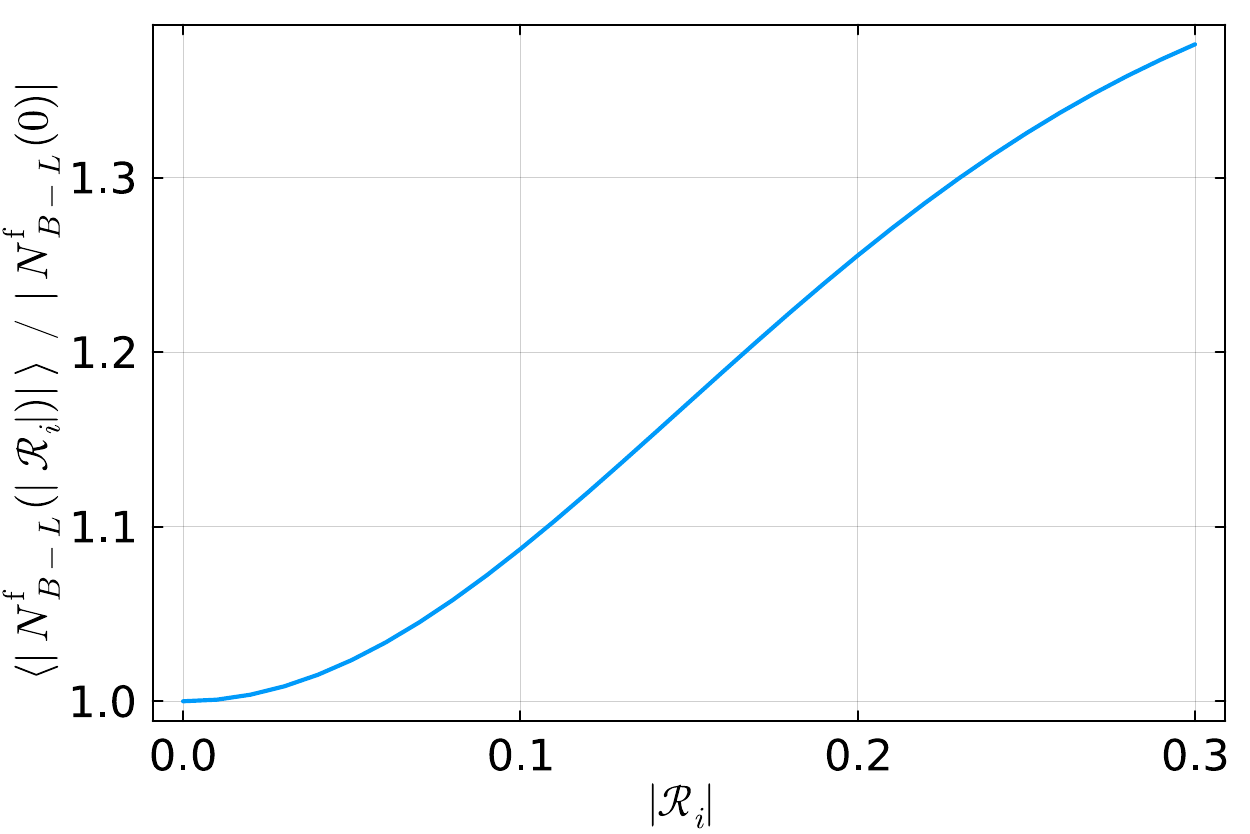}
\caption{\small
Ratio between the spatially averaged freeze-out value $\langle |N_{B-L}^{\rm f}|\rangle_{\rm space}$ with and without perturbations.
Parameter values are the same as Figure~\ref{fig:Ri-zH1} except for $\bar{z}_H=0.5$.
}
\label{fig:Ri-zH05}
\end{figure}

\begin{figure}
\centering
\includegraphics[width=\linewidth]{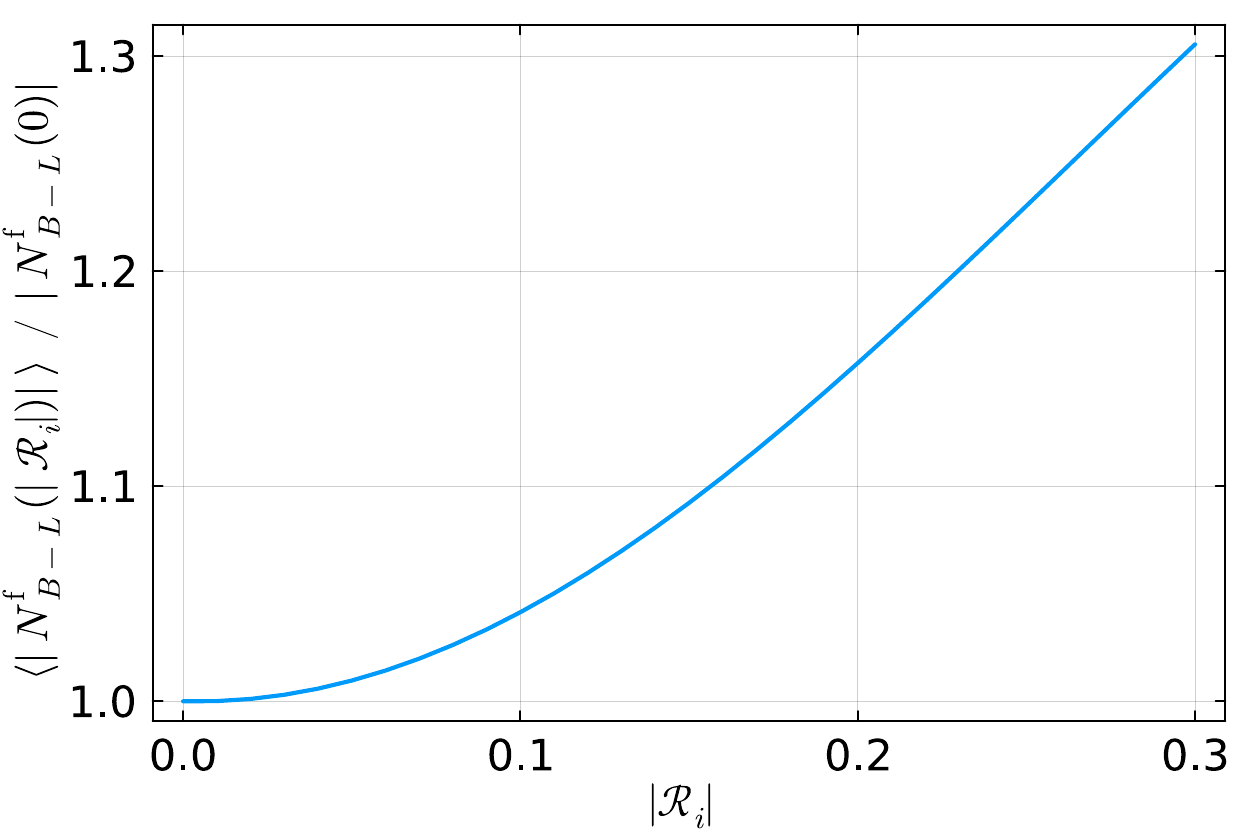}
\caption{\small
Ratio between the spatially averaged freeze-out value $\langle |N_{B-L}^{\rm f}|\rangle_{\rm space}$ with and without perturbations.
Parameter values are the same as Figure~\ref{fig:Ri-zH1} except for $\bar{z}_H=2$.
}
\label{fig:Ri-zH2}
\end{figure}

Figures~\ref{fig:zH05} and \ref{fig:zH2} show the time evolution of $N_1$ and $B-L$, Figures~\ref{fig:2-zH05} and \ref{fig:2-zH2} show the freeze-out values of $B-L$, and Figures~\ref{fig:Ri-zH05} and \ref{fig:Ri-zH2} show the dependence on $|\mathcal{R}_i|$ for $\bar{z}_H=0.5$ and $2$.
These figures correspond to Figures~\ref{fig:1}--\ref{fig:Ri-zH1} in the main text, in which $\bar{z}_H=1$ is adopted.

\section{Derivation of the Boltzmann equation}
\label{Appendix:A}

\subsection{The Boltzmann equation}

We begin with the Boltzmann equation~\cite{Senatore:2008vi}:
\beq
\label{eq:II-101}
\dfrac{df}{d\lambda}=C[f]
\quad\Rightarrow\quad
P^0\dfrac{df}{d\eta}=C[f],
\eeq
where $f$ is the distribution function, $\lambda$ is the affine parameter, $C[f]$ is the collision term, and $P^0$ is the zeroth component of the energy-momentum 4-vector $P^\mu$.
We integrate the former equation using the integral measure in momentum space defined as
\beq
\label{eq:II-102}
\pi=g_{\rm deg}\dfrac{d^3p}{E},
\eeq
where $g_{\rm deg}$ represents the internal degrees of freedom of the species, and $E$ and $p^i$ are the energy and 3-momentum of the particle evaluated in the local inertial frame, respectively.

\subsection{Equivalence of integral measures}

The integral measure (\ref{eq:II-102}) evaluated in an arbitrary coordinate system is given by~\cite{Senatore:2008vi}
\beq
\label{eq:A-1}
\pi\equiv g_{\rm deg}\dfrac{\sqrt{-g}\,d^3P}{P_0}
=g_{\rm deg}\dfrac{d^3P}{P^0\sqrt{-g}}.
\eeq
To check this, note that the relation between the 4-momenta evaluated in the general coordinate system and in the local inertial system is
\beq
\label{eq:A-2}
P^{\mu}={e^{\mu}}_{\nu'}p^{\nu'},
\eeq
where ${e^{\mu}}_{\nu'}$ is the tetrad, and $p^0=-p_0=E$ is the instantaneous energy of the particle with respect to the local observer.
The tetrad satisfies ${e^{\mu}}_{\alpha'}{e^{\nu}}_{\beta'}g_{\mu\nu}=\eta_{\alpha'\beta'}$.
Using (\ref{eq:A-2}) and $E\;dE=p_j\;dp^j$, we may rewrite $dP$ as
\begin{align}
dP^i&={e^i}_0\dfrac{\partial p^0}{\partial p^j}\;dp^j
+{e^i}_k\dfrac{\partial p^k}{\partial p^j}\;dp^j\notag\\
&={e^i}_0\dfrac{p_j}{E}\;dp^j+{e^i}_j\;dp^j
=\left({e^i}_0\dfrac{p_j}{E}+{e^i}_j\right)dp^j.
\end{align}
Substituting this into $dP$ in (\ref{eq:A-1}) and using
\beq
P_0
=g_{0\mu}{e^{\mu}}_{\nu'}p^{\nu'}
=\eta_{\nu'\sigma'}{(e^{-1})^{\sigma'}}_0p^{\nu'},
\eeq
the integral measure indeed reduces to
\beq
\label{eq:A-3}
\pi=g_{\rm deg}\dfrac{d^3p}{E}.
\eeq

\subsection{Equivalence of (\ref{eq:II-103}) and the integral of (\ref{eq:II-101})}

In this subsection we show the equivalence of equation (\ref{eq:II-103}) and the integral of (\ref{eq:II-101}) following Ref.~\cite{Cercignani:2002}.
The momentum integral of the Boltzmann equation (\ref{eq:II-101}) is
\beq
\label{eq:A-4}
\dfrac{g_{\rm deg}}{(2\pi)^3}\int\dfrac{\sqrt{-g}\;d^3P}{P_0}\;\dfrac{df}{d\lambda}
=\dfrac{g_{\rm deg}}{(2\pi)^3}\int\dfrac{d^3p}{E}\;C[f],
\eeq
where we used the equivalence of the integral measures (\ref{eq:A-1}) and (\ref{eq:A-3}).
We aim to show that this equation is equivalent to equation (\ref{eq:II-103}).
For this purpose it is sufficient to consider the collisionless Boltzmann equation, since the collision terms are already identical.
Let $f$ denote the distribution function, and consider
\beq
\label{eq:A-5}
\dfrac{df}{d\lambda}
=\dfrac{dx^{\mu}}{d\lambda}\dfrac{\partial f}{\partial x^{\mu}}
+\dfrac{dP^i}{d\lambda}\dfrac{\partial f}{\partial P^i}=0,
\eeq
that is
\beq
\label{eq:A-6}
P^{\mu}\dfrac{\partial f}{\partial x^{\mu}}
-{\Gamma^i}_{\mu\nu}P^{\mu}P^{\nu}\dfrac{\partial f}{\partial P^i}=0.
\eeq
Integrating both sides of equation (\ref{eq:A-6}) with the integral measure
$(\sqrt{-g}^2/P_0)\;d^3P\;d^4x$ gives
\begin{align}
&\int\left(P^{\mu}\dfrac{\partial f}{\partial x^{\mu}}
-{\Gamma^i}_{\mu\nu}P^{\mu}P^{\nu}\dfrac{\partial f}{\partial P^i}\right)
\dfrac{\sqrt{-g}^2}{P_0}\;d^3P\;d^4x\notag\\
&=\int\dfrac{\partial}{\partial x^{\mu}}
\left(P^{\mu}f\dfrac{\sqrt{-g}^2}{P_0}\right)\;d^3P\;d^4x\notag\\
&\quad
-\int\dfrac{\partial}{\partial P^i}
\left({\Gamma^i}_{\mu\nu}P^{\mu}P^{\nu}f\dfrac{\sqrt{-g}^2}{P_0}\right)\;
d^3P\;d^4x=0.\label{eq:A-7}
\end{align}
The second line of equation (\ref{eq:A-7}) can be rewritten as
\begin{align}
&\int\dfrac{\partial}{\partial x^{\mu}}
\left(P^{\mu}f\dfrac{\sqrt{-g}^2}{P_0}\right)\;d^3P\;d^4x\notag\\
&=\int\biggl[\dfrac{\partial}{\partial x^{\mu}}
\left(\int P^{\mu}f\sqrt{-g}\;\dfrac{d^3P}{P_0}\right)\notag\\
&\hspace{3em}
+{\Gamma^{\kappa}}_{\mu\kappa}
\int P^{\mu}f\sqrt{-g}\;\dfrac{d^3P}{P_0}\biggr]\sqrt{-g}\;d^4x\notag\\
&=\int\left(\int P^{\mu}f\sqrt{-g}\;\dfrac{d^3P}{P_0}\right)_{;\mu}
\sqrt{-g}\;d^4x,\label{eq:A-8}
\end{align}
where we used
\begin{align}
{A^{\mu}}_{;\mu}
&=\dfrac{\partial A^{\mu}}{\partial x^{\mu}}+{\Gamma^{\mu}}_{\mu\sigma} A^{\sigma}
\\[0.25em]
&=\dfrac{\partial A^{\mu}}{\partial x^{\mu}}+\dfrac{\partial \ln \sqrt{-g}}{\partial x^{\mu}} A^{\mu}
=\dfrac{1}{\sqrt{-g}} \dfrac{\partial(\sqrt{-g}\,A^{\mu})}{\partial x^{\mu}},
\end{align}
which follows from
\beq
{\Gamma^{\nu}}_{\mu\nu}
=\dfrac{1}{2} g^{\nu\sigma} \dfrac{\partial g_{\nu\sigma}}{\partial x^{\mu}}
=\dfrac{\partial \ln \sqrt{-g}}{\partial x^{\mu}}.
\eeq
Also, the left-hand side in the third line of equation (\ref{eq:A-7}) vanishes because the volume integral in momentum space can be rewritten as a surface term at infinity where the distribution function $f$ approaches zero.
Thus, equation (\ref{eq:A-7}) can be simplified as
\beq
\int\left(\int P^{\mu}f\sqrt{-g}\;\dfrac{d^3P}{P_0}\right)_{;\mu}
\sqrt{-g}\;d^4x=0.
\eeq
Since this holds for any integral over spacetime,
the integrand must vanish.
Thus we get
\beq
\label{eq:A-9}
\left(\int P^{\mu}f\sqrt{-g}\;\dfrac{d^3P}{P_0}\right)_{;\mu}=0.
\eeq
On the other hand, number density current $\mathcal{N}^{\mu}=nU^{\mu}$ is defined as
\beq
\label{eq:A-10}
\mathcal{N}^{\mu}\equiv
\dfrac{g_{\rm deg}}{(2\pi)^3}\int P^{\mu}f\sqrt{-g}\;\dfrac{d^3P}{P_0},
\eeq
and hence the momentum integral of the collisionless Boltzmann equation (\ref{eq:A-5}) becomes
\beq
\label{eq:A-11}
{\mathcal{N}^{\mu}}_{;\mu}=0.
\eeq
In other words, the left-hand side of equation (\ref{eq:A-4}) can be replaced with ${\mathcal{N}^{\mu}}_{;\mu}$, which yields the Boltzmann equation in the form of equation (\ref{eq:II-103}).

\subsection{Calculation of the covariant derivative of the number density current}

In this subsection we calculate equation (\ref{eq:II-105}) in detail.
In the following we omit the index $N_1$ in $n_{N_1}$.
By using equation (\ref{eq:II-104}) and formulas in Appendix~\ref{Appendix:B},
\begin{align}
&(nU^{\mu})_{;\mu}\notag\\
&=(nU^{\mu})_{,\mu}+{\Gamma^{\mu}}_{\mu\nu}(nU^{\nu})\notag\\
&=(nU^0)_{,0}+{\Gamma^{0}}_{0\nu}(nU^{\nu})
+(nU^i)_{,i}+{\Gamma^{i}}_{i\nu}(nU^i)\notag\\
&=\left[\dfrac{n}{a}(1-A)\right]'
+{\Gamma^0}_{00}(nU^0)
+{\Gamma^0}_{0i}(nU^i)\notag\\
&\quad
+\left(\dfrac{n}{a}v^i\right)_{,i}
+{\Gamma^i}_{i0}(nU^0)
+{\Gamma^i}_{ij}(nU^j)\notag\\
&=\dfrac{n'}{a}(1-A)
-\dfrac{a'}{a^2}n(1-A)
-\dfrac{n}{a}A'\notag\\
&\quad
+\left(\dfrac{a'}{a}+A'\right)\dfrac{n}{a}(1-A)
-\left(A_{,i}-\dfrac{a'}{a}B_{,i}\right)\dfrac{n}{a}v^i
+\dfrac{n}{a}{v^i}_{,i}\notag\\
&\quad
+\left[\dfrac{a'}{a}{\delta^i}_i
+C'{\delta^i}_i
+\left({E^{,i}}_{,i}-\dfrac{1}{3}{\delta^i}_i\nabla^2E\right)'\right]
\dfrac{n}{a}(1-A)\notag\\
&\quad
+(\mbox{1st-order})\times(\mbox{1st-order}).
\end{align}
Dropping the terms of second or higher order in perturbations,
\begin{align}
&(nU^{\mu})_{;\mu}\notag\\
&\simeq
\dfrac{n'}{a}(1-A)
-\dfrac{a'}{a^2}n(1-A)
-\dfrac{n}{a}A'\notag\\
&\hspace{4.5em}\,
+\dfrac{a'}{a}\dfrac{n}{a}(1-A)
+A'\dfrac{n}{a}
+2\dfrac{n}{a}{v^i}_{,i}\notag\\
&\hspace{4.5em}\,
+3\left(\dfrac{a'}{a}+C'\right)\dfrac{n}{a}
-3\dfrac{a'}{a}A\dfrac{n}{a}\notag\\
&=\dfrac{n'}{a}(1-A)
+\dfrac{n}{a}{v^i}_{,i}
+3\dfrac{a'}{a}\dfrac{n}{a}
+3C'\dfrac{n}{a}
-3\dfrac{a'}{a}A\dfrac{n}{a}\notag\\
&=(1-A)\biggl(a^{-1}n'+3\dfrac{a'}{a^2}n\biggr)
+\dfrac{n}{a}({v^i}_{,i}+3C')\notag\\
&=\bar{z}H(1-A)\biggl(\dfrac{dn}{d\bar{z}}+\dfrac{3}{\bar{z}}n\biggr)
+\dfrac{n}{a}({v^i}_{,i}+3C'),
\label{eq:A-12}
\end{align}
is obtained.
Here we used
\begin{align}
\dfrac{d}{d\eta}
&=a\dfrac{d}{dt}
=a\bar{z}H\dfrac{d}{d\bar{z}},
\label{eq:A-add-1}
\end{align}
that follows from entropy conservation.

\subsection{Calculation of the collision term}

In this subsection we calculate equation (\ref{eq:II-106}) in detail.
The distribution function of right-handed neutrinos, defined with the normalization constant $\mathcal{C}$, is given by
\beq
\label{eq:A-13}
f_{N_1}=\mathcal{C}
\exp\left[-\dfrac{(p_{N_1}^2+M_1^2)^{1/2}-\mu_{N_1}}{\bar{T}+\delta T}\right].
\eeq
By integrating this equation over the entire momentum space,
we obtain the particle number density
\begin{align}
n_{N_1}(\bar{z})&=\dfrac{g_{N_1}\mathcal{C}}{2\pi^2}\int_{0}^{\infty}dp_{N_1}\;p_{N_1}^2\notag\\
&\quad\times
\exp\left[-\dfrac{(p_{N_1}^2+M_1^2)^{1/2}-\mu_{N_1}}{\bar{T}+\delta T}\right].
\end{align}
Changing the variable to $p_{N_1}=M_1\sinh\theta$ on the right-hand side, and introducing $z_T(\bar{z})=\bar{z}/(1+\delta_T)$, the integral reduces to
\begin{align}
n_{N_1}(\bar{z})=\dfrac{g_{N_1}\mathcal{C}}{8\pi^2}M_1^3
\int_{0}^{\infty}d\theta\;(\cosh 3\theta-\cosh\theta)\,
\mathrm{e}^{-z_T\cosh\theta}.
\end{align}
Here note that the modified Bessel function of the second kind $K_{\nu}$
is given by the integral
\beq
\label{eq:A-14}
K_{\nu}(z)=\int_{0}^{\infty}d\theta\;\cosh(\nu\theta)\,
\mathrm{e}^{-z\cosh\theta},
\eeq
from which follows
\beq
\label{eq:A-15}
K_{\nu+1}(z)-K_{\nu-1}(z)
=\dfrac{2\nu}{z}K_{\nu}(z).
\eeq
Then the normalization constant is determined to be
\beq
\mathcal{C}
=\dfrac{2\pi^2z_T}{g_{N_1}M_1^3K_2(z_T)}\,
n_{N_1}(\bar{z})\,\mathrm{e}^{-\mu_{N_1}/(\bar{T}+\delta T)},
\eeq
and the distribution function of right-handed neutrinos becomes
\beq
\label{eq:A-16}
f_{N_1}=\dfrac{2\pi^2 z_T n_{N_1}(\bar{z})}{g_{N_1} M_1^3 K_2(z_T)}
\exp\left[-\dfrac{(p_{N_1}^2 + M_1^2)^{1/2}}{\bar{T} + \delta T}\right],
\eeq
which rewrite as $f_{N_1}=\mathcal{C}'n_{N_1}(\bar{z})\,\mathrm{e}^{-E_{N_1}/(\bar{T}+\delta T)}$.
On the other hand, assuming that lepton $l$ and Higgs $\phi$ are in thermal equilibrium and denoting their distribution functions by $f_l^{\rm eq}$ and $f_{\phi}^{\rm eq}$, respectively, the momentum integral of the collision term in the Boltzmann equation for $N_1$ is
\begin{align}
&\dfrac{g_{N_1}}{(2\pi)^3}\int\dfrac{d^3p_{N_1}}{E_{N_1}}\;C[f_{N_1}]\notag\\
&=\int d\Pi_{N_1}\int d\Pi_l\int d\Pi_{\phi}\;
(2\pi)^4|\mathcal{M}|^2\delta^4(p_{N_1}-p_l-p_{\phi})\notag\\
&\quad\times
(f_l^{\rm eq}f_{\phi}^{\rm eq}-f_{N_1})\notag\\
&=-\int d\Pi_{N_1}\int d\Pi_l\int d\Pi_{\phi}\;
(2\pi)^4|\mathcal{M}|^2\delta^4(p_{N_1}-p_l-p_{\phi})\notag\\
&\quad\times
\mathcal{C}'\left[n_{N_1}(\bar{z})\,\mathrm{e}^{-E_{N_1}/(\bar{T}+\delta T)}
-n_l^{\rm eq}n_{\phi}^{\rm eq}\,\mathrm{e}^{-(E_l+E_{\phi})/(\bar{T}+\delta T)}
\right]\notag\\
&\simeq -\int d\Pi_{N_1}\int d\Pi_l\int d\Pi_{\phi}\;
(2\pi)^4|\mathcal{M}|^2\delta^4(p_{N_1}-p_l-p_{\phi})\notag\\
&\quad\times
\mathcal{C}'\,\mathrm{e}^{-E_{N_1}/(\bar{T}+\delta T)}
\bigl[n_{N_1}(\bar{z})-n_{N_1}^{\rm eq}(z_T)\bigr],
\label{eq:A-17}
\end{align}
where we used the energy conservation law $E_{N_1}=E_l+E_{\phi}$ and detailed balance at high temperatures $n_{N_1}^{\rm eq}\simeq n_l^{\rm eq}n_{\phi}^{\rm eq}$.
Note that the matrix elements for $N_1\to l_{\alpha}\phi$ and $N_1\to\bar{l}_{\alpha}\phi^{\ast}$ do not depend on $p_{N_1}$, $p_l$, and $p_{\phi}$ at the tree level.
Let us proceed to perform the integrals over $d\Pi_{\phi}$, $d\Pi_l$, and $d\Pi_{N_1}$ sequentially.
First, for the $d\Pi_{\phi}$ integral we get
\begin{align}
&\dfrac{g_{N_1}}{(2\pi)^3}\int\dfrac{d^3p_{N_1}}{E_{N_1}}\;C[f_{N_1}]\notag\\
&=-8\pi^2|\mathcal{M}|^2
\int\dfrac{d^3p_{N_1}}{(2\pi)^3}\;\dfrac{g_{N_1}}{2E_{N_1}}
\int\dfrac{d^3p_l}{(2\pi)^3}\;\dfrac{g_l}{2p_l}
\dfrac{g_{\phi}}{2(E_{N_1}-p_l)}\notag\\
&\times
\delta(E_{N_1}-p_l-p_{\phi})\,
\mathcal{C}'\,\mathrm{e}^{-E_{N_1}/(\bar{T}+\delta T)}
\bigl[n_{N_1}(\bar{z})-n_{N_1}^{\rm eq}(z_T)\bigr].
\end{align}
Here we approximated the lepton $l$ and Higgs $\phi$ to be massless, assuming that their masses are sufficiently smaller than that of the right-handed neutrino $N_1$.
Thus, performing the integral over $d^3p_l=4\pi p_l^2\;dp_l$ we get
\begin{align}
&\dfrac{g_{N_1}}{(2\pi)^3}\int\dfrac{d^3p_{N_1}}{E_{N_1}}\;C[f_{N_1}]\notag\\
&=-|\mathcal{M}|^2
\int\dfrac{d^3p_{N_1}}{(2\pi)^3}\;\dfrac{g_{N_1}}{2E_{N_1}}
\dfrac{g_lg_{\phi}(E_{N_1}-p_{\phi})}{E_{N_1}-p_l}\notag\\
&\quad\times
\mathcal{C}'\,\mathrm{e}^{-E_{N_1}/(\bar{T}+\delta T)}
\bigl[n_{N_1}(\bar{z})-n_{N_1}^{\rm eq}(z_T)\bigr].
\end{align}
Recalling that $l$ and $\phi$ are in thermal equilibrium, we take thermal average over these species.
Since both $l$ and $\phi$ are massless we can set $\langle p_l\rangle=\langle p_{\phi}\rangle$, which yields
\begin{align}
&\left\langle\dfrac{g_{N_1}}{(2\pi)^3}\int\dfrac{d^3p_{N_1}}{E_{N_1}}\;C[f_{N_1}]\right\rangle\notag\\
&=-g_{N_1}g_lg_{\phi}|\mathcal{M}|^2
\int\dfrac{d^3p_{N_1}}{(2\pi)^3}\notag\\
&\quad\times
\dfrac{\mathcal{C}'\,\mathrm{e}^{-E_{N_1}/(\bar{T}+\delta T)}}{2E_{N_1}}
\bigl[n_{N_1}(\bar{z})-n_{N_1}^{\rm eq}(z_T)\bigr].
\label{eq:A-18}
\end{align}
By changing the variable to $p_{N_1}=M_1\sinh\theta$,
equation (\ref{eq:A-18}) becomes
\begin{align}
&\mathcal{C}'\int_{0}^{\infty}
\dfrac{4\pi p_{N_1}^2\;dp_{N_1}}{2(p_{N_1}^2+M_1^2)^{1/2}}\,
\mathrm{e}^{-(p_{N_1}^2+M_1^2)^{1/2}/(\bar{T}+\delta T)}\notag\\
&=2\pi\mathcal{C}'M_1^2
\int_{0}^{\infty}d\theta\;\sinh^2\theta\,
\mathrm{e}^{-z_T\cosh\theta}\notag\\
&=\dfrac{4\pi^3}{g_{N_1}M_1}\dfrac{K_1(z_T)}{K_2(z_T)},
\label{eq:A-19}
\end{align}
where we used equations (\ref{eq:A-14}) and (\ref{eq:A-15}).
Substituting equation (\ref{eq:A-19}) into equation (\ref{eq:A-18}), we obtain
\begin{align}
&\left\langle\dfrac{g_{N_1}}{(2\pi)^3}\int\dfrac{d^3p_{N_1}}{E_{N_1}}\;C
\right\rangle\notag\\
&=-\dfrac{4\pi^3g_lg_{\phi}|\mathcal{M}|^2}{M_1}
\dfrac{K_1(z_T)}{K_2(z_T)}\bigl[n_{N_1}(\bar{z})-n_{N_1}^{\rm eq}(z_T)\bigr]
\notag\\
&\equiv -\Gamma_D(\bar{z}=\infty)
\biggl\langle\dfrac{1}{\gamma}\biggr\rangle(z_T)
\bigl[n_{N_1}(\bar{z})-n_{N_1}^{\rm eq}(z_T)\bigr].
\label{eq:A-20}
\end{align}

\subsection{Derivation of \texorpdfstring{$N_{N_1}^{\rm eq}$}{NN1eq}}

In this subsection we discuss several expression of $N_{N_1}^{\rm eq}(z_T)$ which holds for different ranges of $z_T$.
Approximating with the Boltzmann distribution, the number density $n_{N_1}^{\rm eq}(z_T)$ can be approximated using the Boltzmann factor as
\beq
\label{eq:A-21}
n_{N_1}^{\rm eq}(z_T)
=\mathcal{C}_0\int\dfrac{d^3p_{N_1}}{(2\pi)^3}\;
\exp\biggl(-\dfrac{E_{N_1}}{\bar{T}+\delta T}\biggr),
\eeq
where $\mathcal{C}_0$ is a normalization constant.
After changing the variable using $E_{N_1}=(p_{N_1}^2+M_1^2)^{1/2}$ and $t=E_{N_1}/M_1$, integration by parts yields
\beq
\label{eq:A-22}
n_{N_1}^{\rm eq}(z_T)
=\mathcal{C}_0\dfrac{M_1^3z_T}{3}
\int_{1}^{\infty}dt\;(t^2-1)^{3/2}\mathrm{e}^{-z_Tt}.
\eeq
By using the integral expression for the modified Bessel function of the second kind
\beq
\label{eq:A-23}
K_{\nu}(z)
=\dfrac{\sqrt{\pi}\,(z/2)^{\nu}}{\varGamma\bigl(\nu+\frac{1}{2}\bigr)}
\int_{1}^{\infty}dt\;(t^2-1)^{\nu-\frac{1}{2}}\mathrm{e}^{-zt},
\eeq
equation (\ref{eq:A-22}) becomes
\beq
\label{eq:A-24}
n_{N_1}^{\rm eq}(z_T)
=\mathcal{C}_0\dfrac{M_1^3z_T}{3}\cdot\dfrac{3}{z_T^2}K_2(z_T)
=\mathcal{C}_0\dfrac{M_1^3}{z_T}K_2(z_T).
\eeq
If the photon number density is normalized $n_{\gamma}^{\rm eq}(z_T)\propto (\bar{T}+\delta T)^3$,
\beq
\label{eq:A-25}
N_{N_1}^{\rm eq}(z_T)
=\dfrac{n_{N_1}^{\rm eq}(z_T)}{n_{\gamma}^{\rm eq}(z_T)}
=\mathcal{C}_0'z_T^2K_2(z_T).
\eeq
with some normalization constant $\mathcal{C}_0'$.
For equation (\ref{eq:A-25}) to hold in the relativistic limit $z_T\ll 1$, it must reduce to (\ref{eq:I-7}) in this limit.
Since
\beq
\label{eq:A-26}
K_2(z)\simeq\dfrac{2}{z}K_1(z)\simeq\dfrac{2}{z^2},\qquad
z\ll 1,
\eeq
the normalization constant is fixed from
\beq
\label{eq:A-27}
\dfrac{3}{4}=N_{N_1}^{\rm eq}(z_T\ll 1)
\simeq\mathcal{C}_0'z_T^2\cdot\dfrac{2}{z_T^2}
=2\mathcal{C}_0',
\eeq
as $\mathcal{C}_0'=3/8$.
From this and equation (\ref{eq:A-25}), we obtain
\beq
\label{eq:A-28}
N_{N_1}^{\rm eq}(z_T)
=\dfrac{3}{8}z_T^2K_2(z_T).
\eeq

\subsection{Derivation of the washout factor \texorpdfstring{$W_{\rm ID}$}{WID}}

In this subsection we derive the expression for the washout factor $W_{\rm ID}$.
The decay rate of $N_1$ is given by
\beq
\label{eq:A-32}
\Gamma_{\rm D}(z)
=\tilde{\Gamma}_{\rm D}\biggl\langle\dfrac{1}{\gamma}\biggr\rangle
=\tilde{\Gamma}_{\rm D}\dfrac{K_1(z_T)}{K_2(z_T)},
\eeq
where $\tilde{\Gamma}_{\rm D}=\Gamma_{\rm D}(z=\infty)$ from $\lim_{z\to\infty}K_1(z)/K_2(z)=1$.

The inverse decay rate $\Gamma_{\rm ID}$ is related to the decay rate $\Gamma_{\rm D}$ through detailed balance
\beq
\label{eq:A-29}
\Gamma_{\rm ID}(z)
=\Gamma_{\rm D}(z)
\dfrac{N_{N_1}^{\rm eq}(z_T)}{N_l^{\rm eq}N_{\phi}^{\rm eq}}.
\eeq
The equilibrium number densities of heavy neutrinos, lepton doublets, and Higgs doublets are, as already derived,
\beq
\label{eq:A-30}
N_{N_1}^{\rm eq}(z_T)
=\dfrac{3}{8}z_T^2K_2(z_T),\quad
N_l^{\rm eq}=\dfrac{3}{4},\quad
N_{\phi}^{\rm eq}=1.
\eeq
Hence the contribution of inverse decays to the washout term $W_{\rm ID}$ is
\beq
\label{eq:A-31}
W_{\rm ID}(z_T)
\equiv\dfrac{1}{2}\dfrac{\Gamma_{\rm ID}(z_T)}{\bar{z}H(\bar{z})}
=\dfrac{1}{4}z_T^2K_2(z_T)\dfrac{\Gamma_{\rm D}(z_T)}{\bar{z}H(\bar{z})}.
\eeq
Substituting equation (\ref{eq:A-32}) and $K\equiv\tilde{\Gamma}_{\rm D}/\bar{z}^2H(\bar{z})$ into equation (\ref{eq:A-31}), the washout factor is found to be
\begin{align}
W_{\rm ID}(z_T)
&=\dfrac{1}{4}z_T^2K_2(z_T)
\dfrac{\tilde{\Gamma}_{\rm D}}{\bar{z}H(\bar{z})}
\cdot\dfrac{K_1(z_T)}{K_2(z_T)}\notag\\
&=\dfrac{1}{4}K\bar{z}z_T^2K_1(z_T).\label{eq:A-41}
\end{align}

\section{Perturbative Christoffel symbols}
\label{Appendix:B}

In this appendix we summarize the metric (\ref{eq:II-8})--(\ref{eq:II-10}) and its inverse, together with the Christoffel symbols.
The metric and its inverse are given by
\begin{align}
g_{00}&=-a^2(1+2A),\qquad
g^{00}=-a^{-2}(1-2A),\label{eq:B-1}\\
g_{0i}&=-a^2B_{,i},\hspace{4.5em}
g^{0i}=-a^{-2}B^{,i},\label{eq:B-2}\\
g_{ij}&=a^2\biggl[(1+2C)\delta_{ij}+2\biggl(E_{,ij}-\dfrac{1}{3}\delta_{ij}\nabla^2E\biggr)\biggr],\\
g^{ij}&=a^{-2}\biggl[(1-2C)\delta^{ij}-2\biggl(E^{,ij}-\dfrac{1}{3}\delta^{ij}\nabla^2E\biggr)\biggr].\label{eq:B-3}
\end{align}
The Christoffel symbols up to first order in perturbations are given by
\begin{align}
{\Gamma^0}_{00}
&=\dfrac{a'}{a}+A',\label{eq:B-4}\\
{\Gamma^0}_{0i}
&=A_{,i}-\dfrac{a'}{a}B_{,i},\label{eq:B-5}\\[0.25em]
{\Gamma^0}_{ij}
&=\dfrac{a'}{a}\delta_{ij}
+\biggl[-2\dfrac{a'}{a}A+\dfrac{(a^2C)'}{a^2}\biggr]\delta_{ij}\notag\\[0.25em]
&\quad
+B_{,ij}
+\dfrac{1}{a^2}
\biggl[a^2\biggl(E_{,ij}-\frac{1}{3}\delta_{ij}\nabla^2E\biggr)\biggr]',
\label{eq:B-6}\\
{\Gamma^i}_{00}
&=A^{,i}-\biggl(B'+\dfrac{a'}{a}B\biggr)^{,i},\label{eq:B-7}\\
{\Gamma^i}_{0j}
&=\dfrac{a'}{a}{\delta^i}_j
+C'{\delta^i}_j
+\biggl({E^{,i}}_{,j}-\dfrac{1}{3}{\delta^i}_j\nabla^2E\biggr)',\label{eq:B-8}\\
{\Gamma^i}_{jk}
&=\left(C_{,k}{\delta^i}_j+C_{,j}{\delta^i}_k-C^{,i}\delta_{jk}\right)
+\dfrac{a'}{a}\delta_{jk}B^{,i}\notag\\
&\quad
+{E^{,i}}_{,jk}
-\dfrac{1}{3}\nabla^2
\bigl({\delta^i}_jE_{,k}+{\delta^i}_kE_{,j}-\delta_{jk}E^{,i}\bigr).\label{eq:B-9}
\end{align}

\section{Calculation of CP asymmetry \texorpdfstring{$\varepsilon_1$}{varepsilon1}}
\label{Appendix:C}

In this appendix we calculate the CP asymmetry $\varepsilon_1$ from the interference between tree and one-loop diagrams in the decay of $N_1$.

\subsection{Prerequisites}

The relevant Yukawa terms in the Lagrangian are
\begin{align}
\mathcal{L}
&\supset
-y_{\alpha\beta}^{(e)}(\bar{l}_{\rm L})_{\alpha}\phi(e_{\rm R})_{\beta}
-y_{\alpha\beta}^{(\nu)}[(\bar{l}_{\rm L})_{\alpha}\circ\phi^{\dagger}]N_{\beta}
+{\rm h.c.},\label{eq:C-1}
\end{align}
where $\alpha,\beta=1,2,3$ denote the lepton flavor, and repeated indices are implicitly summed over.
In addition, the indices L and R represent the left-chiral and right-chiral eigenvectors of the eigenspace (\ref{eq:E-5}) of chirality (\ref{eq:E-2}).

Left-chiral field $l_{\rm L}$ and Higgs field $\phi$ form $\mathrm{SU}(2)$ doublets as
\beq
\label{eq:C-2}
l_{\rm L}
=\begin{pmatrix}\nu_{\rm L}\\e_{\rm L}\end{pmatrix}_{\mathrm{SU}(2)},
\qquad
\phi
=\begin{pmatrix}\phi^+\\\phi^0\end{pmatrix}_{\mathrm{SU}(2)}.
\eeq
The Dirac conjugate of the former is defined as
\beq
\label{eq:C-3}
\bar{l}_{\rm L}=\begin{pmatrix}\bar{\nu}_{\rm L}&\bar{e}_{\rm L}\end{pmatrix}_{\mathrm{SU}(2)},
\eeq
where $\nu_{\rm L}$, $\nu_{\rm R}$, $e_{\rm L}$, $e_{\rm R}$ are all Weyl fields.
The Dirac conjugates for them are defined by promoting to Dirac fields and then projecting them onto the eigenspaces of the chirality $\gamma_5$.
To do so we introduce dummy fields $\nu_{\rm R}'$ and $e_{\rm R}'$ as
\beq
\label{eq:C-4}
\nu=\begin{pmatrix}\nu_{\rm L}\\[0.25em]
\nu_{\rm R}'\end{pmatrix}_{\rm Dirac},
\qquad
e=\begin{pmatrix}e_{\rm L}\\[0.25em]
e_{\rm R}'\end{pmatrix}_{\rm Dirac},
\eeq
and then $\bar{\nu}_{\rm L}$ and $\bar{e}_{\rm L}$ satisfy
\begin{align}
\bar{\nu}_{\rm L}
&=\overline{P_{\rm L}\nu}
=(P_{\rm L}\nu)^{\dagger}\gamma^0
=\nu^{\dagger}P_{\rm L}\gamma^0
=\nu^{\dagger}\gamma^0P_{\rm R},\label{eq:C-5}\\
\bar{e}_{\rm L}
&=\overline{P_{\rm L}e}
=(P_{\rm L}e)^{\dagger}\gamma^0
=e^{\dagger}P_{\rm L}\gamma^0
=e^{\dagger}\gamma^0P_{\rm R},\label{eq:C-6}
\end{align}
where $P_{\rm L}$ and $P_{\rm R}$ are projection operators of chirality defined in equation (\ref{eq:E-4}).
These definitions give
\begin{align}
l
&=\begin{pmatrix}\nu\\e\end{pmatrix}_{\mathrm{SU}(2)}
=\begin{pmatrix}\begin{pmatrix}\nu_{\rm L}\\[0.25em]\nu_{\rm R}'\end{pmatrix}_{\rm Dirac}\\[1em]\begin{pmatrix}e_{\rm L}\\[0.25em]e_{\rm R}'\end{pmatrix}_{\rm Dirac}\end{pmatrix}_{\mathrm{SU}(2)}\notag\\
\Rightarrow\quad
l_{\rm L}
&=P_{\rm L}l
=\begin{pmatrix}P_{\rm L}\nu\\P_{\rm L}e\end{pmatrix}_{\mathrm{SU}(2)}
=\begin{pmatrix}\nu_{\rm L}\\e_{\rm L}\end{pmatrix}_{\mathrm{SU}(2)}.\label{eq:C-7}
\end{align}

The product $\bar{l}_{\rm L}\circ\phi^{\dagger}$ is defined so that it is invariant under $\mathrm{SU}(2)$ [Note that $\mathrm{i}\sigma^2=\bigl(\begin{smallmatrix}0&1\\-1&0\end{smallmatrix}\bigr)$]:
\begin{align}
\bar{l}_{\rm L}\circ\phi^{\dagger}
&=\begin{pmatrix}\bar{\nu}_{\rm L}&\bar{e}_{\rm L}
\end{pmatrix}_{\mathrm{SU}(2)}
\circ\begin{pmatrix}(\phi^+)^{\dagger}\\(\phi^0)^{\dagger}
\end{pmatrix}_{\mathrm{SU}(2)}\notag\\
&=\begin{pmatrix}\bar{\nu}_{\rm L}&\bar{e}_{\rm L}
\end{pmatrix}_{\mathrm{SU}(2)}
\begin{pmatrix}0&1\\-1&0\end{pmatrix}
\begin{pmatrix}(\phi^+)^{\dagger}\\(\phi^0)^{\dagger}
\end{pmatrix}_{\mathrm{SU}(2)}\notag\\
&=\bar{\nu}_{\rm L}(\phi^0)^{\dagger}-\bar{e}_{\rm L}(\phi^+)^{\dagger},\label{eq:C-8}
\end{align}
which can be rewritten as
\beq
\label{eq:C-9}
\bar{l}_{\rm L}\circ\phi^{\dagger}
=\overline{P_{\rm L}l}\circ\phi^{\dagger}
=(\bar{l}P_{\rm R})\circ\phi^{\dagger}
=(\bar{l}\circ\phi^{\dagger})P_{\rm R}.
\eeq
The conjugate of the above gives $\phi\circ l_{\rm L}$:
\begin{align}
\phi\circ l_{\rm L}
&=\begin{pmatrix}\phi^+&\phi^0\end{pmatrix}_{\mathrm{SU}(2)}\circ
\begin{pmatrix}\nu_{\rm L}\\e_{\rm L}\end{pmatrix}_{\mathrm{SU}(2)}
\notag\\
&=\begin{pmatrix}\phi^+&\phi^0\end{pmatrix}_{\mathrm{SU}(2)}
\begin{pmatrix}0&1\\-1&0\end{pmatrix}
\begin{pmatrix}\nu_{\rm L}\\e_{\rm L}\end{pmatrix}_{\mathrm{SU}(2)}
\notag\\
&=\nu_{\rm L}\phi^0-e_{\rm L}\phi^+,\label{eq:C-10}
\end{align}
which can be rewritten as
\beq
\label{eq:C-11}
\phi\circ l_{\rm L}
=P_{\rm L}(\phi\circ l).
\eeq
Note that the sign of the operation $\circ$ changes depending on whether the fields are $\mathrm{SU}(2)$ fundamental or $\mathrm{SU}(2)$ anti-fundamental.

The Majorana field $N$ is the Majorana representation of the right-chiral Weyl field $\nu_{\rm R}$ given by
\beq
\label{eq:C-12}
N=\begin{pmatrix}(\nu_{\rm R})^c\\\nu_{\rm R}\end{pmatrix}_{\rm Dirac},
\eeq
where the charge conjugation operator $c$ is defined as
\beq
\label{eq:C-13}
\begin{array}{l}
(\mbox{left-chiral})^c
=\mathrm{i}\sigma^2(\mbox{left-chiral})^{\dagger},\\[0.75em]
(\mbox{right-chiral})^c
=-\mathrm{i}\sigma^2(\mbox{right-chiral})^{\dagger},
\end{array}
\eeq
thereby $(\nu_{\rm R})^c=-\mathrm{i}\sigma^2\nu_{\rm R}^{\dagger}$.
Thus the relevant Yukawa couplings become
\begin{align}
\mathcal{L}_{\rm int}
&=-y_{\alpha\beta}^{(\nu)}[(\bar{l}_{\rm L})_{\alpha}\circ\phi^{\dagger}]N_{\beta}+{\rm h.c.}\notag\\
&=-y_{\alpha\beta}^{(\nu)}(\bar{l}_{\alpha}\circ\phi^{\dagger})P_{\rm R}N_{\beta}
-y_{\alpha\beta}^{(\nu)\ast}\bar{N}_{\beta}P_{\rm L}(\phi\circ l_{\alpha}).\label{eq:C-14}
\end{align}
Henceforth, we denote $y_{\alpha\beta}^{(\nu)}$ by $y_{\alpha\beta}$.

\subsection{Feynman diagrams}

We list the Feynman diagrams that appear in the one-loop calculation of $\varepsilon_1$.
For the Feynman rules for Majorana fermions, refer to Refs.~\cite{DENNER1992278,DENNER1992467}.

\vspace{-1em}
\subsubsection{Tree diagrams}

\begin{center}
\begin{tikzpicture}[baseline=-0.1cm]
\begin{feynhand}
\vertex (i1) at (-3,0);
\vertex (w2) at (0,0);
\vertex (f3) at (3,1.5);
\vertex (f4) at (3,-1.5);
\vertex (w5) at (-0.25,0.3) {$y_{\alpha1}P_{\rm R}$};
\propag [plain, black, mom'={[arrow style=black] $p_1+p_2$}] (i1) to [edge label=$N_1$] (w2);
\propag [fer, black, mom'={[arrow style=black] $p_1$}] (w2) to [edge label=$l_{\alpha}$] (f3);
\propag [chasca, black, mom={[arrow style=black] $p_2$}] (w2) to [edge label'=$\phi$] (f4);
\end{feynhand}
\end{tikzpicture}\\[1em]
\begin{tikzpicture}[baseline=-0.1cm]
\begin{feynhand}
\vertex (i6) at (-3,0);
\vertex (w7) at (0,0);
\vertex (f8) at (3,1.5);
\vertex (f9) at (3,-1.5);
\vertex (w10) at (-0.25,0.3) {$y_{\alpha1}^{\ast}P_{\rm L}$};
\propag [plain, black, mom'={[arrow style=black] $p_1+p_2$}] (i6) to [edge label=$N_1$] (w7);
\propag [antfer, black, mom'={[arrow style=black] $p_1$}] (w7) to [edge label=$\bar{l}_{\alpha}$] (f8);
\propag [antsca, black, mom={[arrow style=black] $p_2$}] (w7) to [edge label'=$\phi^{\ast}$] (f9);
\end{feynhand}
\end{tikzpicture}
\end{center}

\subsubsection{Vertex diagrams}

\begin{center}
\begin{tikzpicture}[baseline=-0.1cm]
\begin{feynhand}
\vertex (i1) at (-3,0);
\vertex (w2) at (0,0);
\vertex (f3) at (3.75,1.875);
\vertex (f4) at (3.75,-1.875);
\vertex (w5) at (-0.25,0.3) {$y_{\beta1}^{\ast}P_{\rm L}$};
\vertex (w6) at (2.25,1.125);
\vertex (w7) at (2.25,-1.125);
\vertex (w8) at (1.9,1.4) {$y_{\alpha\gamma}P_{\rm R}$};
\vertex (w9) at (1.9,-1.4) {$y_{\beta\gamma}P_{\rm R}$};
\propag [plain, black, mom'={[arrow style=black] $p_1+p_2$}] (i1) to [edge label=$N_1$] (w2);
\propag [fer, black, mom'={[arrow style=black] $p_1$}] (w6) to [edge label=$l_{\alpha}$] (f3);
\propag [chasca, black, mom={[arrow style=black] $p_2$}] (w7) to [edge label'=$\phi$] (f4);
\propag [chasca, black, mom'={[arrow style=black] $-p_1+q$}] (w6) to [edge label=$\phi$] (w2);
\propag [antfer, black, mom'={[arrow style=black] $p_2+q$}] (w2) to [edge label=$\bar{l}_{\beta}$] (w7);
\propag [plain, black, mom'={[arrow style=black] $q$}] (w7) to [edge label=$N_{\gamma}$] (w6);
\end{feynhand}
\end{tikzpicture}\\[1em]
\begin{tikzpicture}[baseline=-0.1cm]
\begin{feynhand}
\vertex (i10) at (-3,0);
\vertex (w11) at (0,0);
\vertex (f12) at (3.75,1.875);
\vertex (f13) at (3.75,-1.875);
\vertex (w14) at (-0.25,0.3) {$y_{\beta1}P_{\rm R}$};
\vertex (w15) at (2.25,1.125);
\vertex (w16) at (2.25,-1.125);
\vertex (w17) at (1.9,1.4) {$y_{\alpha\gamma}^{\ast}P_{\rm L}$};
\vertex (w18) at (1.9,-1.4) {$y_{\beta\gamma}^{\ast}P_{\rm L}$};
\propag [plain, black, mom'={[arrow style=black] $p_1+p_2$}] (i10) to [edge label=$N_1$] (w11);
\propag [antfer, black, mom'={[arrow style=black] $p_1$}] (w15) to [edge label=$\bar{l}_{\alpha}$] (f12);
\propag [antsca, black, mom={[arrow style=black] $p_2$}] (w16) to [edge label'=$\phi^{\ast}$] (f13);
\propag [antsca, black, mom'={[arrow style=black] $-p_1+q$}] (w15) to [edge label=$\phi^{\ast}$] (w11);
\propag [fer, black, mom'={[arrow style=black] $p_2+q$}] (w11) to [edge label=$l_{\beta}$] (w16);
\propag [plain, black, mom'={[arrow style=black] $q$}] (w16) to [edge label=$N_{\gamma}$] (w15);
\end{feynhand}
\end{tikzpicture}
\end{center}

\subsubsection{Wavefunction diagrams I}

\begin{center}
\begin{tikzpicture}[baseline=-0.1cm]
\begin{feynhand}
\vertex (i1) at (0,0);
\vertex (w2) at (1.5,0);
\vertex (w3) at (4,0);
\vertex (w4) at (5.5,0);
\vertex (f5) at (7.5,1.75);
\vertex (f6) at (7.5,-1.75);
\vertex (w7) at (2.05,0) {$y_{\beta1}^{\ast}P_{\rm L}$};
\vertex (w8) at (3.45,0) {$y_{\alpha\gamma}P_{\rm R}$};
\vertex (w9) at (6.3,0) {$y_{\beta\gamma}P_{\rm R}$};
\propag [plain, black, mom'={[arrow style=black] $p_1+p_2$}] (i1) to [edge label=$N_1$] (w2);
\propag [plain, black, mom'={[arrow style=black] $p_1+p_2$}] (w3) to [edge label=$N_{\gamma}$] (w4);
\propag [fer, black, mom={[arrow style=black] $p_1$}] (w4) to [edge label'=$l_{\alpha}$] (f5);
\propag [chasca, black, mom'={[arrow style=black] $p_2$}] (w4) to [edge label=$\phi$] (f6);
\propag [antsca, black, mom'={[arrow style=black] $p_1+p_2+q$}] (w2) to [edge label=$\phi^{\ast}$, half right, looseness=1.5] (w3);
\propag [antfer, black, mom={[arrow style=black] $-q$}] (w2) to [edge label'=$\bar{l}_{\beta}$, half left, looseness=1.5] (w3);
\end{feynhand}
\end{tikzpicture}\\[1em]
\begin{tikzpicture}[baseline=-0.1cm]
\begin{feynhand}
\vertex (i10) at (0,0);
\vertex (w11) at (1.5,0);
\vertex (w12) at (4,0);
\vertex (w13) at (5.5,0);
\vertex (f14) at (7.5,1.75);
\vertex (f15) at (7.5,-1.75);
\vertex (w16) at (2.05,0) {$y_{\beta1}P_{\rm R}$};
\vertex (w17) at (3.45,0) {$y_{\alpha\gamma}^{\ast}P_{\rm L}$};
\vertex (w18) at (6.3,0) {$y_{\beta\gamma}^{\ast}P_{\rm L}$};
\propag [plain, black, mom'={[arrow style=black] $p_1+p_2$}] (i10) to [edge label=$N_1$] (w11);
\propag [plain, black, mom'={[arrow style=black] $p_1+p_2$}] (w12) to [edge label=$N_{\gamma}$] (w13);
\propag [antfer, black, mom={[arrow style=black] $p_1$}] (w13) to [edge label'=$\bar{l}_{\alpha}$] (f14);
\propag [antsca, black, mom'={[arrow style=black] $p_2$}] (w13) to [edge label=$\phi^{\ast}$] (f15);
\propag [chasca, black, mom'={[arrow style=black] $p_1+p_2+q$}] (w11) to [edge label=$\phi$, half right, looseness=1.5] (w12);
\propag [fer, black, mom={[arrow style=black] $-q$}] (w11) to [edge label'=$l_{\beta}$, half left, looseness=1.5] (w12);
\end{feynhand}
\end{tikzpicture}
\end{center}

\subsubsection{Wavefunction diagrams II}

\begin{center}
\begin{tikzpicture}[baseline=-0.1cm]
\begin{feynhand}
\vertex (i1) at (0,0);
\vertex (w2) at (1.5,0);
\vertex (w3) at (4,0);
\vertex (w4) at (5.5,0);
\vertex (f5) at (7.5,1.75);
\vertex (f6) at (7.5,-1.75);
\vertex (w7) at (2.05,0) {$y_{\beta1}P_{\rm R}$};
\vertex (w8) at (3.45,0) {$y_{\alpha\gamma}P_{\rm R}$};
\vertex (w9) at (6.3,0) {$y_{\beta\gamma}^{\ast}P_{\rm L}$};
\propag [plain, black, mom'={[arrow style=black] $p_1+p_2$}] (i1) to [edge label=$N_1$] (w2);
\propag [plain, black, mom'={[arrow style=black] $p_1+p_2$}] (w3) to [edge label=$N_{\gamma}$] (w4);
\propag [fer, black, mom={[arrow style=black] $p_1$}] (w4) to [edge label'=$l_{\alpha}$] (f5);
\propag [chasca, black, mom'={[arrow style=black] $p_2$}] (w4) to [edge label=$\phi$] (f6);
\propag [chasca, black, mom'={[arrow style=black] $p_1+p_2+q$}] (w2) to [edge label=$\phi$, half right, looseness=1.5] (w3);
\propag [fer, black, mom={[arrow style=black] $-q$}] (w2) to [edge label'=$l_{\beta}$, half left, looseness=1.5] (w3);
\end{feynhand}
\end{tikzpicture}\\[1em]
\begin{tikzpicture}[baseline=-0.1cm]
\begin{feynhand}
\vertex (i10) at (0,0);
\vertex (w11) at (1.5,0);
\vertex (w12) at (4,0);
\vertex (w13) at (5.5,0);
\vertex (f14) at (7.5,1.75);
\vertex (f15) at (7.5,-1.75);
\vertex (w16) at (2.05,0) {$y_{\beta1}^{\ast}P_{\rm L}$};
\vertex (w17) at (3.45,0) {$y_{\alpha\gamma}^{\ast}P_{\rm L}$};
\vertex (w18) at (6.3,0) {$y_{\beta\gamma}P_{\rm R}$};
\propag [plain, black, mom'={[arrow style=black] $p_1+p_2$}] (i10) to [edge label=$N_1$] (w11);
\propag [plain, black, mom'={[arrow style=black] $p_1+p_2$}] (w12) to [edge label=$N_{\gamma}$] (w13);
\propag [antfer, black, mom={[arrow style=black] $p_1$}] (w13) to [edge label'=$\bar{l}_{\alpha}$] (f14);
\propag [antsca, black, mom'={[arrow style=black] $p_2$}] (w13) to [edge label=$\phi^{\ast}$] (f15);
\propag [antsca, black, mom'={[arrow style=black] $p_1+p_2+q$}] (w11) to [edge label=$\phi^{\ast}$, half right, looseness=1.5] (w12);
\propag [antfer, black, mom={[arrow style=black] $-q$}] (w11) to [edge label'=$\bar{l}_{\beta}$, half left, looseness=1.5] (w12);
\end{feynhand}
\end{tikzpicture}
\end{center}

\subsection{Calculation of tree diagrams}

In this subsection we calculate the tree diagrams.
Hereafter we assume that the masses of the leptons and the Higgs boson are negligible compared to the mass of the heavy neutrino $M_1$.
We often use
\beq
\label{eq:C-15}
p_1^2=p_2^2=0,\qquad
p_1\cdot p_2=\dfrac{M_1^2}{2},
\eeq
for the momenta of the lepton $p_1$ and the Higgs boson $p_2$.
Here the second equation is derived from $(p_1+p_2)^2=M_1^2$.

First we consider $N_1\to l_{\alpha}\phi$.
Let $s$ and $t$ be the spins of the initial and final states of $N_1$ and $l_{\alpha}$, respectively.
The Lagrangian in equation (\ref{eq:C-14}) can be written as
\begin{align}
\mathcal{L}_{\rm int}&\supset
-y_{\alpha\beta}[\bar{\nu}_{\alpha}(\phi^0)^{\dagger}
-\bar{e}_{\alpha}(\phi^+)^{\dagger}]P_{\rm R}N_{\beta}.
\end{align}
Thus there are two possibilities for the final state: $\nu_{\alpha}\phi^0$ and $e_{\alpha}\phi^+$, originating from the $\mathrm{SU}(2)$ doublet.
Here the matrix element $\mathrm{i}\mathcal{M}_{N_1\to l_{\alpha}\phi}^{(s,t)}$ is defined for each channel, and hence the result obtained below must be multiplied by a factor of two when calculating the total decay rate.
The total matrix element is the sum of the tree and the one-loop:
\beq
\label{eq:C-16}
\mathrm{i}\mathcal{M}_{N_1\to l_{\alpha}\phi}^{\mbox{\scriptsize (tree+1-loop)}(s,t)}
=\mathrm{i}\mathcal{M}_{N_1\to l_{\alpha}\phi}^{\mbox{\scriptsize (tree)}(s,t)}
+\mathrm{i}\mathcal{M}_{N_1\to l_{\alpha}\phi}^{\mbox{\scriptsize (1-loop)}(s,t)}.
\eeq
The squared magnitude of the matrix element at the tree level is obtained by taking the sum over spins $s$ and $t$ and then averaging over the spin $s$ of the initial state
\begin{align}
\Bigl|\mathrm{i}\overline{\mathcal{M}}_{N_1\to l_{\alpha}\phi}^{\mbox{\scriptsize (tree)}}\Bigr|^2
&\equiv\dfrac{1}{2}\sum_{s,t}
\Bigl|\mathrm{i}\mathcal{M}_{N_1\to l_{\alpha}\phi}^{\mbox{\scriptsize (tree)}(s,t)}\Bigr|^2\notag\\
&=\dfrac{1}{2}\sum_{s,t}|\bar{u}_{l_{\alpha}}^t(p_1)(-\mathrm{i}y_{\alpha1}P_{\rm R})u_{N_1}^s(p_1+p_2)|^2\notag\\
&=\dfrac{1}{2}|y_{\alpha1}|^2\tr[(\slashed{p}_1+\slashed{p}_2+M_1)P_{\rm L}\slashed{p}_1P_{\rm R}]
\notag\\
&=|y_{\alpha1}|^2(p_1\cdot p_2)
=\dfrac{1}{2}|y_{\alpha1}|^2M_1^2,\label{eq:C-17}
\end{align}
where we used the spin sum formula (\ref{eq:E-23}), the trace formulas (\ref{eq:E-10}), (\ref{eq:E-11}), (\ref{eq:E-14}), and (\ref{eq:E-3}), as well as the approximation in equation (\ref{eq:C-15}).

Next, for $N_1\to\bar{l}_{\alpha}\phi^{\ast}$, a similar calculation yields
\begin{align}
\Bigl|\mathrm{i}\overline{\mathcal{M}}_{N_1\to\bar{l}_{\alpha}\phi^{\ast}}^{\mbox{\scriptsize (tree)}}\Bigr|^2
&\equiv\dfrac{1}{2}\sum_{s,t}
\Bigl|\mathrm{i}\mathcal{M}_{N_1\to\bar{l}_{\alpha}\phi^{\ast}}^{\mbox{\scriptsize (tree)}(s,t)}\Bigr|^2\notag\\
&=\dfrac{1}{2}\sum_{s,t}|\bar{v}_{l_{\alpha}}^t(p_1)(-\mathrm{i}y_{\alpha1}^{\ast}P_{\rm L})u_{N_1}^s(p_1+p_2)|^2\notag\\
&=\dfrac{1}{2}|y_{\alpha1}|^2\tr[(\slashed{p}_1+\slashed{p}_2-M_1)P_{\rm R}\slashed{p}_1P_{\rm L}]\notag\\
&=|y_{\alpha1}|^2(p_1\cdot p_2)
=\dfrac{1}{2}|y_{\alpha1}|^2M_1^2.\label{eq:C-18}
\end{align}

\subsection{Calculation of vertex diagrams}

In this subsection we calculate vertex diagrams and the resulting interference with the tree-level diagrams.
We first consider $N_1\to l_{\alpha}\phi$.
The combined contribution of the tree and one-loop diagrams is
\begin{align}
&\Bigl|\mathrm{i}\overline{\mathcal{M}}_{N_1\to l_{\alpha}\phi}^{\mbox{\scriptsize (tree+1-loop)}}\Bigr|^2
\equiv\dfrac{1}{2}\sum_{s,t}\Bigl|\mathrm{i}\mathcal{M}_{N_1\to l_{\alpha}\phi}^{\mbox{\scriptsize (tree+1-loop)}(s,t)}\Bigr|^2\notag\\
&\simeq\dfrac{1}{2}\sum_{s,t}\Bigl|\mathrm{i}\mathcal{M}_{N_1\to l_{\alpha}\phi}^{\mbox{\scriptsize (tree)}(s,t)}\Bigr|^2\notag\\
&\quad
+\dfrac{1}{2}\sum_{s,t}\Bigl[\Bigl(\mathrm{i}\mathcal{M}_{N_1\to l_{\alpha}\phi}^{\mbox{\scriptsize (tree)}(s,t)}\Bigr)^{\ast}
\Bigl(\mathrm{i}\mathcal{M}_{N_1\to l_{\alpha}\phi}^{\mbox{\scriptsize (1-loop)}(s,t)}\Bigr)+{\rm c.c.}\Bigr].\label{eq:C-19}
\end{align}
The interference part is
\begin{align}
&I_{\rm vertex}\notag\\
&\equiv\dfrac{1}{2}\sum_{s,t}\Bigl(\mathrm{i}\mathcal{M}_{N_1\to l_{\alpha}\phi}^{\mbox{\scriptsize (tree)}(s,t)}\Bigr)^{\ast}\Bigl(\mathrm{i}\mathcal{M}_{N_1\to l_{\alpha}\phi}^{\mbox{\scriptsize (1-loop)}(s,t)}\Bigr)\notag\\
&=\mathrm{i}y_{\alpha1}^{\ast}y_{\beta1}^{\ast}y_{\alpha\gamma}y_{\beta\gamma}\notag\\
&\quad\times
\int\!\!\dfrac{d^4q}{(2\pi)^4}\;\dfrac{-\frac{1}{2}M_1^3M_{\gamma}-M_1M_{\gamma}(p_1\cdot q)}{(q^2-M_{\gamma}^2)(-p_2-q)^2(-p_1+q)^2}\notag\\
&\equiv y_{\alpha1}^{\ast}y_{\beta1}^{\ast}y_{\alpha\gamma}y_{\beta\gamma}D_{\rm vertex},\label{eq:C-20}
\end{align}
where we used equation (\ref{eq:E-6}), the spin sum formula (\ref{eq:E-23}), and the trace formulas (\ref{eq:E-10}), (\ref{eq:E-11}), (\ref{eq:E-13}).
Furthermore, performing the $q$-integration using Feynman parametrization (\ref{eq:E-26}) and the dimensional regularization formulas (\ref{eq:E-27}) and (\ref{eq:E-28}), we get
\begin{align}
&D_{\rm vertex}
=\dfrac{1}{2}\int\!\!\dfrac{d^4q}{\mathrm{i}(2\pi)^4}\notag\\
&\quad\times
\dfrac{M_1^3M_{\gamma}+2M_1M_{\gamma}(p_1\cdot q)}{(q^2-M_{\gamma}^2+\mathrm{i}\epsilon)(-p_2-q+\mathrm{i}\epsilon)^2(-p_1+q+\mathrm{i}\epsilon)^2}\notag\\
&=\dfrac{M_1^2}{32\pi^2}\int_{0}^{1}dx\int_{0}^{1-x}dy\;\dfrac{M_1M_{\gamma}(1-y)}{M_{\gamma}^2(1-x-y)-M_1^2xy-\mathrm{i}\epsilon},\label{eq:C-21}
\end{align}
where $-\mathrm{i}\epsilon$ in the denominator comes from the $\mathrm{i}\epsilon$ prescription.

We define $m\equiv M_1/M_{\gamma}$ and perform the integration over $y$ in equation (\ref{eq:C-21}).
Considering the complex $y$-plane, there exists a pole at $y=(1-x)/(1+m^2x)\equiv a$ within the range $0\leqslant y\leqslant 1-x$ on the real axis, which we avoid by taking an upper semicircle.
We divide the integration range of $y$ into three intervals $[0,a-\epsilon']$, $[a-\epsilon',a+\epsilon']$, and $[a+\epsilon',1]$.
\begin{figure}[h]
\centering
\includegraphics{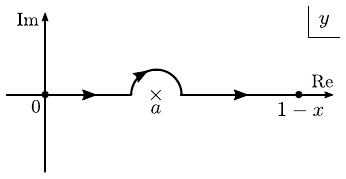}
\caption{\small
Contour of $D_{\rm vertex}$.
}
\label{fig:contour}
\end{figure}
In the middle interval $[a-\epsilon',a+\epsilon']$, we perform the variable transformation $y=a+\epsilon'\mathrm{e}^{\mathrm{i}\theta}$ (where $\epsilon'>0$ is a small quantity and $0\leqslant\theta\leqslant\pi$).
The imaginary part of $D_{\rm vertex}$ is evaluated as
\begin{flalign}
&\mathrm{i}\Im(D_{\rm vertex})\notag\\
&=\dfrac{M_1^2}{32\pi^2}\int_{0}^{1}dx\int_{a-\epsilon'}^{a+\epsilon'}dy\;\dfrac{m(1-y)}{1-x-y-m^2xy-\mathrm{i}\epsilon}\notag\\
&=-\dfrac{M_1^2}{32\pi^2}\int_{0}^{1}dx\;\dfrac{m}{1+m^2x}\int_{a-\epsilon'}^{a+\epsilon'}dy\;\dfrac{y-1}{y-\frac{1-x}{1+m^2x}+\mathrm{i}\epsilon}\notag\\
&=-\dfrac{M_1^2}{32\pi^2}\int_{0}^{1}dx\;\dfrac{m}{1+m^2x}\int_{\pi}^{0}\mathrm{i}\epsilon'\mathrm{e}^{\mathrm{i}\theta}\;d\theta\;\dfrac{a-1+\mathcal{O}(\epsilon')}{\epsilon'\mathrm{e}^{\mathrm{i}\theta}}\notag\\
&\xrightarrow[\epsilon'\downarrow 0]{}-\dfrac{M_1^2}{32\pi^2}\int_{0}^{1}dx\;(-\mathrm{i}\pi)\dfrac{m}{1+m^2x}\biggl(\dfrac{1-x}{1+m^2x}-1\biggr),\label{eq:C-22}
\end{flalign}
where the $+\mathrm{i}\epsilon$ in the denominator specifies that the integration contour lies in the upper half-plane.

Upon performing the $x$-integration, we find
\beq
\label{eq:C-23}
\Im(D_{\rm vertex})
=\dfrac{M_1^2}{32\pi}\biggl\{\dfrac{1}{m}\biggl[1-\biggl(1+\dfrac{1}{m^2}\biggr)\ln(1+m^2)\biggr]\biggr\}.
\eeq
Thus equation (\ref{eq:C-20}) becomes
\begin{align}
I_{\rm vertex}
&=\mathrm{i}y_{\alpha1}^{\ast}y_{\beta1}^{\ast} y_{\alpha\gamma}y_{\beta\gamma}\Im(D_{\rm vertex}).\label{eq:C-24}
\end{align}

We next consider $N_1\to\bar{l}_{\alpha}\phi^{\ast}$.
Similar calculation as above gives
\begin{align}
&I'_{\rm vertex}\notag\\
&\equiv\dfrac{1}{2}\sum_{s,t}
\Bigl(\mathrm{i}\mathcal{M}_{N_1\to\bar{l}_{\alpha}\phi^{\ast}}^{\mbox{\scriptsize (tree)}(s,t)}\Bigr)^{\ast}
\Bigl(\mathrm{i}\mathcal{M}_{N_1\to\bar{l}_{\alpha}\phi^{\ast}}^{\mbox{\scriptsize (1-loop)}(s,t)}\Bigr)\notag\\
&=\cdots\notag\\
&=\mathrm{i}y_{\alpha1}y_{\beta1}y_{\alpha\gamma}^{\ast}y_{\beta\gamma}^{\ast}\dfrac{M_1^2}{32\pi}\biggl\{\dfrac{1}{m}\biggl[1-\biggl(1+\dfrac{1}{m^2}\biggr)\ln(1+m^2)\biggr]\biggr\}.\label{eq:C-25}
\end{align}
Regarding the external lines, while the anti-lepton part changed from $u_{l_{\alpha}}^t$ to $v_{\bar{l}_{\alpha}}^t$, the $N_1$ part remained as $u_{N_1}^s$.
This is due to the Majorana nature of $N_1$.

Combining above results, the CP asymmetry $(\varepsilon_1)_{\rm vertex}$ is given by
\begin{align}
(\varepsilon_1)_{\rm vertex}
&=\dfrac{\sum_{\alpha}\Bigl(\Bigl|\mathrm{i}\overline{\mathcal{M}}_{N_1\to l_{\alpha}\phi}^{\mbox{\scriptsize (tree+1-loop)}}\Bigr|^2-\Bigl|\mathrm{i}\overline{\mathcal{M}}_{N_1\to\bar{l}_{\alpha}\phi^{\ast}}^{\mbox{\scriptsize (tree+1-loop)}}\Bigr|^2\Bigr)}{\sum_{\alpha}\Bigl(\Bigl|\mathrm{i}\overline{\mathcal{M}}_{N_1\to l_{\alpha}\phi}^{\mbox{\scriptsize (tree)}}\Bigr|^2+\Bigl|\mathrm{i}\overline{\mathcal{M}}_{N_1\to\bar{l}_{\alpha}\phi^{\ast}}^{\mbox{\scriptsize (tree)}}\Bigr|^2\Bigr)}\notag\\
&=\dfrac{\Im(y_{\alpha1}y_{\beta1}y_{\alpha\gamma}^{\ast}y_{\beta\gamma}^{\ast})}{8\pi\sum_{\alpha}|y_{\alpha1}|^2}F(m),\qquad
m\equiv\dfrac{M_1}{M_{\gamma}},\notag\\
F(m)
&\equiv
\dfrac{1}{m}\biggl[1-\biggl(1+\dfrac{1}{m^2}\biggr)\ln(1+m^2)\biggr].\label{eq:C-26}
\end{align}
As seen from this expression, in order for $\varepsilon_1$ to be non-zero, $D_{\rm vertex}$ must have an imaginary part.

\subsection{Calculation of wavefunction diagrams}

Regarding the wavefunction diagrams, there appears two types of contributions: wavefunction I and wavefunction II.
However, only the former contributes to the CP asymmetry (this will be verified later).
Additionally, be mindful of the factor two which appears below.
This arises when two different particle species belonging to the $\mathrm{SU}(2)$ group appear in the loop.
For example, when fix the decay channel to $N_{\alpha}\to\nu_{\alpha}\phi^0$,
the particles appearing in the triangular loop in the vertex diagrams are fixed to $\phi^0$ and $\nu_{\beta}$, since the vertex is connected to the final state.
On the other hand, in the wavefunction diagrams, there is no constraint on the particles entering the loop from the final state, and hence both $\nu_{\alpha}\phi^0$ and $e_{\alpha}\phi^+$ contribute.

The interference term of the wavefunction diagram I is
\begin{align}
&I_{\rm wave}^{(1)}\notag\\
&=\dfrac{1}{2}\sum_{s,t}\Bigl(\mathrm{i}\mathcal{M}_{N_1\to l_{\alpha}\phi}^{\mbox{\scriptsize (tree)}(s,t)}\Bigr)^{\ast}\Bigl(\mathrm{i}\mathcal{M}_{N_1\to l_{\alpha}\phi}^{\mbox{\scriptsize (1-loop)}(s,t)}\Bigr)\notag\\
&=\mathrm{i}y_{\alpha1}^{\ast}y_{\beta1}^{\ast}y_{\alpha\gamma}y_{\beta\gamma}\dfrac{2M_1M_{\gamma}}{M_1^2-M_{\gamma}^2}\int\!\!\dfrac{d^4q}{(2\pi)^4}\;\dfrac{p_1\cdot q}{q^2(p_1+p_2+q)^2}\notag\\
&\equiv y_{\alpha1}^{\ast}y_{\beta1}^{\ast}y_{\alpha\gamma}y_{\beta\gamma}D_{\rm wave}^{(1)},\label{eq:C-27}
\end{align}
where we used (\ref{eq:E-6}), the spin sum formula (\ref{eq:E-23}), and the trace formulas (\ref{eq:E-10}), (\ref{eq:E-11}) and (\ref{eq:E-14}). 
Furthermore, applying Feynman parametrization (\ref{eq:E-25}) and the dimensional regularization formula (\ref{eq:E-28}), we perform the $q$-integration and extract the imaginary part, obtaining
\begin{align}
&D_{\rm wave}^{(1)}\notag\\
&=-\dfrac{2M_1M_{\gamma}}{M_1^2-M_{\gamma}^2}\int\!\!\dfrac{d^4q}{\mathrm{i}(2\pi)^4}\;\dfrac{p_1\cdot q}{q^2(q+p_1+p_2)^2}\notag\\
&=-\dfrac{1}{16\pi^2}\dfrac{M_1^3M_{\gamma}}{M_1^2-M_{\gamma}^2}\int_{0}^{1}dx\;x\log\bigl[-M_1^2x(1-x)-\mathrm{i}\epsilon\bigr],\label{eq:C-28}
\end{align}
i.e.,
\beq
\label{eq:C-29}
\Im(D_{\rm wave}^{(1)})
=\dfrac{1}{32\pi}\dfrac{M_1^3M_{\gamma}}{M_1^2-M_{\gamma}^2}.
\eeq
Here note that $\log\bigl[-M_1^2x(1-x)-\mathrm{i}\epsilon\bigr]$ is specified with the integration contour lying in the lower half-plane, thus yielding $-\mathrm{i}\pi$ instead of $+\mathrm{i}\pi$.
From this and equation (\ref{eq:C-28}), we obtain
\beq
\label{eq:C-30}
I_{\rm wave}^{(1)}
=\mathrm{i}y_{\alpha1}^{\ast}y_{\beta1}^{\ast}y_{\alpha\gamma}y_{\beta\gamma}\dfrac{M_1^2}{32\pi}\dfrac{m}{m^2-1},\quad
m\equiv\dfrac{M_1}{M_{\gamma}}.
\eeq

The interference term of the wavefunction diagram II is
\begin{align}
&I_{\rm wave}^{(2)}
=\dfrac{1}{2}\sum_{s,t}
\Bigl(\mathrm{i}\mathcal{M}_{N_1\to l_{\alpha}\phi}^{\mbox{\scriptsize (tree)}(s,t)}\Bigr)^{\ast}\Bigl(\mathrm{i}\mathcal{M}_{N_1\to l_{\alpha}\phi}^{\mbox{\scriptsize (1-loop)}(s,t)}\Bigr)\notag\\
&=\mathrm{i}y_{\alpha1}^{\ast}y_{\beta1}y_{\alpha\gamma}
y_{\beta\gamma}^{\ast}\dfrac{2M_1^2}{M_1^2-M_{\gamma}^2}
\int\!\!\dfrac{d^4q}{(2\pi)^4}\;
\dfrac{p_2\cdot q}{q^2(p_1+p_2+q)^2},
\label{eq:C-31}
\end{align}
where we have taken into account equation (\ref{eq:E-6}), the spin sum formula (\ref{eq:E-23}) and the trace formulas (\ref{eq:E-12}) and (\ref{eq:E-15}).
Applying Feynman parametrization (\ref{eq:E-25}) and the dimensional regularization formula (\ref{eq:E-28}), we perform the $q$-integration in a similar way to those in equation (\ref{eq:C-30}), obtaining
\beq
\label{eq:C-32}
I_{\rm wave}^{(2)}
=\mathrm{i}y_{\alpha1}^{\ast}y_{\beta1}y_{\alpha\gamma} y_{\beta\gamma}^{\ast}\dfrac{M_1^2}{32\pi}\dfrac{m^2}{m^2-1},\quad
m\equiv\dfrac{M_1}{M_{\gamma}}.
\eeq
By summing equations (\ref{eq:C-30}) and (\ref{eq:C-32}),
similarly to equation (\ref{eq:C-26}), we find
\begin{align}
(\varepsilon_1)_{\rm wave}
&=\dfrac{\sum_{\alpha}\Bigl(\Bigl|\mathrm{i}\overline{\mathcal{M}}_{N_1\to l_{\alpha}\phi}^{\mbox{\scriptsize (tree+1-loop)}}\Bigr|^2-\Bigl|\mathrm{i}\overline{\mathcal{M}}_{N_1\to\bar{l}_{\alpha}\phi^{\ast}}^{\mbox{\scriptsize (tree+1-loop)}}\Bigr|^2\Bigr)}{\sum_{\alpha}\Bigl(\Bigl|\mathrm{i}\overline{\mathcal{M}}_{N_1\to l_{\alpha}\phi}^{\mbox{\scriptsize (tree)}}\Bigr|^2+\Bigl|\mathrm{i}\overline{\mathcal{M}}_{N_1\to\bar{l}_{\alpha}\phi^{\ast}}^{\mbox{\scriptsize (tree)}}\Bigr|^2\Bigr)}\notag\\
&=\dfrac{\Im(y_{\alpha1}y_{\beta1}y_{\alpha\gamma}^{\ast}y_{\beta\gamma}^{\ast})}{8\pi\sum_{\alpha}|y_{\alpha1}|^2}\cdot\dfrac{m}{m^2-1}\notag\\
&\quad
+\dfrac{\Im(y_{\alpha1}y_{\beta1}^{\ast}y_{\alpha\gamma}^{\ast}y_{\beta\gamma})}{8\pi\sum_{\alpha}|y_{\alpha1}|^2}\cdot\dfrac{m^2}{m^2-1}.\label{eq:C-33}
\end{align}
Here note that
\begin{align}
\Im(y_{\alpha1}y_{\beta1}^{\ast}y_{\alpha\gamma}^{\ast}y_{\beta\gamma})
&=\Im\Biggl[\sum_{\alpha}(y_{\alpha1}y_{\alpha\gamma}^{\ast})
\sum_{\beta}(y_{\beta1}^{\ast}y_{\beta\gamma})\Biggr]\notag\\
&=\Im\Biggl(\biggl|\sum_{\alpha}y_{\alpha1}y_{\alpha\gamma}^{\ast}
\biggr|^2\Biggr)=0,
\label{eq:C-34}
\end{align}
from which follows that the contribution from wavefunction diagram II is zero.
Thus we find
\beq
\label{eq:C-35}
(\varepsilon_1)_{\rm wave}
=\dfrac{\Im(y_{\alpha1}y_{\beta1}y_{\alpha\gamma}^{\ast}y_{\beta\gamma}^{\ast})}{8\pi\sum_{\alpha}|y_{\alpha1}|^2}\cdot\dfrac{m}{m^2-1},\qquad
m\equiv\dfrac{M_1}{M_{\gamma}}.
\eeq

\subsection{Final result and approximate expression}

By combining equations (\ref{eq:C-26}) and (\ref{eq:C-35}), the expression for the CP asymmetry is obtained as
\begin{align}
\varepsilon_1
&=\dfrac{\Im(y_{\alpha1}y_{\beta1}y_{\alpha\gamma}^{\ast}y_{\beta\gamma}^{\ast})}{8\pi\sum_{\alpha}|y_{\alpha1}|^2}\notag\\
&\quad\times
\dfrac{1}{m}\biggl[1+\dfrac{m^2}{m^2-1}-\biggl(1+\dfrac{1}{m^2}\biggr)\ln(1+m^2)\biggr].\label{eq:C-36}
\end{align}
At this point we switch to $x\equiv M_{\gamma}^2/M_1^2$ in place of $m\equiv M_1/M_{\gamma}$ to match the literature
\begin{align}
\varepsilon_1
&=\dfrac{\Im(y_{\alpha1}y_{\beta1}y_{\alpha\gamma}^{\ast}y_{\beta\gamma}^{\ast})}{8\pi\sum_{\alpha}|y_{\alpha1}|^2}f(x),\qquad
x\equiv\dfrac{M_{\gamma}^2}{M_1^2},\notag\\
f(x)&\equiv\sqrt{x}\,\biggl[1+\dfrac{1}{1-x}-(1+x)\ln\biggl(1+\dfrac{1}{x}\biggr)\biggr].\label{eq:C-37}
\end{align}
Note that $\Im(y_{\alpha1}y_{\beta1}y_{\alpha\gamma}^{\ast}y_{\beta\gamma}^{\ast})=0$ and hence $\varepsilon_1=0$ for $\gamma=1$.
Thus the contributions from the second and third generations are necessary to generate nonzero $\varepsilon_1$.

\begin{figure}[t]
\centering
\includegraphics[width=\linewidth]{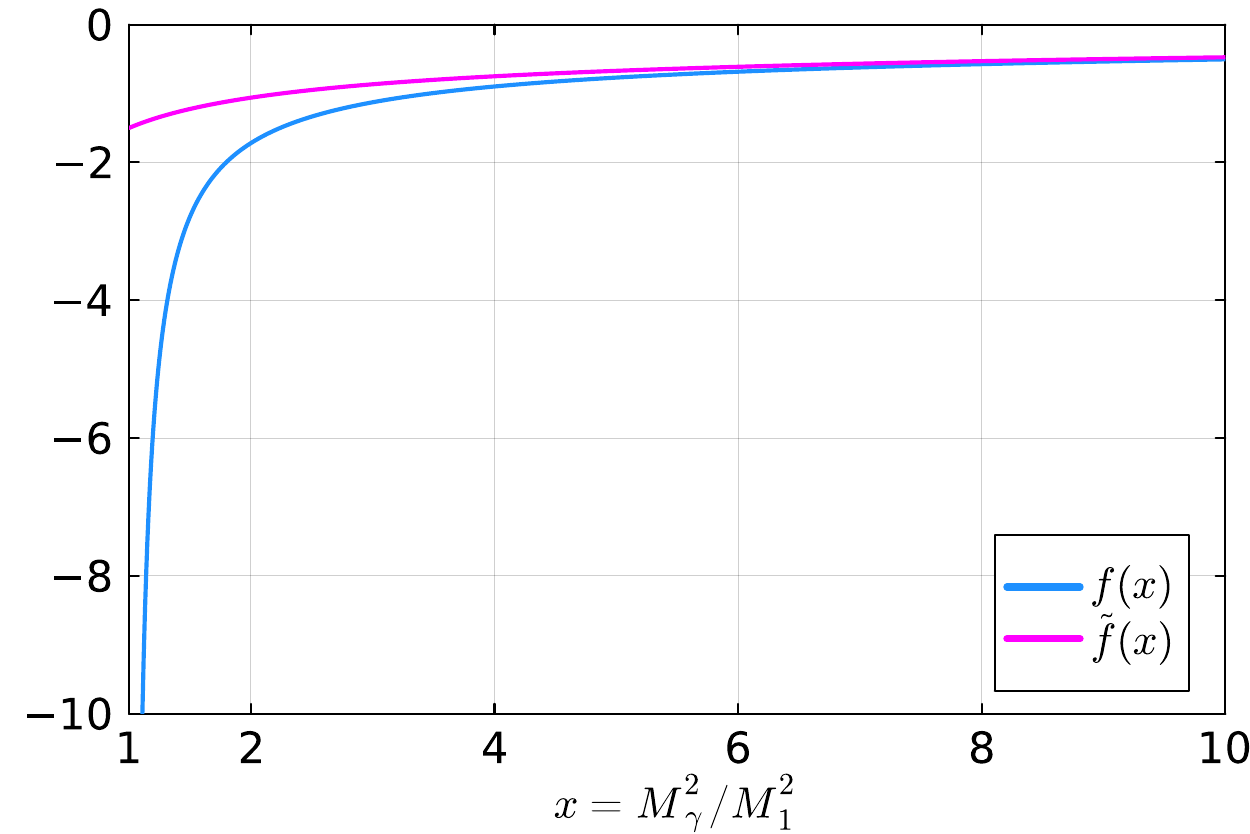}
\caption{\small
Functions $f(x)$ (blue) and $\tilde{f}(x)\equiv -3/(2\sqrt{x})$ (magenta).}
\label{fig:4}
\end{figure}

When the magnitude of the CP asymmetry $\varepsilon_1$ is much smaller than one, $x=M_{\gamma}^2/M_1^2$ is expected to be much larger than unity.
In this case, we perform a Taylor expansion of $f(x)$ for $1/x\ll 1$:
\beq
\label{eq:C-38}
f(x)
=-\biggl[\dfrac{3}{2\sqrt{x}}+\dfrac{5}{6x^{3/2}}+\mathcal{O}\biggl(\dfrac{1}{x^{5/2}}\biggr)\biggr],
\eeq
from which we define $\tilde{f}$ taking the leading term
\beq
\label{eq:C-39}
\tilde{f}(x)
\equiv-\dfrac{3}{2\sqrt{x}}.
\eeq
The plots for $f(x)$ and $\tilde{f}(x)$ are shown in Figure~\ref{fig:4}.
When approximating $f(x)$ with $\tilde{f}(x)$, the CP asymmetry (\ref{eq:C-37}) is approximated as
\beq
\label{eq:C-40}
\varepsilon_1
\simeq-\dfrac{3M_1}{16\pi\sum_{\alpha}|y_{\alpha1}|^2}\Im\biggl(y_{\alpha1}y_{\alpha\gamma}^{\ast}\dfrac{1}{M_{\gamma}}y_{\beta1}y_{\beta\gamma}^{\ast}\biggr).
\eeq

\section{Derivation of the maximal CP asymmetry \texorpdfstring{$|\varepsilon_1|^{\rm max}$}{varepsilon1absmax}}
\label{Appendix:D}

From the Lagrangian (\ref{eq:I-4}), the neutrino masses are given by
\cite{Buchmuller:2002rq}
\beq
\label{eq:D-1}
\mathcal{L}_m
=-\dfrac{M_{\alpha\beta}}{2}\bar{N}_{\alpha}^cN_{\beta}-y_{\alpha\beta}\phi^{\ast}\bar{l}_{\alpha}\bar{N}_{\beta}+{\rm h.c.},
\eeq
where $M$ is the heavy neutrino mass matrix.
Here $(m_{\rm D})_{\alpha\beta}=y_{\alpha\beta}v$ from the Higgs mechanism, where $m_{\rm D}$ is the Dirac mass matrix, $y_{\alpha\beta}$ is the Yukawa coupling, and $v=\langle\phi\rangle$ is the VEV of the Higgs field $\phi$.
We adopt the mass eigenstate basis for the heavy neutrinos, in which $M$ is diagonal with real positive eigenvalues $M_1\leqslant M_2\leqslant M_3$.

Through the seesaw mechanism, the light neutrino mass matrix $m_{\nu}$ can be expressed in terms of $m_{\rm D}$ and the inverse of $M$,
\beq
\label{eq:D-2}
m_{\nu}=-m_{\rm D}\dfrac{1}{M}m_{\rm D}^{\intercal},
\eeq
where higher order terms in $1/M$ are neglected.
$m_{\nu}$ can be diagonalized by a unitary matrix $U^{(\nu)}$
\beq
\label{eq:D-3}
U^{(\nu)\dagger}m_{\nu}U^{(\nu)\ast}
=-\begin{pmatrix}m_1&0&0\\0&m_2&0\\0&0&m_3\end{pmatrix}
\equiv -D_m,
\eeq
with real positive eigenvalues $m_1\leqslant m_2\leqslant m_3$.
Substituting equation (\ref{eq:D-2}) into equation (\ref{eq:D-3}), we get
\begin{align}
&U^{(\nu)\dagger}
\bigl(-m_{\rm D}M^{-1}m_{\rm D}^{\intercal}\bigr)
U^{(\nu)\ast}\notag\\
&=-U^{(\nu)\dagger}(yv)D_M^{-1}(yv)^{\intercal}U^{(\nu)\ast}\notag\\
&=-v^2U^{(\nu)\dagger}yD_M^{-1}y^{\intercal}U^{(\nu)\ast}=-D_m,\label{eq:D-4}
\end{align}
where $D_M$ is the diagonalized heavy mass matrix $M$.
Equation (\ref{eq:D-4}) means that
\beq
\label{eq:D-5}
\Omega
\equiv vD_m^{-1/2}U^{(\nu)\dagger}yD_M^{-1/2}
\eeq
is an orthogonal matrix $\Omega\Omega^{\intercal}=\bm{1}$.
This implies that $\Im(\Omega^{\intercal}\Omega)_{11}=0$, which yields
\begin{align}
0
&=\Im(\Omega^{\intercal}\Omega)_{11}\notag\\
&=\Im\bigl(vD_M^{-1/2}y^{\intercal}U^{(\nu)\ast}D_m^{-1/2}\cdot vD_m^{-1/2}U^{(\nu)\dagger}yD_M^{-1/2}\bigr)_{11}\notag\\
&=v^2\dfrac{1}{M_1}\Im\bigl(D_m^{-1}y^{\intercal}U^{(\nu)\ast}\cdot U^{(\nu)\dagger}y\bigr)_{11}\notag\\
&=v^2\dfrac{1}{M_1}\sum_{\alpha=1,2,3}\dfrac{1}{m_{\alpha}}\Im\bigl(U^{(\nu)\dagger}y\bigr)_{\alpha1}^2,\label{eq:D-6}
\end{align}
that is,
\beq
\label{eq:D-7}
\dfrac{1}{m_1}\Im\bigl(U^{(\nu)\dagger}y\bigr)_{11}^2
=-\sum_{\alpha\neq 1}\dfrac{1}{m_{\alpha}}\Im\bigl(U^{(\nu)\dagger}y\bigr)_{\alpha1}^2.
\eeq

On the other hand, the CP asymmetry $\varepsilon_1$ is obtained from equation (\ref{eq:C-40})
\beq
\label{eq:D-8}
\varepsilon_1
\simeq-\dfrac{3}{16\pi}\dfrac{M_1}{(y^{\dagger}y)_{11}}\Im\biggl(y^{\dagger}y\dfrac{1}{M}y^{\intercal}y^{\ast}\biggr)_{11},
\eeq
where $M$ is the heavy neutrino mass matrix.
Substituting equation (\ref{eq:D-3}) and equation (\ref{eq:D-7}) into (\ref{eq:D-8}), and expressing it in terms of
\beq
\label{eq:D-9}
\tilde{y}=U^{(\nu)\dagger}y
\eeq
instead of $y$, we obtain
\begin{align}
\Im\biggl(\tilde{y}^{\dagger}\tilde{y}\dfrac{1}{M}\tilde{y}^{\intercal}\tilde{y}^{\ast}\biggr)_{11}
&=-\dfrac{1}{v^2}\Im\biggl(\tilde{y}^{\dagger}U^{(\nu)\dagger}m_{\nu}U^{(\nu)\ast}\tilde{y}^{\ast}\biggr)_{11}\notag\\
&=-\dfrac{m_1}{v^2}\Im(\tilde{y}_{11}^2)-\sum_{\alpha\neq 1}\dfrac{m_{\alpha}}{v^2}\Im(\tilde{y}_{\alpha1}^2)\notag\\
&=-\dfrac{1}{v^2}\sum_{\alpha\neq 1}\dfrac{\Delta m_{\alpha1}^2}{m_{\alpha}}\Im(\tilde{y}_{\alpha1}^2),
\label{eq:D-10}
\end{align}
that translates into
\beq
\label{eq:D-11}
\varepsilon_1
=\dfrac{3}{16\pi}\dfrac{M_1}{v^2}\sum_{\alpha\neq 1}\dfrac{\Delta m_{\alpha1}^2}{m_{\alpha}}\dfrac{\Im(\tilde{y}_{\alpha1}^2)}{(\tilde{y}^{\dagger}\tilde{y})_{11}},
\eeq
where $\Delta m_{\alpha1}^2\equiv m_{\alpha}^2-m_1^2$.

Now, consider the normalized Yukawa couplings~\cite{Buchmuller:2003gz}
\beq
\label{eq:D-14}
z_{\alpha}=\dfrac{\tilde{y}_{\alpha1}^2}{(\tilde{y}^{\dagger}\tilde{y})_{11}}
=X_{\alpha}+\mathrm{i}Y_{\alpha},
\eeq
which follow
\beq
\label{eq:D-15}
0\leqslant|z_{\alpha}|\leqslant 1,\qquad
\sum_{\alpha}|z_{\alpha}|=1.
\eeq
From the orthogonality condition $(\Omega^{\intercal}\Omega)_{11}=1$, the additional constraint is given by
\beq
\label{eq:D-16}
\sum_{\alpha}\dfrac{\tilde{m}_1}{m_{\alpha}}z_{\alpha}=1,
\eeq
where $\tilde{m}_1$ is the effective neutrino mass (\ref{eq:III-0}).
From equation (\ref{eq:D-11}), the CP asymmetry reads in the new variables
\beq
\label{eq:D-17}
\varepsilon_1
=\dfrac{3}{16\pi}\dfrac{M_1}{v^2}
\biggl(\dfrac{\Delta m_{21}^2}{m_2}Y_2+\dfrac{\Delta m_{31}^2}{m_3}Y_3\biggr).
\eeq
Since $m_3>m_2$,
\begin{align}
\dfrac{\Delta m_{31}^2}{m_3}-\dfrac{\Delta m_{21}^2}{m_2}
&=\dfrac{m_3^2-m_1^2}{m_3}-\dfrac{m_2^2-m_1^2}{m_2}\notag\\
&=\dfrac{(m_3-m_2)(m_1^2+m_2m_3)}{m_2m_3}>0\notag\\
\Rightarrow\quad
\dfrac{\Delta m_{31}^2}{m_3}
&>\dfrac{\Delta m_{21}^2}{m_2}.\label{eq:D-18}
\end{align}
From this, we can presume that the maximal CP asymmetry
is achieved when $|Y_3|$ is maximal.

The additional condition (\ref{eq:D-16}) yields
\begin{align}
&\dfrac{Y_1}{m_1}+\dfrac{Y_2}{m_2}+\dfrac{Y_3}{m_3}=0,
\label{eq:D-19}\\
&\dfrac{\tilde{m}_1}{m_1}X_1+\dfrac{\tilde{m}_1}{m_2}X_2+\dfrac{\tilde{m}_1}{m_3}X_3=1.\label{eq:D-20}
\end{align}
Since $m_3>m_2>m_1$, we can set $X_2=X_3=Y_2=0$ for maximal $|Y_3|$.
Then, it follows that
\beq
\label{eq:D-21}
Y_1=-\dfrac{m_1}{m_3}Y_3,\qquad
X_1=\dfrac{m_1}{\tilde{m}_1}.
\eeq
The second equation in (\ref{eq:D-15}) gives
\beq
\label{eq:D-22}
\sqrt{X_1^2+Y_1^2}+|Y_3|=1.
\eeq
The conditions (\ref{eq:D-21}) and (\ref{eq:D-22}) determine $|Y_3|$ as a function of $m_1$, $m_3$, and $\tilde{m}_1$:
\begin{align}
&\biggl(1-\dfrac{m_1^2}{m_3^2}\biggr)|Y_3|^2-2|Y_3|+1-\dfrac{m_1^2}{\tilde{m}_1^2}=0\notag\\[0.25em]
&\Rightarrow\quad|Y_3|=1-\dfrac{m_1}{m_3}\sqrt{1+\dfrac{m_3^2-m_1^2}{\tilde{m}_1^2}},
\label{eq:D-23}
\end{align}
where $|Y_3|<1$ is taken into account.
Substituting $Y_2=0$ and equation (\ref{eq:D-23}) into equation (\ref{eq:D-17}) yields
\beq
\label{eq:D-24}
|\varepsilon_1|^{\rm max}
=\dfrac{3}{16\pi}\dfrac{M_1m_3}{v^2}\Biggl[1-\dfrac{m_1}{m_3}\biggl(1+\dfrac{m_3^2-m_1^2}{\tilde{m}_1^2}\biggr)^{1/2}\Biggr],
\eeq
where $m_3^2-m_1^2<m_3^2$ is used.
This gives the upper bound for $|\varepsilon_1|$.

\section{Formulas for Feynman diagrams}

\subsection{Formulas for \texorpdfstring{$\gamma$}{gamma} matrices}

\subsubsection{Definition and properties of \texorpdfstring{$\gamma$}{gamma} matrices}

The $\gamma$ matrices consist of four $n\times n$ matrices $\gamma^{\mu}$ (where $\mu=0,1,2,3$) and satisfy the following anticommutation relations:
\beq
\label{eq:E-1}
\{\gamma^{\mu},\gamma^{\nu}\}
=\gamma^{\mu}\gamma^{\nu}+\gamma^{\nu}\gamma^{\mu}
=2\eta^{\mu\nu}\bm{1}_{n\times n}.
\eeq
Equation (\ref{eq:E-1}) is called the Clifford algebra.

\subsubsection{Definition and properties of chirality \texorpdfstring{$\gamma_5$}{gamma5}}

We define {\it chirality} $\gamma_5$ as
\beq
\label{eq:E-2}
\gamma_5\equiv\mathrm{i}\gamma^0\gamma^1\gamma^2\gamma^3.
\eeq
$\gamma_5$ satisfies
\beq
\label{eq:E-3}
(\gamma_5)^2=\bm{1},\qquad
(\gamma_5)^{\dagger}=\gamma_5,\qquad
\gamma_5\gamma^{\mu}=-\gamma^{\mu}\gamma_5,
\eeq
for $\mu=0,1,2,3$.

The eigenspaces of $\gamma_5$ with eigenvalues $\pm 1$ are referred to as left-chiral and right-chiral, respectively.
Let $P_{\rm L}$ and $P_{\rm R}$ denote the projection operators onto the left-chiral and right-chiral eigenspaces, respectively.
These satisfy
\beq
\label{eq:E-4}
P_{\rm R}=\dfrac{\bm{1}+\gamma_5}{2},\qquad
P_{\rm L}=\dfrac{\bm{1}-\gamma_5}{2},
\eeq
and decompose the Dirac spinor $\psi$ into
\beq
\label{eq:E-5}
\psi_{\rm R}=P_{\rm R}\psi,\qquad
\psi_{\rm L}=P_{\rm L}\psi.
\eeq
These projection operators satisfy
\beq
\label{eq:E-6}
P_{\rm R}^2=P_{\rm R},\quad
P_{\rm L}^2=P_{\rm L},\quad
P_{\rm R}P_{\rm L}=P_{\rm L}P_{\rm R}=0.
\eeq

\subsubsection{Feynman slash notation}

The contraction between the 4-vector $a^{\mu}$ and the $\gamma$ matrix $\gamma^{\mu}$ is denoted by
\beq
\label{eq:E-7}
\slashed{a}\equiv\gamma^{\mu}a_{\mu}
=\gamma_{\mu}a^{\mu},
\eeq
from which it follows that
\beq
\label{eq:E-8}
\slashed{a}^2=a^2.
\eeq

\subsubsection{Trace formulas}

Traces of gamma matrices are given by
\begin{align}
\tr(\bm{1})
&=4,\label{eq:E-9}\\
\tr(\slashed{a})
&=\tr(\slashed{a}\slashed{b}\slashed{c})=\cdots=0,\label{eq:E-10}\\
\tr(\slashed{a}\slashed{b})
&=(a\cdot b)\tr(\bm{1})=4(a\cdot b),\label{eq:E-11}\\
\tr(\slashed{a}\slashed{b}\slashed{c}\slashed{d})
&=4[(a\cdot b)(c\cdot d)\notag\\
&\qquad-(a\cdot c)(b\cdot d)+(a\cdot d)(b\cdot c)],\label{eq:E-12}\\
\tr(\gamma_5)
&=0,\label{eq:E-13}\\
\tr(\slashed{a}\slashed{b}\gamma_5)
&=0,\label{eq:E-14}\\
\tr(\slashed{a}\slashed{b}\slashed{c}\slashed{d}\gamma_5)
&=-4\mathrm{i}\epsilon^{\mu\nu\rho\sigma}a_{\mu}b_{\nu}c_{\rho}d_{\sigma}.\label{eq:E-15}
\end{align}

\subsection{Properties of Dirac fermions}

The positive frequency solution $u^s(p)$ of the Dirac equation satisfies
\beq
\label{eq:E-16}
(\slashed{p}-m)u^s(p)=0,
\eeq
while the negative frequency solution $v^s(p)$ satisfies
\beq
\label{eq:E-17}
(-\slashed{p}-m)v^s(p)=0,
\eeq
where $p$ is the four-momentum, and $s$ denotes the spin eigenvalue.
From the normalization condition of the solution $u^s(p)$
\beq
\label{eq:E-18}
\bar{u}^s(p)u^{s'}(p)=2m\delta_{ss'},
\eeq
the orthonormality follows
\begin{align}
\bar{v}^s(p)v^{s'}(p)
&=-2m\delta_{ss'},\label{eq:E-19}\\
\bar{u}^s(p)v^{s'}(p)
&=\bar{v}^s(p)u^{s'}(p)=0,\label{eq:E-20}\\
u^{s\dagger}(\bm{p})v^{s'}(-\bm{p})
&=v^{s\dagger}(\bm{p})u^{s'}(-\bm{p})=0,\label{eq:E-21}\\
\bar{u}^s(p)\gamma^{\mu}u^{s'}(p)
&=\bar{v}^s(p)\gamma^{\mu}v^{s'}(p)=2p^{\mu}\delta_{ss'}.\label{eq:E-22}
\end{align}
Also, $u^{s=\pm}(p)$ and $v^{s=\pm}(p)$ form complete two-dimensional solution spaces satisfying $\slashed{p}=m$ and $-\slashed{p}=m$, respectively.
The projection operators onto these solution spaces are given by $(\pm\slashed{p}+m)/2m$, which yield
\begin{align}
\sum_{s}u^s(p)\bar{u}^s(p)&=\slashed{p}+m,\label{eq:E-23}\\
\sum_{s}v^s(p)\bar{v}^s(p)&=\slashed{p}-m.\label{eq:E-24}
\end{align}
The proportionality coefficients are determined by taking into account the normalization conditions given in equations (\ref{eq:E-18}) and (\ref{eq:E-19}).

\subsection{Formulas for Feynman integrals}

\subsubsection{Feynman parametrization}

Feynman parametrization is given by
\begin{align}
\dfrac{1}{ab}
&=\int_{0}^{1}dx\;\dfrac{1}{[ax+b(1-x)]^2},\label{eq:E-25}
\\[0.25em]
\dfrac{1}{abc}
&=\int_{0}^{1}dx\int_{0}^{1}dy\int_{0}^{1}dz\;
\dfrac{2\,\delta(x+y+z-1)}{(ax+by+cz)^3}\notag\\[0.25em]
&=\int_{0}^{1}dx\int_{0}^{1-x}dy\;
\dfrac{2}{[ax+by+c(1-x-y)]^3}.\label{eq:E-26}
\end{align}

\subsubsection{Momentum integrals with dimensional regularization}

Momentum integrals with dimensional regularization are given by
\begin{align}
&\int\!\!\dfrac{d^nk}{(2\pi)^n}\;\dfrac{1}{(m^2+2k\cdot p-k^2)^{\alpha}}
\notag\\
&\hspace{8em}
=\dfrac{\mathrm{i}\varGamma\bigl(\alpha-\frac{n}{2}\bigr)}{(4\pi)^{\frac{n}{2}}\varGamma(\alpha)}\dfrac{1}{(m^2+p^2)^{\alpha-\frac{n}{2}}},\label{eq:E-27}\\
&\int\!\!\dfrac{d^nk}{(2\pi)^n}\;
\dfrac{k^{\mu}}{(m^2+2k\cdot p-k^2)^{\alpha}}\notag\\
&\hspace{8em}
=\dfrac{\mathrm{i}\varGamma\bigl(\alpha-\frac{n}{2}\bigr)}
{(4\pi)^{\frac{n}{2}}\varGamma(\alpha)}\dfrac{p^{\mu}}{(m^2+p^2)^{\alpha-\frac{n}{2}}}.\label{eq:E-28}
\end{align}
Setting $\varepsilon\equiv(4-n)/2\ll 1$, the gamma function gives
\beq
\label{eq:E-34}
\varGamma(\varepsilon)
=\dfrac{1}{\varepsilon}-\gamma+\dfrac{1}{2}\biggl(\gamma^2+\dfrac{\pi^2}{6}\biggr)\varepsilon+\mathcal{O}(\varepsilon^2),
\eeq
where $\gamma$ is the Euler-Mascheroni constant.

\bibliography{main}

\end{document}